\documentclass{pasa}%

\usepackage{supertabular,booktabs}
\usepackage{graphicx}
\usepackage{multirow}
\usepackage{graphicx}
\usepackage{pdflscape}
\usepackage{txfonts}
\usepackage{natbib}
\usepackage{float}
\usepackage{xcolor}
\usepackage{rotating} 
\usepackage{adjustbox}
\usepackage{subfig}
\usepackage{etoolbox}
\usepackage{longtable}
\usepackage[toc]{appendix}
\usepackage{ulem}

\title{Investigation of the double-lobed sources of the Cygnus constellation core}

\author[J. Saponara et al.]{J. Saponara$^1$, P. Benaglia$^1$, I. Andruchow$^{2,3}$, C. H. Ishwara-Chandra$^{4}$, H. T. Intema$^{5,6}$

\affil{$^1$ Instituto Argentino de Radioastronom\'{\i}a, CONICET-CICPBA-UNLP, CC5 (1897) Villa Elisa, Prov. de Buenos Aires, Argentina}

\affil{$^2$ Facultad de Cs. Astron\'{o}micas y Geof\'{\i}sicas, Universidad Nacional de La Plata}

\affil{$^3$ Instituto de Astrof\'{\i}sica de La Plata (UNLP - CONICET, CCT La Plata)}

\affil{$^4$ National Centre for Radio Astrophysics, Tata Institute of Fundamental Research, Pune University Campus, Pune, 411007, India}

\affil{$^5$ International Centre for Radio Astronomy Research, Curtin University, Bentley, WA 6102, Australia}

\affil{$^6$ Leiden Observatory, Leiden University, Niels Bohrweg 2, 2333 CA Leiden, the Netherlands}
}

\jid{PASA}
\doi{10.1017/pas.\the\year.xxx}
\jyear{\the\year}

\usepackage{aas_macros}
\usepackage{hyperref} 
\hypersetup{colorlinks,citecolor=blue,linkcolor=blue,urlcolor=blue}

\begin{document}

\begin{frontmatter}
\maketitle

\begin{abstract}

We present a collection of double-lobed sources towards a 20~sq~deg area of the Cygnus region at the northern sky, observed at 325 and 610~MHz with the Giant Metrewave Radio Telescope. The 10$''$ resolution achieved at 325~MHz is 5.5 times better than previous studies, while at 610~MHz these are the first results ever of such a large area, mapped with 6$''$ angular resolution. After a thorough visual inspection of the images at the  two bands, we found 43 double-lobed source candidates, proposed as such due to the presence of two bright peaks, within a few arcminutes apart, joined by a bridge or a central nucleus. All but two are presented here as a double-lobed candidates for the first time. 
Thirty-nine of the candidates were covered at both bands, and we provide the spectral index information for them.  We have searched for positional coincidences between the detected sources/components and other objects from the literature, along the electromagnetic spectrum. Twenty-three candidates possess radio counterpart(s), 12 present infrared counterparts, and one showed an overlapping X-ray source. We analysed each  candidate considering morphology, counterparts, and spectral indices. 
Out of the 43 candidates, 37 show characteristics compatible with an extragalactic nature, two of probably Galactic origin, three remain as dubious cases, though with feature(s) compatible with an extragalactic nature, and the remaining one, evidence of physically unrelated components. The median spectral index of the 40 putative extragalactic sources is $-1.0$. Their celestial surface density at 610~MHz resulted in 1.9 per sq deg, across a region lying at the Galactic plane.
\end{abstract}

\end{frontmatter}

\section{INTRODUCTION }
\label{sec:intro}

The Cygnus constellation covers a relevant area of the northern  sky and  harbours many  stellar  associations and clusters.
The Cygnus OB2 association, at its centre, is a well studied rich region with several massive stars, as known since the early work of \cite{Munch-1953A,Schulte-1956a,Schulte-1956-b,Schulte-1958,Reddish-1966,K-2002,Albacete-2007}; see also the review of \cite{Reipurth-2008}. Recent radio results at arcsecond resolution were presented in \cite{Benaglia-2020b}, further including Cyg\,OB8 and OB9, using Giant Metrewave  Radio Telescope  (GMRT) data.
Some massive stars towards these regions show significant radio emission,  like the Wolf-Rayet stars WR\,146, 147 for instance, with their colliding winds, that generate non-thermal (NT) radio fluxes \citep[e.g., ][and many references therein]{Williams1997,Benaglia-2020a}. 
Besides massive, early-type stars, other types of objects were studied in this region, like the proto-planetary disk-like sources \citep{Isequilla-2019} and young stellar objects \citep[YSO,][]{Isequilla-2020}. Some types of Galactic sources, related to stellar systems at different stages, present jet/outflows morphology. Interesting examples are those related to protostars, YSO and Herbig-Haro objects \citep[HH, see][the latest review]{Anglada2018}. In  the  case of HH80/81 and jets, besides NT  spectral indices as in other cases, \cite{carrasco2010} measured important linear polarisation, confirming synchrotron  emission. We inspect the GMRT images used by \cite{Benaglia-2020b} to build their catalogue. We identified several faint radio sources with double-lobed mophologies; many of them observed for the first time. And, even though they are towards the Galactic plane,  some of them might have an extragalactic nature. 

Regular members of the extragalactic zoo, the Active Galactic Nuclei (AGN) are notably strong sources with a spectral energy distribution (SED) ranging from gamma to radio wavelengths \citep{Brown-2019}; generally the SED presents a peak at UV and significant luminosity contribution at the infrared and X-rays wavelengths. Radio galaxies are a subclass of AGNs. The radio continuum maps usually reveal a bright centre (core), jets and lobes whose extension can reach Mpc.
The radio galaxies were classified into two main kinds, the Fanaroff-Riley I and II \cite[FRI, FRII,][]{Fanaroff-1974}. The FRIs present their low brightness regions further from the core than their high brightness regions, while it is in the other way round for the FRIIs. At the FRII high brightness regions, the jets are decelerated by the interaction with the circumgalactic medium. Both classic examples are M84 and 3C175, respectively \citep[e.g.,][]{Laing1983}.
But the classifications account for the standard type of radio galaxies. Many others with different morphologies, such us X-shapes \citep{Leahy-1984}, collimated jets, and also some of them with inner double-lobed radio structure as well as a larger outer double-lobed structure, called  ‘double-double’ radio galaxies (DDRGs)
\citep{Arno-2000,Brocksopp-2011}, were also reported.
In addition, the majority of the studies were performed out of the Galactic plane. The optical identification is difficult against a region crowded with emitters at that spectral range, and the lack of high resolution and deep radio images are a problem as well.\\ 

An investigation of double-lobed radio sources in the Cygnus region was lacking. In this work, we present the study of the double-lobed type sources in the central part of the Cygnus region, near the Galactic plane, to characterise those with extragalactic features;  many of them are reported for the first time in this work. GMRT observations are introduced in Section 2.  Section 3 describes the identification processes, and Section 4 the search for counterparts. The discussion is presented in Section 5, and the main conclusions are listed in Section 6.

\section{RADIO OBSERVATIONS}

Along the past years, different radio surveys were focused on the Galactic plane and thus on the Cygnus area. \cite{Garwood-1988}
using the Very Large Array (VLA) carried out continuum observations at 1.4~GHz ($b=0^\circ$). The resolution achieved in this survey was up to 4$''$ and the findings complete to about 30-mJy peak in flux density. The results were complemented by \cite{zoon-1990}, for $|b|<0.8^\circ$, with similar angular resolution and flux limit. In 1996, with the Texas Interferometer at 365~MHz, \cite{Doublges-1996} imaged the area at arcminutes scale above flux densities of $0.25-0.4$~Jy; and \cite{Taylor-1996}  carried out Westerbork Synthesis Radio Telescope (WSRT) observations at 327~MHz for $|b|<1.6^\circ$ (WSRT  Galactic Plane survey or WSRTGP), achieving an angular resolution of $\sim1'$ and detecting sources brighter than 10~mJy~beam$^{-1}$. This same interferometer was used to observe the Cyg~OB2 region and build the 325 and 1400~MHz continuum survey, published by \cite{Gunawan-2003}. The attained angular resolutions were 13$''$ and 55$''$ with $5\sigma$ flux density limits of $\sim$2~mJy and ~$10-15$~mJy, respectively.
Very recently, \cite{morford2020} published results of a deep field mapping of the Cyg\,OB2 association at 21~cm, attaining an angular  resolution of 180~mas and rms as low as 21~$\mu$Jy.\\

\begin{figure}
    \centering 
    \includegraphics[width=0.5\textwidth]{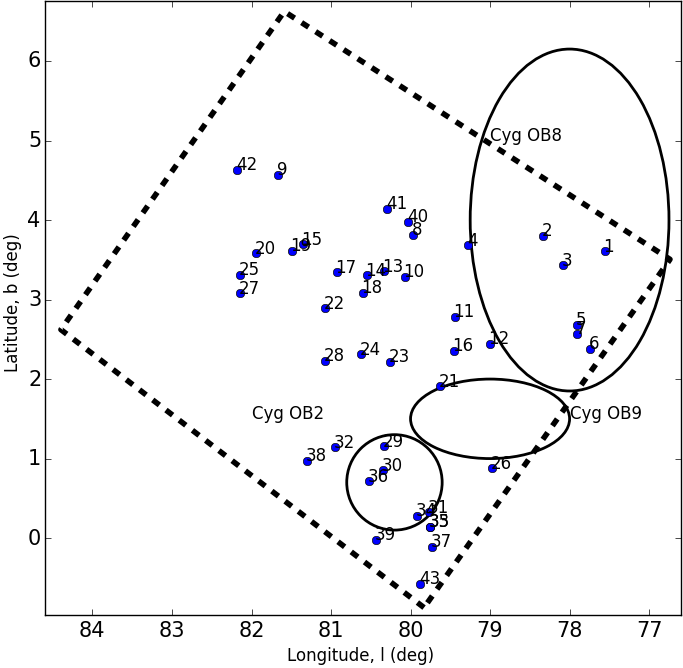}
   \caption{Observed area of the Cygnus constellation marked with a black dashed contour box, over the identified stellar associations Cygnus\,OB2 and Cygnus\,OB8 as well as Cygnus\,OB9. The blue dots are the double-lobed sources presented in this paper, with  ID numbers as in Table\,\ref{long}.}
    \label{fig:reg-obs}
\end{figure}

In order to achieve both high-angular resolution and sensitivity below the mJy threshold, the central 19.7~sq deg of the Cygnus constellation (see Fig.~\ref{fig:reg-obs}) were observed using two bands, centred at 325 and 610~MHz, with the Giant Metrewave Radio Telescope. The data were collected during 172~hours along 2013-2017, using a 32~MHz bandwidth, and were calibrated and processed uniformly, using the SPAM routines \citep{Intema-2014}. For full details on observations, calibration and imaging, see \cite{Benaglia-2020b}. The SPAM routines provided also the final  mosaics, one at  each band, for which a robust weighting  of  $-1$ was chosen. We have made use here of 
the images that are based on  those observations. The mosaics present a mean rms of 0.5~mJy~beam$^{-1}$ and 0.2 mJy~beam$^{-1}$ at 325 and 610~MHz, but we note that the rms value locally varies depending on the presence of extended and/or diffuse emission, up to  0.9 and 0.5~mJy~beam$^{-1}$. The resultant synthesised beams at each frequency at  full resolution were $10'' \times 10''$, and 6$'' \times 6''$. The final mosaic image sizes resulted in ($6487  \times 6573$), and ($12580 \times 13837$) pixels, of sizes 2.5$''$ and 1.5$''$, respectively. We note that this paper is the fourth of a series  derived from the analysis of the Cygnus region through GMRT observations, see \cite{Benaglia-2020a,Benaglia-2020b,Benaglia-2021a}.
Besides the SPAM  mosaics, we used, for some specific regions, complementary images built with the astronomical image processing system \citep[AIPS,][]{greisen2003}, following standard procedures, and different weightings \citep[see][for details]{Benaglia-2021a}.

\section{IDENTIFYING DOUBLE-LOBED SOURCES IN THE CYGNUS REGION}

For the present study, we visually searched in the image mosaics for objects with two bright peaks, within a few arcminutes apart, joined by a bridge or path of emission, either diffuse or intense, or with a central nucleus.
We proposed them as `double-lobed' candidates\footnote{In \cite{Benaglia-2020b}, the authors grouped candidates for double sources (not necessarily bi-lobed), using a different approach to build the sample: the pairs of sources were selected from the Gaussian fits resultant of running a source finding routine, and the selection process constraints consisted of a limiting ratio of component integrated flux densities and a limiting separation (<=2’).}. To take into consideration the diverse local rms, we made individual images of each source, showing simultaneously the continuum emission at both bands  when possible, see Figs.~\ref{fig:galaxitas1}, \ref{fig:galaxitas2}, \ref{fig:galaxitas3}, \ref{fig:galaxitas4}, \ref{fig:galaxitas5} and \ref{fig:galaxitas6}.

The spectral index ($\alpha$) of a source is a key parameter to reveal its nature.
It is defined in terms of the integrated flux density $S$  at frequency $\nu$; here we used  the convention $S_{\nu} \propto \nu^\alpha$. To build the spectral index map of sources, we convolved and regridded the 610~MHz image to exactly the same synthesised beam and grid as the 325~MHz one (thus the pixel size in both maps is the same, $2.5''\times2.5''$). The {\sc miriad} task {\sc cgcurs} was used to estimate the integrated flux densities of the identified sources, within the region defined by 3 times the local rms. The main radio properties of the identified double-lobed sources are listed in Table~\ref{long}, being

\begin{itemize}
\item[] {\it Column (1)} -- Source/candidate number (\#).
\item[] {\it Column (2)} -- Source ID. The sources are listed in increasing order  of right ascension: first those with GMRT data at both bands, and second the ones with only 610~MHz data.
\item[] {\it Column (3)} -- Label corresponding to the type of source centre,
as (B): detected bridge; (Nu): detected nucleus candidate; (G) measured geometric centre. The orientation of the lobes is quoted by means of cardinal celestial points (N, S, E, W, NW, SE, NE and SW).

\item[] {\it Columns (4,5)} -- J2000 position in right ascension and declination of the components identified in Column 3. 

\item[] {\it Columns (6)} -- The total extension of the sources derived considering the distance between the extreme of the outermost contour (=3 rms) of each lobe at 325~MHz. Estimated errors: $\pm$2~arcsec.

\item[] {\it Columns (7)-(9)} -- Integrated flux density of each component, at the 325~MHz image, at the 610~MHz image convolved to the 325~MHz beam, and at the 610~MHz full resolution image, respectively.  The error corresponds to 3$\sigma$ (with $\sigma$, local rms).  When a source was detected at one frequency only, the upper limit --at the other frequency band-- corresponds to 3 times the local rms noise. Whenever was possible to calculate the integrated flux-density of the bridge emission, we measured its flux above 3$\sigma$, checking not to introduce the emission from the lobes. Since all nuclei sizes were encompassed by one beam, their integrated flux densities were calculated over  the beam-size area.
\item[] {\it Column (10)} -- Spectral index $\alpha$ of the component identified in Column 3.

\item[] {\it Column (11)} -- Total spectral index $\alpha_T$ of the double-lobed source candidate; calculated using the flux sum of each source component.

\end{itemize}

\onecolumn

\begin{landscape}
\begin{longtable}[c]{ccrrrcrrrcc}
 \caption{Position, integrated flux density and spectral index of double-lobed sources components, and total spectral index. Nu: nucleus, G: geometric centre, B: bridge. $SC_{610\,{\rm MHz}}$ refers to the integrated flux density obtained from the 610~MHz image convolved to the 325~MHz  beam.}\\
\hline\hline
 \# &ID& &$RA$ (J2000) & $Dec$  (J2000) & Total &$S_{325\,{\rm MHz}}$& $SC_{610\,{\rm MHz}}$& $S_{610\,{\rm MHz}}$ & $\alpha_{610\rm MHz}^{325\rm MHz}$ &$\alpha_T$\\
& & &(hms)& (dms)& size ($''$) &(mJy)& (mJy) & (mJy) &     \\

\hline
\endfirsthead
\caption{continued.}\\

\hline\hline
\# &ID& &$RA$ (J2000) & $Dec$  (J2000) & Total &$S_{325\,{\rm MHz}}$& $SC_{610\,{\rm MHz}}$& $S_{610\,{\rm MHz}}$ & $\alpha_{610\rm MHz}^{325\rm MHz}$ &$\alpha_T$\\
& & &(hms)& (dms)& size ($''$) &(mJy)& (mJy) & (mJy) &     \\
\hline

\endhead
\hline
\endfoot


\#1 & J201255+404423 & B & 20:12:55.00     &  40:44:23.0  &  60  & 4.7 $\pm$ 1 & 3.5 $\pm$ 1.5 & 1.8 $\pm$ 0.4  &  &\multirow{3}{4em}{$-$0.8$\pm$0.4}              \\
  & & N  &      :55.44     &       :33.3  &   & 36  $\pm$ 4   & 22  $\pm$ 4   & 24 $\pm$ 4   &  $-0.8\pm$0.3 &\\    
 &  & S  &      :54.88     &       :01.5  &   & 72  $\pm$ 12  & 42  $\pm$ 7   & 47 $\pm$4    &  $-0.8\pm$0.4 &\\
  
\#2 &J201418+412931     &B? &  20:14:18.40     &  41:29:31.0  &  45  & 5.4 $\pm$ 2   & <0.3          &<0.8         &     &  \multirow{3}{4em}{$-$1.2$\pm$0.3}              \\
 & & NE    &       :19.13     &       :37.6  &  &  51  $\pm$ 1   & 27  $\pm$ 5  & 30 $\pm$ 4  & $-1.1\pm$0.3&  \\
 & & SW   &       :17.43     &       :23.5  &   & 77  $\pm$ 1   & 35  $\pm$ 7  & 39 $\pm$ 5  &  $-1.2\pm$0.3&   \\
                                                    
\#3& J201512+410457    &G  &  20:15:11.20     &  41:04:57.0  & 55  & <0.2       &<0.7       &   <0.08        &                &  \multirow{3}{4em}{$-$0.9$\pm$0.4}      \\
 & & N   &       :11.39     &   :05:10.2  &   & 29  $\pm$ 6   &  17 $\pm$ 2  &   18 $\pm$ 2   &  $-0.8\pm$0.4 &    \\
 & & S   &       :11.07     &    :04:38.5  &  &  14  $\pm$ 2   &  7  $\pm$ 1  &    8 $\pm$ 1   & $-1.1\pm$0.3 &    \\
                                                    
\#4& J201739+421243   & Nu  &  20:17:39.24     &  42:12:43.3  &  66  &2 $\pm$ 0.5   &  3  $\pm$ 1  &   3 $\pm$1   & & \multirow{3}{4em}{$-$2.0$\pm$0.4}   \\
 & & N  &       :40.00     &       :13:12.0  &  &  20 $\pm$ 2    &  5  $\pm$ 2  &  6 $\pm$ 1  &  $-2.2\pm$0.6 &\\
 & & S  &       :38.60     &       :12:26.6  &   & 36 $\pm$ 4    &  8  $\pm$ 1  &  9 $\pm$ 1  &  $-2.4\pm$0.3 &\\

\#5 &J201758+403100   &  Nu  &  20:17:58.00    &  40:31:00.0   &   72  &35 $\pm$ 4    & 14 $\pm$ 2   &  15 $\pm$ 1 && \multirow{3}{4em}{$-$1.3$\pm$0.1}   \\
 & & NE  &       :18:00.16    &	     :31:07.8   &  &   278 $\pm$ 15  &  131 $\pm$ 4  &  138 $\pm$ 4 & $-1.2\pm$0.1 &\\
 & & SW  &       :17:54.96    &      :30:33.5   &  &   217 $\pm$ 11  &  84 $\pm$ 3  &  93 $\pm$ 3 &  $-1.5\pm$0.1  &\\

\#6& J201829+402714    & B  &   20:18:28.61   & 40:27:14.0 &25 & 2  $\pm$ 1    &   7  $\pm$ 3    & 3 $\pm$ 1  &  &\multirow{3}{4em}{$-$0.9$\pm$0.5}     \\  
&  & N  &        :28.46   &      :20.5 &  &48  $\pm$ 8     &  24  $\pm$ 6    & 20 $\pm$ 1 & $-1.1\pm$0.5&  \\
&  & S  &        :28.62   &      :05.6 &  &24  $\pm$ 4     &  10  $\pm$ 2    &  8 $\pm$ 1  & $-1.4\pm$0.4   &  \\

\#7 &J201847+401230    & B  & 20:18:46.64     & 40:12:30.0    &45    &15 $\pm$ 5      &     4 $\pm$ 1   & 3 $\pm$ 1  & &\multirow{3}{4em}{$-$2.2$\pm$0.4} \\
&  & N  &      :46.76     &      :38.9    &  &  100 $\pm$ 20   &  25  $\pm$ 4   & 24 $\pm$ 3 &  $-2.2\pm$0.4& \\  
&  & S  &      :47.05     &      :19.4    &   & 45  $\pm$ 10    &  10  $\pm$ 2   & 9 $\pm$1  &  $-2.4\pm$0.5& \\

\#8& J201910+425136   &  Nu  &  20:19:09.33  & 42:51:36.25 & 120 &   <0.6  &  <0.03   & 0.8 $\pm$ 0.3 &   &\multirow{3}{4em}{$-$1.3$\pm$0.4}                  \\
 & & N  &       :08.01  &   :52:28.4 & & 9 $\pm$ 3  & 3 $\pm$ 1  &  4  $\pm$ 1  & $-1.7\pm$0.7  & \\
 & & S  &       :10.25  &   :50:37.9 & &  180 $\pm$ 30  &  81 $\pm$ 18     & 84 $\pm$12 & $-1.3\pm$0.4   &  \\

\#9& J202058+444050    & G  &  20:20:58.00  & 44:40:50.0    &120 &  <1  & <0.01  & <0.15&   &\multirow{3}{4em}{$-$0.6$\pm$0.1}    \\
 & & N  &       :57.65  & :41:30.5    &  &610 $\pm$ 35 & 404 $\pm$ 16 & 403 $\pm$ 48 & $-0.6\pm$0.1 &  \\
 & & S  &       :57.90  & :40:05.0    & & 20 $\pm$1    & 19 $\pm$1   & 22 $\pm$ 2   & $-0.1\pm$0.1 & \\

\#10& J202151+423847 & G &  20:21:51.00  & 42:38:47.0    & 60 &  0.3 $\pm$ 0.1  &   1.6 $\pm$ 0.2 & 1.3 $\pm$ 0.2 & &\multirow{3}{4em}{$-$0.9$\pm$0.2}  \\
 & & E &       :54.20  &      :32.8    & & 28  $\pm$ 3   & 15  $\pm$ 1     & 19 $\pm$ 1 & $-0.9\pm$0.2  & \\
  && W &       :48.30  &      :57.6    &  &16  $\pm$ 1     &   8  $\pm$ 1  & 8  $\pm$ 0.4   & $-1.1\pm$0.2  &  \\
                   
\#11& J202209+415033     & Nu & 20:22:09.00 &  41:50:33.0     & 60 &  0.7 $\pm$ 0.4 &  1 $\pm$ 0.4  & 0.7 $\pm$ 0.1 &  &\multirow{3}{4em}{$-$0.6$\pm$0.4}   \\ 
&  & N &      :09.30 &       :47.8     & &  9   $\pm$ 1  &  6  $\pm$ 1   & 5 $\pm$ 1      &  $-0.6\pm$0.4   & \\
&  & S &      :09.60 &       :10.3     &   &9.5   $\pm$ 1  &  6  $\pm$ 1   & 5 $\pm $1    &  $-0.7\pm$0.3   &  \\
                  
\#12& J202218+411728   & B   & 20:22:18.00 & 41:17:28.0     &60 &  2   $\pm$ 1   & 3  $\pm$ 1     & 2 $\pm$ 0.4  &  &\multirow{3}{4em}{$-$0.7$\pm$0.3}   \\                  
 & & NW &      :17.00 &      :40.7     & & 16  $\pm$ 2  & 9   $\pm$ 1   & 9 $\pm$1    &  $-0.9\pm$0.3    &  \\
 & & SE &      :19.28 &      :12.4     & & 10  $\pm$ 1  & 6   $\pm$ 0.5  & 5 $\pm$ 0.5  &  $-0.8\pm$0.2      &  \\

\#13& J202222+425409   &  B  & 20:22:22.00 & 42:54:09.0     & 72 & 5   $\pm$ 3         &   4.6 $\pm$1    & 2.6 $\pm$ 0.4 &&\multirow{3}{4em}{$-$0.9$\pm$0.4}   \\
 & & E &      :24.06 &      :07.6     &  &191 $\pm$ 33 	&   106  $\pm$ 17 & 106 $\pm$ 10 &   $-0.9\pm$0.4  &  \\
 & & W &      :19.14 &      :10.1     & & 329 $\pm$ 54 	&   179  $\pm$ 26 & 180 $\pm$ 17 &   $-1.0\pm$0.3      & \\

\#14& J202315+430252    & B  & 20:23:15.00 & 43:02:52.0 & 45 & 4 $\pm$ 2  & 1.7 $\pm$ 1  & 1 $\pm$ 0.4 & &\multirow{3}{4em}{$-$1.0$\pm$0.4}     \\
&  & NE &      :14.79 &      :03:03.0 &   &20 $\pm$ 4  & 12 $\pm$ 2 & 12 $\pm$ 1   & $-0.8\pm$0.4 &  \\ 
 & & SW &      :16.17 &      :02:42.1 &  &20 $\pm$ 3  & 10 $\pm$ 1 & 10 $\pm$ 1    & $-1.1\pm$0.3     & \\
 
\#15& J202359+435525    &  Nu & 20:23:58.61 &  43:55:25.0   &72 &  1.7 $\pm$ 0.3  & 2   $\pm$ 0.3 & 2 $\pm$ 0.2  &  &\multirow{3}{4em}{$-$1.0$\pm$0.2}  \\
 & & NE &      24:02.38 &       :52.8   &  &40  $\pm$ 3    & 21  $\pm$ 2   & 22 $\pm$1&   $-1.0\pm$0.2   & \\
 & & SW &      23:55.87 &       :03.0   & & 46  $\pm$ 4    & 23  $\pm$ 2   & 24 $\pm$1 &  $-1.1\pm$0.2     & \\
                               
\#16& J202403+413611   & Nu   & 20:24:03.00  & 41:36:11.0   & 90 &   20$ \pm$  2 &  12  $\pm$ 2 &  11 $\pm$ 2 & &\multirow{3}{4em}{$-$0.7$\pm$0.2}    \\
 & & NW &      :01.05  &   :36:43.1   &  &  35  $\pm$ 3 &  22  $\pm$ 1 &  25 $\pm$ 1 & $-0.7\pm$0.1  & \\
 & & SE &      :04.15  &   :35:37.2   &   & 37  $\pm$ 2 &  25  $\pm$ 1 &  28 $\pm$ 1 & $-0.6\pm$0.1 & \\

\#17& J202413+432220   & B   & 20:24:13.00  & 43:22:20.0    &  72 & 65 $\pm$  13  & 19 $\pm$ 3    & 22  $\pm$ 2   & &\multirow{3}{4em}{$-$0.8$\pm$0.3}    \\
 & & NE &      :15.84  &      :33.5    &  & 488 $\pm$ 70  & 258 $\pm$ 25  & 243 $\pm$ 23 & $-1.0\pm$0.3 & \\
&  & SW &      :10.76  &      :04.8    &  & 1164 $\pm$ 212 & 729 $\pm$ 69  & 724 $\pm$ 56 & $-0.7\pm$0.3 &  \\
                      
\#18& J202422+425722 &B & 20:24:22.00  & 42:57:22.0    &  50 &  20 $\pm$ 3 &   12  $\pm$ 4   & 8 $\pm$ 2 &  &\multirow{3}{4em}{$-$1.0$\pm$0.3}  \\
&  & W &      :20.86  &      :27.7    &   & 176 $\pm$ 26 &   99  $\pm$ 4   & 95$\pm$ 9 &   $-0.9\pm$0.2  & \\	
&  & E &      :23.48  &      :18.5    &   & 90 $\pm$ 20 &   44  $\pm$ 8   & 43$\pm$ 5 &    $-1.1\pm$0.4 &  \\  
                   
\#19& J202448+435956   & B   & 20:24:47.88  & 43:59:56.0   & 45 &    1.4 $\pm$ 0.7     &  0.7 $\pm$ 0.3 & 0.6 $\pm$ 0.2 &   &\multirow{3}{4em}{$-$0.8$\pm$0.2}  \\
 & & E &      :49.68  &   :59:50.0   &  &  47  $\pm$ 6  	&  29  $\pm$ 2   & 29 $\pm$ 2   &  $-0.8\pm$0.2  &\\
&  & W &      :46.76  &   :00:00.4   &   & 15  $\pm$ 1  	&  9   $\pm$ 1  & 9 $\pm$ 1    &  $-0.8\pm$0.2   & \\

\#20& J202625+442048  & B    & 20:26:25.00  & 44:20:48.0   & 60 & 2.7 $\pm$ 0.7 & 2 $\pm$1    & 1.4 $\pm$ 0.3  &  &\multirow{3}{4em}{$-$1.0$\pm$0.4}\\
&  & NE &      :25.93  &   :21:02.0   &  &46  $\pm$ 10  & 25  $\pm$ 4 & 25  $\pm$ 4    & $-0.9\pm$0.4 & \\ 
 & & SW &      :23.83  &   :20:39.6   &  &34  $\pm$ 6   & 17  $\pm$ 2 & 18  $\pm$ 2    & $-1.1\pm$0.3  &  \\
                  
\#21 &J202632+412942   &  G & 20:26:32.62  & 41:29:42.6   & 45 & 6   $\pm$ 1  & 0.8 $\pm$ 0.2 & 0.8 $\pm$ 1 &  &\multirow{3}{4em}{$-$1.4$\pm$0.4} \\
 & & E &      :34.48  &      :47.0   &  &114 $\pm$ 25 & 54  $\pm$ 10  & 55 $\pm$7    &  $-1.2\pm$0.4 &\\
 & & W &      :30.21  &      :41.2   &  &107 $\pm$ 20 & 39  $\pm$ 7   & 41 $\pm$ 5   &  $-1.6\pm$0.4 &\\

\#22 &202645+431435    &  B  & 20:26:45.00 & 43:14:35.0   & 138 & <0.2   & <0.5  & <0.5 &  &\multirow{3}{4em}{$-$1.6$\pm$0.1} \\
 & & N &      :47.18 &   :16:09.4   &  & 8   $\pm$ 1 	&  2   $\pm$  0.3 & 2 $\pm$ 0.3  &   $-2.2\pm$0.3 &\\
 & & S &      :43.84 &   :14:05.5   &  & 20  $\pm$ 1 	&  8  $\pm$  0.3 & 8 $\pm$ 0.2 &    $-1.4\pm$0.1  &\\   
   
\#23& J202710+421113    & G  & 20:27:10.00  & 42:11:13.0   &55 & 1 $\pm$ 0.5 & 0.8 $\pm$ 0.2 & 0.8 $\pm$ 0.1 &&\multirow{3}{4em}{$-$1.3$\pm$0.3}\\  
&  & NE &      :11.36  &      :26.1   & & 59 $\pm$ 9  & 25  $\pm$ 4   & 23  $\pm$3    &  $-1.4\pm$0.3&\\
&  & SW &      :08.25  &      :09.8   &  &18 $\pm $4  & 8.7 $\pm$ 0.2 & 8.7 $\pm$ 0.1  &  $-1.1\pm$0.3&\\

\#24& J202755+423215  & G  & 20:27:55.00  & 42:32:15.0 & 60 & 11  $\pm$5   & 0.9 $\pm$ 0.5 & 0.4 $\pm$ 0.1 &    &\multirow{3}{4em}{$-$1.4$\pm$0.2}                \\
&  & NE &      :57.12  &   :32:30.6 &  &349 $\pm$ 56 & 146 $\pm$ 14  & 337 $\pm$30   & $-1.4\pm$0.3 & \\
&  & SW &      :53.30  &   :31:56.4 & & 800 $\pm$ 14 & 337 $\pm$ 32  & 146 $\pm$12   & $-1.4\pm$0.1  & \\ 
\\                   
\#25 &J202817+442114   & B   & 20:28:17.00  & 44:21:14.0 &60 &  1 $\pm$ 0.3   & 0.5 $\pm$ 0.2  & 0.7 $\pm$ 0.3 &  &\multirow{3}{4em}{$-$0.8$\pm$0.2}  \\ 
 & & NE &      :18.54  &   :21:32.9 &  &19  $\pm$ 2  	& 10  $\pm$ 1   & 11 $\pm$1   & $-1.0\pm$0.2  &\\
 & & SW &      :16.40  &   :20:48.0 &  &25  $\pm$ 2  	& 16  $\pm$ 1   & 15 $\pm$1   & $-0.7\pm$0.2   &\\
                   
\#26 &J202859+402141  & B  & 20:28:59.00   & 40:21:41.0  &60 &  <0.4 &   1.6 $\pm$ 1   & 2.4 $\pm$ 0.4  &&\multirow{3}{4em}{$+$0.4$\pm$0.7}\\     
&  & N &      :58.92     &   :22:05.5  &  &18 $\pm$ 4    &  16  $\pm$  8  & 15  $\pm$1    &  $-0.2\pm$0.8  &\\
&  & S &      :59.00     &   :21:24.9  & & 4  $\pm$ 1       &  10  $\pm$ 1   & 9.6 $\pm$ 1   &  $+1.4\pm$0.4&   \\

\#27 &J202919+441341 & G   & 20:29:18.00  & 44:13:41.8  & 60&  <0.3   &  <0.3  &  <0.1 & &\multirow{3}{4em}{$-$0.9$\pm$0.2} \\
&  & NW &     :14.97  &   :13:39.7  &   &145   $\pm$ 14  &   84  $\pm$ 5 & 85$\pm$ 4  &  $-0.9\pm$0.2 &\\
&  & SE &     :21.28   &   :12:21.4  &   & 10 $\pm$1         &  5 $\pm$ 0.5 &  5  $\pm$ 0.5   &    $-1.1\pm$0.2    & \\
                 
\#28& J202943+425122    & B & 20:29:43.00  & 42:51:22.0   & 40 &  2  $\pm$ 1   & 1   $\pm$ 0.3  & 1 $\pm$ 0.3 &  &\multirow{3}{4em}{$-$0.9$\pm$0.4} \\
&  & E &      :44.84  &      :29.4  & &  17 $\pm$ 3  	& 9   $\pm$  1   & 9 $\pm$ 1   & $-1.0\pm$0.3 & \\		
 & & W &      :42.24  &      :20.7 &  &  23 $\pm$ 4  	& 14  $\pm$ 2   & 13 $\pm$1   & $-0.8\pm$0.3   &  \\	
     
 \#29& J203201+413730  &  G & 20:32:01.00  & 41:37:30.0  &  50 &  10 $\pm$ 3       &    4   $\pm$ 1  & 4 $\pm$ 1& &\multirow{3}{4em}{$-$1.2$\pm$0.1}  \\
&  & N &      :1.92   &      :53.8  & &   159 $\pm$ 10     &    77  $\pm$ 2  & 73 $\pm$ 2 & $-1.1\pm$0.1 & \\	
&  & S &      :0.91   &      :07.6  &  &  172 $\pm$ 10     &    76  $\pm$ 2  & 80 $\pm$2 & $-1.3\pm$0.1  & \\
             
 \#30& J203323+412720   &  B& 20:33:23.00  & 41:27:20.0 & 60 & 14 $\pm$ 6    &	 8 $\pm$ 3   & 7 $\pm$ 1 & &\multirow{3}{4em}{$-$1.1$\pm$0.2}  \\
 & & W &      :22.43  &      :25.3 &  &126 $\pm$ 9  & 56  $\pm$ 3  & 66 $\pm$ 7 & $-1.3\pm$0.1 &\\ 	
&  & E &      :24.42  &      :17.3 &  &130 $\pm$ 16 & 67  $\pm$ 6  & 57 $\pm$ 3 & $-1.0\pm$0.2 &\\	
                  
\#31 & J203345+404015  & Nu  & 20:33:45.58  & 40:40:15.7 &120 &  5 $\pm$ 1       & 3 $\pm$ 1       & 3$\pm$ 1&   &\multirow{3}{4em}{$-$1.3$\pm$0.2}  \\
&  & N &      :47.39  &   :41:00.2  & &  111 $\pm$ 8  	&  46  $\pm$ 2   & 48 $\pm$ 2&  $-1.4\pm$0.1&  \\		
&  & S &      :43.63  &   :39:28.2  &  & 156 $\pm$ 12 	&  71  $\pm$ 4   & 73 $\pm$ 4&  $-1.2\pm$0.1&   \\    

\#32 & J203402+420615   & Nu & 20:34:02.60  & 42:06:15.4   & 72 & 11 $\pm$ 2     &   5 $\pm$ 1    & 9 $\pm$ 1&   &\multirow{3}{4em}{$-$1.2$\pm$0.1}\\
&  & NE &      :04.09  &   :06:31.5   &  & 134 $\pm$ 4  	& 60  $\pm$ 1  & 48 $\pm$ 2 &  $-1.3\pm$0.1 & \\	
&  & SW &      :01.31  &   :05:59.2   &  & 86  $\pm$ 7  	& 42  $\pm$ 2  & 73 $\pm$ 4 &  $-1.1\pm$0.1 &\\

\#33  &J203413+403349    & B&20:34:12.97      & 40:33:49.6  & 60 & 3 $\pm$ 1    & 0.6 $\pm$ 0.1 & 0.6 $\pm$ 0.2&&\multirow{3}{4em}{$-$1.5$\pm$0.3}\\ 
 & & N &      :13.28        &   :34:13.2  &  &25 $\pm$ 4   &  9 $\pm$ 1    &  9 $\pm$ 2   & $-1.6\pm$0.3&\\
 & & S &     :12.28         &   :33:34.8  & & 51$\pm$ 8   &  21 $\pm$ 3   &  21$\pm$ 4  &   $-1.4\pm$0.3 &\\

\#34 &J203425+404605   & Nu & 20:34:25.80     & 40:46:05.9  & 60 &     39 $\pm$ 5       &  22 $\pm$ 4    & 22$\pm$ 3& &\multirow{3}{4em}{$-$1.2$\pm$0.2}  \\
&  & W &      :23.00	    &   :45:56.0  &  &   70  $\pm$ 3  	&  35  $\pm$ 2  &43 $\pm$ 1 &   $-1.1\pm$0.1& \\ 
&  & E &      :27.00	    &   :46:02.0  &  &   54  $\pm$ 2  	&  21  $\pm$ 5  & 28 $\pm$ 1&   $-1.5\pm$0.4 & \\ 

\#35 &J203431+413037    & B & 20:34:30.84  & 41:30:37.8   &  60 &   3 $\pm$ 1     &  3 $\pm$ 1     & 1.5 $\pm$ 0.2&&\multirow{3}{4em}{$-$0.4$\pm$0.3} \\
&  & NW &      :30.47  &      :51.4   &    & 33  $\pm$ 3  &  27  $\pm$ 3  & 29 $\pm$  2        & $-0.3\pm$0.3 &\\ 
&  & SE &      :31.21  &      :25.6   &    & 28  $\pm$ 2  &  18  $\pm$ 1  & 28 $\pm$ 1        & $+$0.7$\pm$0.1 & \\

\#36  &J203431+403332   &  B & 20:34:30.98  & 40:33:32.1   & 78 & 6  $\pm$ 1 & 1  $\pm$ 0.1  & 1  $\pm$ 0.1&&\multirow{3}{4em}{$-$1.2$\pm$0.1}\\                & & N &      :31.08  &   :34:14.2   &  &55 $\pm$4 & 23 $\pm$ 1    & 22 $\pm$ 1&  $-1.4\pm$0.1&\\ 
      &  & S &      :32.04  &   :32:34.8   &  &52 $\pm$4 & 29 $\pm$ 1    & 25 $\pm$ 1&  $-0.9\pm$0.1&\\
\\
\#37&  J203529+403201 & B& 20:35:29.78  & 40:32:01.2  & 70 & 4 $\pm$ 0.4      &   11 $\pm$ 2            & 5 $\pm$ 1&  &\multirow{3}{4em}{$-$0.7$\pm$0.3} \\
 & & NE &      :30.58 & 40:32:06.9	&  &  75  $\pm$ 10  &  35  $\pm$  4  &35$\pm$ 2 &   $-1.2\pm$0.3 & \\ 
 & & SW &      :29.11 & 40:31:59.9	&   & 49  $\pm$ 10  &  39  $\pm$ 4   &12$\pm$ 1 &   $-0.4\pm$0.4  & \\

\#38 & J203556+421744 & B   & 20:35:57.11     & 42:17:44.5   &  78 &  122 $\pm$ 32   &  78 $\pm$ 18   &101 $\pm$ 8&&\multirow{3}{4em}{$-$1.3$\pm$0.3} \\                       
 & & NW &      :55.00     &    :18:06.2	 &  &  3636 $\pm$ 385  &  1556 $\pm$  212 & 1526 $\pm$ 158&     $-1.3\pm$0.3  &\\
&  & SE &      :58.80     &    :17:22.9	 &  &  2863 $\pm$ 227  &  1334 $\pm$ 167 &  1318 $\pm$ 104&     $-1.2\pm$0.2  &  \\
                  
\#39& J203720+405941   &  Nu & 20:37:20.73  & 40:59:41.7    & 60 & 2 $\pm$ 1	        &  4 $\pm$ 1     & 2 $\pm$ 1    & &\multirow{3}{4em}{$-$1.1$\pm$0.2}  \\
 & & N &      :22.00  & 41:00:08.3    & & 90  $\pm$ 10  	& 43 $\pm$ 4    & 45 $\pm$ 4   & $-1.1\pm$0.2 & \\
 & & S &      :19.72  & 40:59:15.1    &  &105 $\pm$ 10 	&  48  $\pm$ 6  & 47 $\pm$ 6   & $-1.2\pm$0.2   & \\
\hline\hline
&&&&&&\\
\hline\hline
\#40 &J201836+430018 & G  & 20:18:36.00 &43:00:18.0 &120&&&&\\
        &    & N  &   :37.03    & :57.9       & &&&117 $\pm$ 5&\\ 
       &     & S  &   :34.86    & 42:59:33.3   &     &&&218 $\pm$ 3&\\  

\#41& J201841+431844 & Nu  & 20:18:41.00 &43:18:44.0&390&&&0.8$\pm$0.1&\\ 
       &     & N  &     :40.00& :22:04.8       &&&&48 $\pm$ 1&\\  
       &     & S  &     :44.06& :16:13.2      &&&&44 $\pm$ 2&\\  
 
\#42 &J202220+450835 & Nu  & 20:22:20.01 &45:08:35.9&60&&&0.7$\pm$ 0.2&\\
       &     & NW  &      :17.59 &   :43.14      &&&&8 $\pm$ 1&\\  
       &     & SE  &      :22.43 &   :25.6      &&&&11 $\pm$ 1&\\  

\#43 &J203755+401328 & B  & 20:37:55.00 &40:13:28.0&60&&&7 $\pm$1&&\\
      &      & W  &      :54.16    &:32.6        &&&&25.7 $\pm$ 2.0&\\  
      &      & E  &     :57.43     &  :29.2      &&&&26.5 $\pm$ 2.0&\\

  \hline
  \label{long}
  \end{longtable}
\end{landscape}
\twocolumn

\section{MULTI-WAVELENGTH CROSS-IDENTIFICATION AND ANALYSIS}\label{section4}

Once we have identified and characterised the position and the integrated flux density of the double-lobed source candidates in the radio images, we made use of NED\footnote{https://ned.ipac.caltech.edu/} and Simbad\footnote{http://simbad.u-strasbg.fr/simbad/} databases to perform multi-wavelength cross-match positional association at other frequencies.  For every double-lobed source, we defined three search regions: one for each lobe and a third for its central point/source/bridge. We consulted the specific catalogues to gather additional information  if necessary. In the following subsections, we describe the results.

\subsection{Radio sources}\label{sec:radio}

We found radio counterparts for many of the double-lobed sources candidates. In most cases, the matches belong to the NRAO VLA Sky Survey \citep[NVSS,][]{NVSS}, performed with the Very Large Array (VLA) at 1.4~GHz, with an angular resolution of 45$''$. Furthermore, many others were reported in surveys carried out with the WSRT, such as the ones mentioned in Sect. 2 
by \cite{Taylor-1996} and by \cite{Gunawan-2003}. To highlight the information at 1400~MHz and emphasise the improvement of the angular resolution at 325~MHz that our observations reached, we present, in Figs.~\ref{fig:galaxitas1} to \ref{fig:galaxitas6}, the counterparts found corresponding to the NVSS and WSRT surveys with circles of the size of their attained beams and labelled as `1.4~GHz' and `327~MHz', respectively. Additional detection, reported by previous studies, were also found (but not shown, for the sake of clarity). We will mention them as we describe the counterparts found for each source. For all double-lobed sources with NVSS counterparts, we gathered the NVSS integrated flux density at 1.4~GHz, and, using the convolved maps (to a 45$''$ beam) at 610~MHz, we derived the spectral index. The results are shown in Table~\ref{Tab:results-nvss}.\\

The spectral index plays an important role in the identification of the dominating process leading to the observed emission. It is well known that negative spectral index values in radio are indicative of a non-thermal emission origin, related to extragalactic sources \citep[e.g. synchrotron,][]{ginzburg1967}, but setting a limit is not straight forward around zero values ($0.5<\alpha<-0.5$) because this range is indicative of optically thin free-free emission (from star-forming galaxies or Galactic ionised objects for instance), and also of thick synchrotron emission (core-dominated AGNs). Positive spectral index values can be associated with young compact sources \citep[e.g.,][]{Saikia2004, Pratap2005}.

Since our celestial region of study is nearby the Galactic plane, to be conservative, we considered a limiting spectral index value of $-$0.7 as the main criterion to discriminate the double-lobed source candidates between extragalactic ($\alpha \leq -0.7$)  or Galactic  ($\alpha > -0.7$) origin \citep[see,  for  instance,][]{ibar2010,ocran2020a}.
Whenever was possible, we did also consider the value of the spectral index $\alpha_{1400~\rm MHz}^{610~\rm MHz}$ in our classification. The detailed information of the flux  densities measured and the spectral indices derived are given in Tables~\ref{long}~$\&$~\ref{Tab:results-nvss}. \\

\begin{table}
 \caption{Integrated flux density of NVSS counterparts, and spectral indices.}
 \label{Tab:results-nvss}
 \centering
 \begin{tabular}{l@{~~}r@{~~}r@{~~}r@{~~}} 
 \hline
 \hline

\# \,\,\,\,\, NVSS ID&$S_{1400\,{\rm MHz}}$& $SC_{610\,{\rm MHz}}$ & $\alpha_{1400\rm MHz}^{610\rm MHz}$\\
 &  (mJy) & (mJy) & \\
\hline
\#6\,\,\,  J201828+402702 & 5.6$\pm$0.7	& 44$\pm$9     & $-2.3\pm$0.3  \\
\#8\,\,\,  J201910+425044  &16.6$\pm$0.1 & 67$\pm$8& $-1.7\pm$0.2\\
\#9* J202057+444130	& \multirow{2}{4em}{109.6$\pm$3.3}	& \multirow{2}{4em}{\,\,\,\,\,353$\pm$98}   & \multirow{2}{4em}{\,\,$-1.4\pm$0.3}\\
\,\,\,\,\,\,\,\,\, J202057+443959 &&&\\
\#13 J202220+425415 	& 71.1$\pm$2.7	& 279$\pm$45    & $-1.6\pm$0.2  \\
\#17*J202410+432202 &\multirow{2}{4em}{\,\,\,\,421$\pm$ 13}  &   \multirow{2}{4em}{ 874$\pm$200} & \multirow{2}{4em}{\,\,$-0.9\pm$0.3}\\
\,\,\,\,\,\,\,\,\,\, J202415+432235 & &&\\
\#18  J202422+425729& 39.6$\pm$1.7&133 $\pm$ 30& $-1.4\pm$0.3\\
\#19 J202449+435958	& 15.8$\pm$2.2	& 42$\pm$8     & $-1.1\pm$0.3  \\
\#20  J202625+442056 	& 12$\pm$0.7	& 41$\pm$8     & $-1.5\pm$0.2  \\
\#21 J202634+412952	& 5.7$\pm$0.8	& 81$\pm$3     & $-3.2\pm$0.2  \\
\#24 J202754+423205	& 124.7$\pm$4.3	& 402$\pm$74    & $-1.4\pm$0.2  \\
\#25  J202816+442100	& 5$\pm$0.7	& 20$\pm$3     & $-1.7\pm$0.2  \\
\#28   J202943+425119 & 7.5$\pm$0.7 & 23$\pm$5&   $-1.3\pm$0.3\\
\#29 J203201+413722	& 15.2$\pm$0.8	& 151$\pm$20    & $-2.8\pm$0.3  \\
\#30  J203323+412723	& 8.4$\pm$0.7	& 96$\pm$21    & $-2.9\pm$0.2  \\
\#32  J203402+420614&11.6$\pm$0.7& 119$\pm$20 & $-2.8\pm$0.3\\
\#35 J203431+413046 &27.4$\pm$1.5& 40$\pm$10& $-0.4\pm$0.2\\
\#38  {J203555+421803} & 604$\pm$21 &     2546$\pm$359&  $-1.7\pm$0.3\\
\#39  J203720+405924	& 22.3$\pm$3.2	& 48$\pm$7     & $-0.8\pm$0.2  \\
\#42   J202219+450837 &11$\pm$2& 20$\pm$3 & $-$0.7$\pm$0.3\\ 
\#43 J203756+401326 &8.7$\pm$ 0.7 & 65$\pm$15 & $-$2.4$\pm$0.3\\
\hline
\end{tabular}
\tabnote{The spectral indices were derived between catalogued NVSS integrated flux densities ($S$) and the convolved  integrated  flux densities ($SC$) of the double-lobed sources at 610~MHz. *In order to estimate the total spectral index we added the NVSS integrated flux densities of both lobes.}
\end{table}

\subsection{Sources at other wavelengths}\label{sec:otherwavel}

Although NED and Simbad databases cover the entire electromagnetic spectrum, besides radio sources, only infrared ones were found, except for one X-ray counterpart. Infrared sources are expected to be frequent in the line of sight towards the Cygnus region.  But we note that as double radio sources are likely extragalactic objects, and the Cygnus OB2 region is in the Galactic plane, it is remarkably challenging to find conclusive evidence of the right counterparts in optical/IR for these radio sources.

Even though we defined three search regions, as is explained at the beginning of the section, we focused the infrared search on the surroundings of the centre (Nu) or the geometric centre (G) of the double-lobed sources, considering that we aim to identify the possible host galaxy for the proposed extragalactic sources. In checking for the positional overlapping in the plane of the  sky, we considered the following extensions of infrared sources for each catalogue using circles with diameters of: 6$''$for WISE \citep{WISE-2010A}, 2$''$ for 2MASS \citep{2mass-2006}, 2$''$ for Spitzer \citep{spitzer-2004}, and 18$''$ for IRAS \citep{IRAS-1984}. Below, we present fully the arguments that led us to the classification of each double-lobed source.\\

\subsection{Extragalactic sources}

\noindent{\bf J201255+404423} (\#1), {\bf J201418+412931} (\#2) and {\bf J201847+401230} (\#7). {At the position of these sources there are no detection at other frequencies, see Fig.~\ref{fig:galaxitas1}. An inclination of the line linking the lobes towards the line of sight, will result in a modification of the actual sizes/shapes as seen in the images. In particular, an inclination angle smaller than 45~deg will imply higher projection effects, including the appreciation of lobes with different sizes. We interpret that this is the case for source \#1.}\\

\noindent {\bf J201739+421243} (\#4) and {\bf J201758+403100} (\#5). For these sources we found no associations at other frequencies, see Fig.~\ref{fig:galaxitas1}. As a complement, we show the spectral index maps as well as the error distribution in Fig.~\ref{fig:sp-galaxies}.\\

\noindent{\bf J201829+402714} (\#6) and {\bf J201910+425136} (\#8). Both sources have positional coincidences with NVSS catalogued records, see Table~\ref{Tab:results-nvss}. For \#8, the NVSS component overlaps only the southern lobe (NVSS ID: J201910+425044). The radio counterpart of \#6, which is a component of the NVSS survey is not resolved, see Fig.~\ref{fig:galaxitas1}.\\

\noindent {\bf  J202151+423847} (\#10). There are two possible associations at infrared wavelengths for this source: WISEA J202150.84$+$423846.9 and 2MASS J20215131$+$4238521, coincident with the position of the geometric centre. Besides, the radio image at 610~MHz reveals two knots in-between the IR sources and the lobes  (see Figs.~\ref{fig:galaxitas2} and ~\ref{fig:IR+radio}). These knots could be part of common bipolar jets (an unseen weak bridge) formed time ago, of an AGN with much activity in the past. The morphology could be also indicative of DDRG, as two double radio galaxies that share a centre.\\

\noindent{\bf J202209+415033} (\#11), {\bf J202218+411728} (\#12), \noindent{\bf J202222+425409} (\#13) and {\bf J202315+430252} (\#14). For these sources, there are no detection at other wavelengths, except for \#13,  which has one radio counterpart, unresolved, at 1.4~GHz, see Fig.~\ref{fig:galaxitas2} and Table~\ref{Tab:results-nvss}.\\

\noindent{\bf  J202359+435525} (\#15). There is one possible association at infrared wavelengths (WISEA J202358.42$+$435519.3), in agreement with the position of the nucleus, see Fig.~\ref{fig:galaxitas2} and Fig.~\ref{fig:IR+radio}. It might be revealing the location of the putative host galaxy. We show the spectral index distribution and error maps; see Fig.~\ref{fig:sp-galaxies}.\\

\noindent{\bf J202403$+$413611} (\#16). For this object, there is one possible association at infrared frequencies (WISEA J202403.03+413611.3), at its geometric centre; see Figs.~\ref{fig:galaxitas2} and \ref{fig:IR+radio}. Considering the spectral index value within the errors, the object might be an extra-galactic source where the core may have reignited, giving it a similar brightness as the southern lobe.\\

\noindent{\bf J202413+432220} (\#17). This source was detected by previous radio surveys: at 365~MHz with the TEXAS telescope \citep[][$S$ = 1.76$\pm0.14$~Jy, 4$'$ beam]{Douglas-1996}, and at 5~GHz with the MIT-green Bank telescope \cite[][$S$ = 69$\pm$11~mJy, beam of 3$'$]{Bennet-1986}. We found two NVSS components coincident with each lobe (NVSS IDs: J202410+432202, J202415+432235; see Fig.~\ref{fig:galaxitas2} and Table~\ref{Tab:results-nvss}.\\

\noindent{\bf J202422+425722} (\#18), {\bf J202448+435956} (\#19) and {\bf J202625+442048} (\#20). Each of these sources have coincident NVSS components, all unresolved  (see  Figs.~\ref{fig:galaxitas2}, \ref{fig:galaxitas3} and Table~\ref{Tab:results-nvss}). \\

\noindent {\bf J202632+412942} (\#21) and {\bf J202755+423215} (\#24). For both sources there are possible associations at the infrared band, WISEA J202634.96$+$412943.8 and WISEA J202755.97$+$423224.3, respectively. Besides, the two sources are also unresolved NVSS components.  Moreover, the source \#21 has positional coincidence with a WSRTGP catalogue record; see Figs.~\ref{fig:galaxitas3} and \ref{fig:IR+radio}.\\

\noindent{\bf J202645+431435} (\#22). There is one possible association at infrared frequencies for this source (WISEA J202645.01+431439.8), which could be pinpointing the host galaxy, see  Figs.~\ref{fig:galaxitas3} and \ref{fig:IR+radio}. The morphology of this source seems to indicate that the nucleus is closer to the southern lobe.\\

\noindent {\bf J202710+421113} (\#23) and {\bf J202817+442114} (\#25). There are just one detection at other frequencies for \#25, for which there is an NVSS component overlapping its southern lobe (NVSS ID: J202816+442100),   Fig.~\ref{fig:galaxitas3}. \\

\noindent{\bf J202919$+$441341} (\#27). There are no associations at other wavelengths. The intensity distribution at both radio bands is complex. We may be looking at a double-lobed source, where projection effects play an important role, see  Fig.~\ref{fig:galaxitas3}. The possibility that we were looking at just one lobe (to the north) and that the southern component is the core seems  unlikely, since this latter source do not present the flat spectral index expected from cores.\\

\noindent{\bf J202943$+$425122} (\#28). There is one counterpart at radio frequencies. It is reported in the NVSS catalogue (NVSS ID: J202943+425119), see Fig.~\ref{fig:galaxitas4} and Table~\ref{Tab:results-nvss}; the source is unresolved.\\

\noindent{\bf J203201$+$413730} (\#29). For this  source, there are several counterparts at radio frequencies. Is also known as NVSS J203201+413722 and is unresolved. \cite{Marti-2007} studied it in detail; they proposed it as a radio galaxy which might suffered different episodes of jet activity (see Fig.\,\ref{fig:galaxitas4}). The authors observed the field using the VLA C+D array configuration at 5~GHz as well as the GMRT at 610~MHz. These observations, along with a marginal Chandra X-ray detection in the vicinity of \#29 reported by \cite{Butt-2007}, made it a radio galaxy candidate with an optically-thick compact core \citep{Marti-2007}. The two lobes of \#29 were also reported by \cite{Gunawan-2003} at their two observing bands; they named them as SBHW-217 ($S_{\rm 350MHz}=85\pm$5~mJy, $SC_{\rm 1400MHz}=36\pm$5~mJy) -northern lobe- and SBHW-218 ($S_{\rm 350MHz}=122\pm$35~mJy, $SC_{\rm 1400MHz}=39\pm$5~mJy) -southern lobe-, where the quoted integrated fluxes $SC$ at 1400~MHz were derived from images convolved to the 325~MHz beam.\\

\noindent{\bf J203323$+$412720} (\#30) and {\bf J203402$+$420615} (\#32). The source \#30 was previously reported by \cite{Marti-2007} and is also an unresolved NVSS component. 
The source \#32 has one infrared counterpart (WISEA J203401.86$+$420618.2), and a radio counterpart which is an unresolved NVSS component, see Fig.~\ref{fig:galaxitas4} and Fig.~\ref{fig:IR+radio}.\\ 

\noindent{\bf J203345$+$404015} (\#31). There is one possible association at infrared frequencies, with the object 2MASS J20334540+4040133. At radio frequencies, the NW lobe was reported in WSTRGP \citep{Taylor-1996} and both lobes were reported by \cite{Gunawan-2003}, at two bands; the ones here called NW, as SBHW-106 ($S_{\rm 350MHz} = 75\pm$12~mJy, $SC_{\rm 1400MHz}=25\pm$4~mJy), and SE, as SBHW-104  ($S_{\rm 350MHz} = 82\pm$4~mJy, $SC_{\rm 1400MHz} = 34\pm$7~mJy); see Figs.~\ref{fig:galaxitas4} and  \ref{fig:IR+radio}. As a complement we show the spectral index map as well as the error maps in Fig.~\ref{fig:sp-galaxies}.\\

\noindent{\bf J203413$+$403349} (\#33). Eastern to this double-lobed source there is a third source ($\sim40"$ apart, and 30$''$ in size); \cite{Gunawan-2003} reported 1.4~GHz WSRT emission at its position, from data with angular resolution of 13$''$. We measured $S_{\rm 325MHz} = 134\pm$16~mJy and $SC_{\rm 610MHz} = 68\pm$8~mJy for this third source; the spectral index is then $\alpha_{325}^{610}$=-1.7$\pm$0.3. The angular resolutions at the three bands (325, 610  and 1400~MHz) are sufficient to clearly detach the three sources; besides, the northern and southern lobes present extension towards each other, challenging  a physical association between this eastern source and the double-lobed \#33. See Fig.~\ref{fig:galaxitas4}.\\ 

{\bf J203425$+$404605} (\#34) and {\bf J203431$+$403332} (\#36). We found no detections at other frequencies for these sources. See Fig.~\ref{fig:galaxitas4}.\\

\noindent{\bf J203556$+$421744} (\#38) and {\bf J203720$+$405941} (\#39). There are counterparts at radio frequencies. Both are reported in the NVSS catalogue. Nevertheless, from the  NVSS cutouts, only \#39 is resolved at 1.4~GHz. The source \#38 has a radio counterpart, which is a component of the WSRTGP surveys \citep{NVSS,Taylor-1996} and also one X-ray possible counterpart \citep[SSTSL2 J203557.22+421744.0,][]{Evans2010}. See Fig.~\ref{fig:galaxitas5}.\\ 

\noindent{\bf J201836$+$430018} (\#40) and {\bf J202220$+$450835} (\#42). There are possible associations at infrared frequencies for both sources; one for \#40 (WISEA~J201836.53$+$430018.7) and two for \#42 (2MASS~J20221985$+$4508338 and WISEA~J202219.75$+$450834.5). Besides, \#42 was also reported as an NVSS component, the source is unresolved. See Figs.~\ref{fig:galaxitas6} and \ref{fig:IR+radio1}.\\

\noindent{\bf J203755$+$401328} (\#43). There is one counterpart at radio frequencies. The source is listed in the NVSS catalogue, it is unresolved. See Fig.~\ref{fig:galaxitas6}.\\

 \subsection{Galactic sources}

In the sample of double-lobed sources studied here, there are a few 
with flat or positive spectrum, for which non-thermal radiation mechanisms do not apply. Their emission can be explained advocating thermal processes, by close-by (Galactic) plasma media, that emit either under optically thick or thin regimes \citep{osterbrock1989}. \\

\noindent {\bf J202859$+$402141} (\#26). This source shows one possible counterpart, nearby the geometric centre and at infrared frequencies (WISEA J202858.96$+$402146.8), and another at radio frequencies (WSRTGP component); See Figs.~\ref{fig:galaxitas3} and \ref{fig:IR+radio}. The spectral index of the northern lobe is close to zero, while that of the southern lobe is $>1$, see Table~\ref{long}. We note that the spectral index error values are high, because the signal-to-noise at the 325~MHz image was low at the position of this source. However, both indices are consistent with those corresponding to Galactic emission. Due to that, we propose that \#26 belongs to the Galaxy.\\

\noindent{\bf J203431$+$413037} (\#35). For this source there is one infrared detection onto the bridge, coincident with the proposed nucleus, WISEA J203433.70+412953.7 and one counterpart at radio frequencies (NVSS component, ID: J203431+413046, unresolved);see Figs.~\ref{fig:galaxitas4} and \ref{fig:IR+radio1}. Besides, there is an HII region at $RA, Dec({\rm J2000})=$20:34:31.9, $+$41:30:52 \cite[G080.522+00.714,][]{Anderson-2015}. 
The spectral index suggests that it is a  Galactic object.\\

\subsection{Dubious cases}

Besides the {\bf (37+2)} double-lobed sources already identified either as extragalactic or Galactic, there remain four cases in which the true nature is not straightforward to unveil, and are analysed below.\\

\noindent {\bf J201512+410457} (\#3) There are no association at other frequencies for this source; see Fig.~\ref{fig:galaxitas1}. The radio image at 610~MHz shows that the components present a kind of extension towards each other. Taking this into  account, and the fact that the spectral index of both components (similar values) is below $-0.7$, we favour an extragalactic origin for  
this object.\\

\noindent{\bf J202058$+$444050} (\#9). This source has counterparts at radio frequencies. Each lobe candidate is coincident with a NVSS component; see Fig.~\ref{fig:galaxitas1}. The northern one presents a negative spectral index ($-0.7\pm0.1$), while the spectral index of the southern one is close to $-0.1$. This led us to propose that instead of a double-lobed source, it could  well be two different objects, an extragalactic (northern) one and the other of Galactic nature. An alternate possibility is that the southern component is a radio core, of which only the northern lobe is being detected. The sizes of each object are  45 and 20\,arcsec, north and south respectively, separated by a distance of 2\,arcmin. The evidence gathered is not enough to confirm this candidate as a double-lobed source. \\
 
\noindent{\bf J203529$+$403201} (\#37). There are no detections at other frequencies at its position; see Fig.~\ref{fig:galaxitas5}.The spectral indices of both lobes, separated 45\,arcsec, are quite different; in particular the one associated with SW lobe, which is close to zero, see Table~\ref{long}. A possible scenario as an extragalactic source, is that the SW component is a bright hotspot, with its flat-spectrum emission being the reason why the spectral index is so different from the NE component.\\
 
\noindent{\bf J201841$+$431844} (\#41) shows emission that resembles two highly collimated jets extending along 7\,arcmin on the sky. It presents a knot in between the two lobes, indicative of a possible nucleus. The source is not covered by our 325~MHz mosaic image and lacks of infrared or radio putative counterparts; see Fig.~\ref{fig:galaxitas6}. Notwithstanding, the morphology of the object is typical AGN-like sources.

\section{Discussion}

\subsection{Generalities on spectral indices}

\begin{figure}
     \centering
     \includegraphics[width=0.5\textwidth]{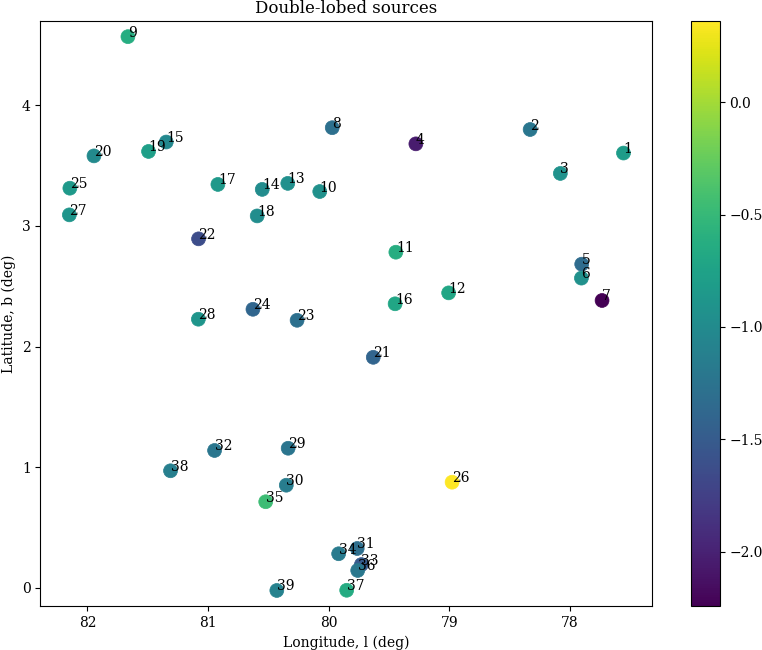}
     \caption{Layout of the observed double-lobed sources in Galactic coordinates, with their spectral indices represented with colours, according to the scale at the right.}
     \label{fig:sp-color}
 \end{figure}

Thirty-nine out of the 43 candidates presented in Table~\ref{long} were observed at the two GMRT bands (325 and 610~MHz), allowing them to retrieve information on their spectral indices. We plotted the layout of those 39 sources in the RA-Dec sky, including a spectral index scale (Fig.~\ref{fig:sp-color}). No trends are recognised, along a $\sim 5 \times 5$ sq deg area ($0^{\circ} \leq b < 5^{\circ}$), both with regards to the indices and the location of the sources relative to the Galactic plane.

 \begin{figure}
 \centering
 \includegraphics[width=0.5\textwidth]{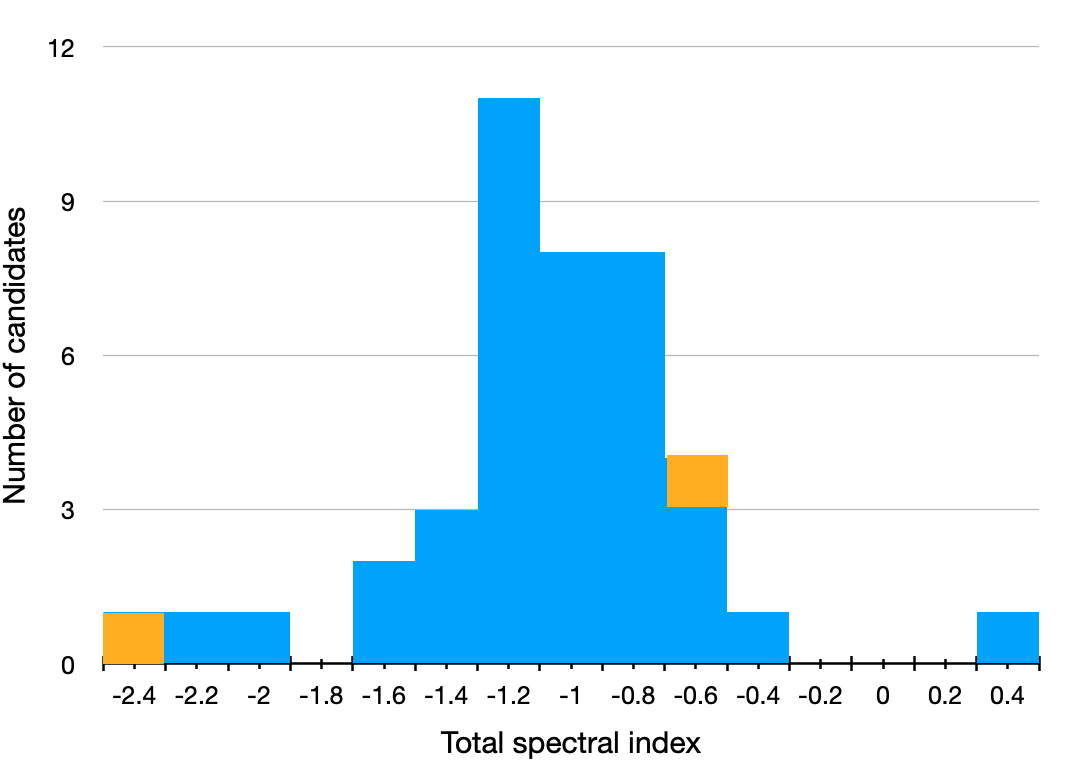}
 \caption{Number of candidates with spectral index information, as a function of their total spectral index ($\alpha_{\rm T}$) between 325~MHz and 610~MHz, for the 39 double-lobed detected at both frequency bands with the GMRT (in blue), and for the sources \#42, \#43 between 610~MHz and 1400~MHz (in orange). 
 }
    \label{fig:histos-spixes}
    \end{figure}
 
The total spectral index values, 
derived using the integrated flux density along the whole candidate including material in-between lobes, remain in the interval  [$-2.5$, $+0.5$] (see Fig.\,\ref{fig:histos-spixes}).
The median spectral index value is $-1.0\pm0.1$. 
This is in line with the fact that at the observed bands, the processes involved in non-thermal emission make sources brighter than those related to thermal sources, especially in the case of non-extended ones. In addition, the strategy followed when {\bf imaging} the original mosaics was to use a  weighting scheme aimed to outline discrete sources against diffuse or extended emission.

For those (among the more extended) sources with a high signal-to-noise ratio ($>$21), we built the spectral index distribution maps; they are shown in Fig.\,\ref{fig:sp-galaxies}. The higher $\alpha$ values in the central region are indicative of the nucleus of an extragalactic source.

\begin{figure}
 \centering
 \includegraphics[width=0.5\textwidth]{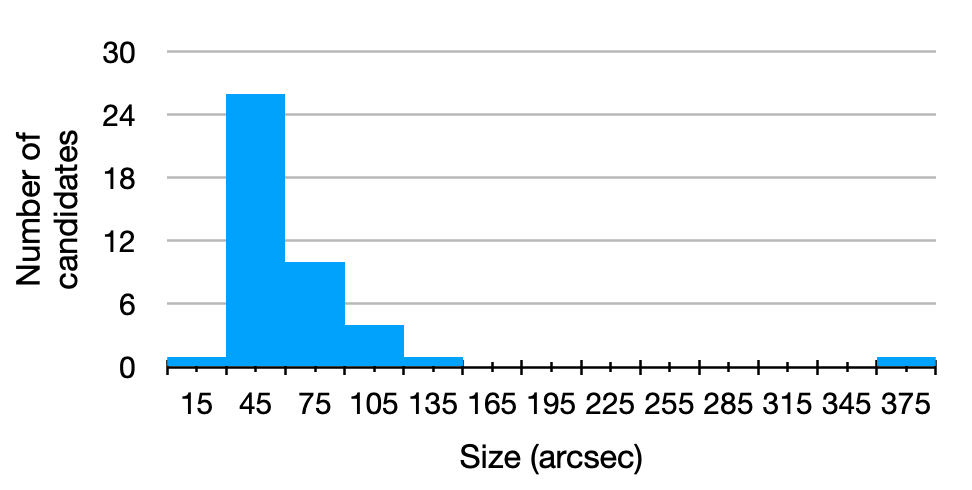}
 \caption{Distribution of sizes of the 43  double-lobed candidates. {\bf The extension of the sources ranged from $\sim$ 25~arcsec up to $\sim$ 375~arcsec ($\sim$ 7~arcmins).}}
 \label{fig:histosizes}
 \end{figure}

For completeness, we present in Fig.\,\ref{fig:histosizes}  the  distribution of the total sizes  of  the 43 candidates (Table~\ref{long}), as  seen in the 610~MHz band.

\subsection{Recap on cross-identifications}

 The cross-identification of the 43 candidates with previous surveys such as the NVSS contributed to examining each candidate's nature, and helped as verification of the detections presented here (see Sect.\,\ref{section4}).

We found fourteen IR detections overlapping the nucleus or proposed geometric centre of twelve radio candidates reported along Sect.~\ref{section4}. Out of these twelve positional coincidences, nine were only WISE sources, one was a 2MASS, and for the remaining two, we found sources in both  the WISE and 2MASS catalogues. The search criteria for the positional  overlapping is given Sect.\,\ref{sec:otherwavel}.
The location of those IR sources are shown in the radio images (Figs. \ref{fig:galaxitas1} to \ref{fig:galaxitas6}), labelled as `WISE' and `2MASS'. We present in Figs.\,\ref{fig:IR+radio}  and \ref{fig:IR+radio1} the infrared images towards the mentioned eleven candidate. We superposed the 325 and 610~MHz emission contours, which allowed us to explore the association between the IR sources and the double-lobed candidates. 


The X-ray counterpart found favours the extragalactic origin of source \#38, and might indicate that this source is at a closer distance than the rest, see e.g.~\cite{1978MNRAS.183..129E,2013A&A...556A.120M, 2013MNRAS.431..978L} and references therein.\\

\subsection{The extragalactic double-lobed sources} 

After the careful inspection of the 43 double-lobed source candidates presented in this work, and taking into account all the information collected, we found that 37 of them are compatible with an extragalactic nature. For those sources that we proposed a WISE source as a possible candidate, which could be associated with the host galaxy, we estimated the WISE colours W1-W2 and W2-W3, following \citet{gurkan2014}. The values that we obtained (W1-W2\,<\,0.2 and W2-W3\,<\,5.8) are compatible with those shown by radio-loud AGNs \citep[see Fig. 3 in][]{gurkan2014}.
 Among the 37 sources we found some of them with peculiar morphology such as:
\begin{itemize}
    \item \#19, \#30 and \#34, that resemble head-tail galaxies like the ones compiled by \citet{pal-2021}, see for example their Fig. 2.
    \item \#29, which seems to be of FRI type.
\end{itemize} 

\subsection{On ultra steep spectrum sources}

\begin{figure}
     \centering
     \includegraphics[width=0.45\textwidth]{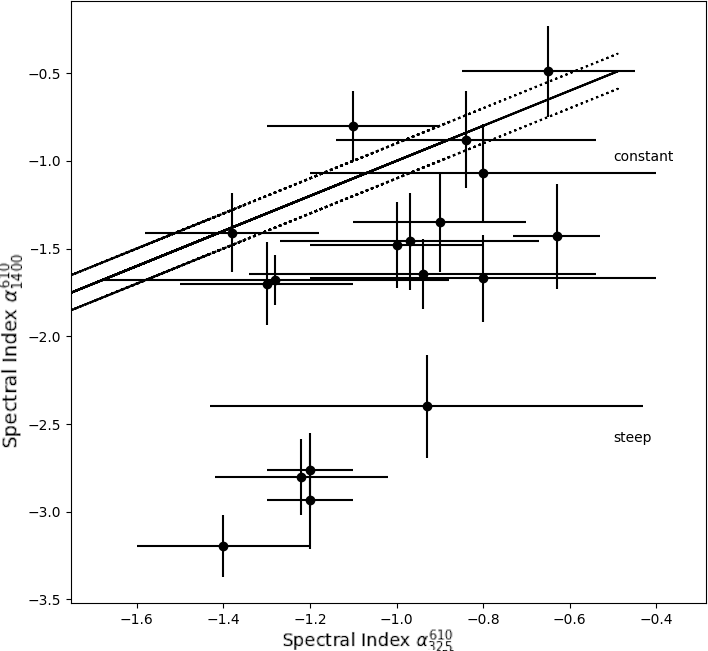}
     \caption{Radio colour-colour diagram. The spectral indices of the detected double-lobed sources between 325 and 610~MHz versus the spectral indices between 1400 and 610~MHz. The solid and dashed lines mark the constant spectral index at both radio colours and the adopted errors of 0.1. The sources with $\alpha_{1400}^{610}<<-1.5$ are considered to be ultra-steep-spectrum sources.}
     \label{fig:steep}
 \end{figure}

The existence a radio counterpart at 1.4~GHz allowed us to test for Ultra-Steep Spectrum sources (USS), i.e., objects with $\alpha \leq -1.3$. \citet{saxena-2018} searched USS as proxies of high-redshift radio galaxies (HzRGs), for which they request additional characteristics. The authors identified HzRG through a radio colour-colour plot (sources with $\alpha_{1400}^{610}<-1.5$, see their figure 5), the absence of optically identified counterparts and angular sizes smaller than 30$''$ (see also references therein). 
Then, following \citet{saxena-2018}, and for the sources detected at both GMRT bands that possess a radio counterpart at 1.4~GHz, we built the radio colour-colour diagram $\alpha_{325}^{610}$ vs $\alpha_{1400}^{610}$; see Fig.~\ref{fig:steep}. 
The plot reveals that the majority of the double-lobed sources have a constant or a steep spectrum. In particular, there are five sources at the loci of USS ones, with $\alpha << -1.5$:
\#6, \#21, \#29, \#30 and \#32. It can be appreciated in Table\,\ref{Tab:results-nvss} that these five sources, together with \#43, are the ones with the steepest spectral indices, $\alpha \leq -2.3$; \#43 was not covered by our 325~MHz image and, it will not be analysed. Only one of the sources has an extension of$\sim30''$ (source \#6).
New observations at 1.4~GHz, with better sensitivity and higher angular resolution, are needed to confirm whether candidate \#6 is a USS source.\\

\section{Statistics and conclusions}

\begin{itemize}
    \item In a region of 19.7 sq deg of the sky, 43 probable double-lobed sources were found at 610 MHz, 39 of them also at 325 MHz.
    \item The angular resolutions of the data sets were 6$''$  to 10$''$, and the average sensitivities, 0.2 and 0.5~mJy~beam$^{-1}$ (at 610 and 325~MHz, respectively).
    \item The extension of the sources ranged from $\sim$ 25 arcsec up to $\sim$ 7 arcmins.
    \item Twenty-one out of the 43 sources have a bridge.
    \item Eleven out of the 43 sources have a detected nucleus.
    \item Thirty-seven out of the 43 sources have spectral indices characteristic of extragalactic objects, probably AGNs.
    \item Nine out of the 37 proposed as extragalactic sources present infrared sources at the position of the a putative nucleus.
    \item Two out of the 43 sources have flat or positive spectral indices, favouring a Galactic origin.
    \item For one of the 43 sources, the two emitting objects are probably not physically related.
    \item Three out of the 43 remains as dubious cases, though with feature(s) compatible with an extragalactic nature.
    \item The median spectral index value of the extragalactic sources is $-1.0\pm0.1$. 
    Five extragalactic sources present an ultra steep spectra. Only one of them (\#6) has a small size ($\leq$30$''$) consistent with a high-redshift radio galaxy.
    \item Although the large angular scale probed at the  highest  observing band was 17$'$, no sources with extension above 7 arcmin were detected. 
     
\end{itemize}

\begin{acknowledgements}
The radio data presented here were obtained with the Giant Metrewave Radio Telescope (GMRT). The GMRT is operated by the National Centre for Radio Astrophysics of the Tata Institute of Fundamental Research. We thank the staff of the GMRT that made these observations possible. This research has made use of the NASA/IPAC extragalactic database (NED) that is  operated  by  the  Jet  Propulsion  Laboratory,  California  Institute of Technology, under contract with the National Aeronautics and Space Administration and also we made use of the SIMBAD database, operated at CDS, Strasbourg, France, and of NASA’s Astrophysics Data System bibliographic series.  J.S. and P.B.  acknowledge support from ANPCyT PICT 0773--2017. 
The authors want to thank the anonymous referee for the careful reading of the manuscript and their many insightful comments and suggestions. C.H.I.C. acknowledges the support of the Department of Atomic Energy, Government of India, under the project  12-R\&D-TFR-5.02-0700.

\end{acknowledgements}

\appendix
\section{Appendix}

In this appendix we present the GMRT images of 39 double-lobed source candidates at 610 and 325~MHz, and the remaining 4 --out of a total of 43-- only observed at 610~MHz (Figs.\,\ref{fig:galaxitas1} to \ref{fig:galaxitas6}). The candidates showed radio emission at rather different scales. The one with the lowest $\sigma$ value is \#12 ($\sigma_{12} = 0.1$~mJy\,beam$^{-1}$). To allow comparison among sub-figures, we added in the bottom right corner of each one a scale number (preceded with an `x'), that represents the factor to multiply $\sigma_{12}$ to get the $\sigma$ of the candidate sub-figure.

For some of the candidates (\#4, \#5, \#15, \#31) we show the spectral index distribution and error maps (Fig.\,\ref{fig:sp-galaxies}). And for those probable galaxies with putative infrared counterparts we present the radio contours on top of the IR images (Figs.~\ref{fig:IR+radio}  and \ref{fig:IR+radio1}).

\setcounter{figure}{0}
\renewcommand{\thefigure}{A\arabic{figure}}

\begin{figure*}
\begin{center}
    \includegraphics[width=0.33\textwidth]{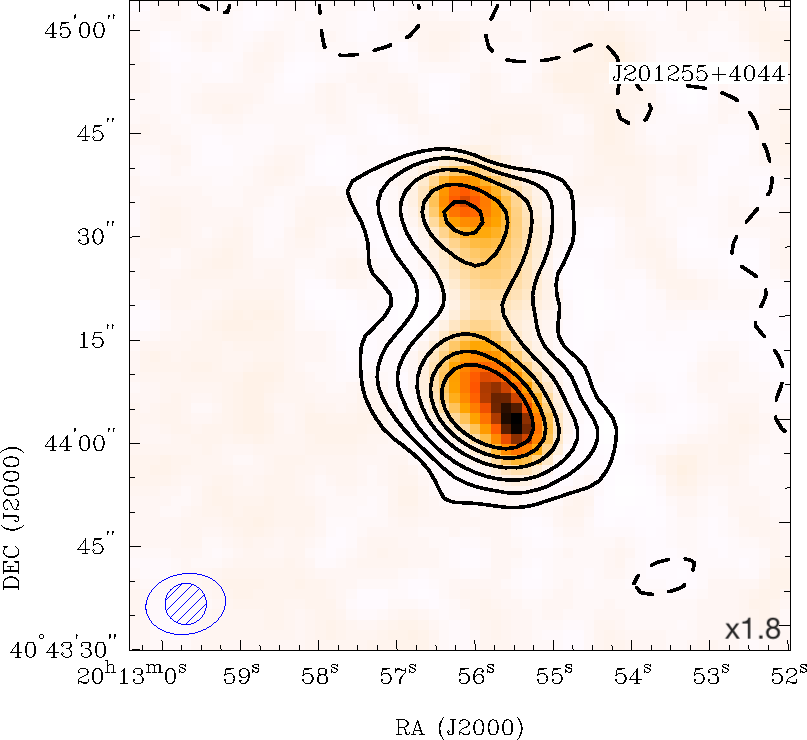} 
    \includegraphics[width=0.27\textwidth]{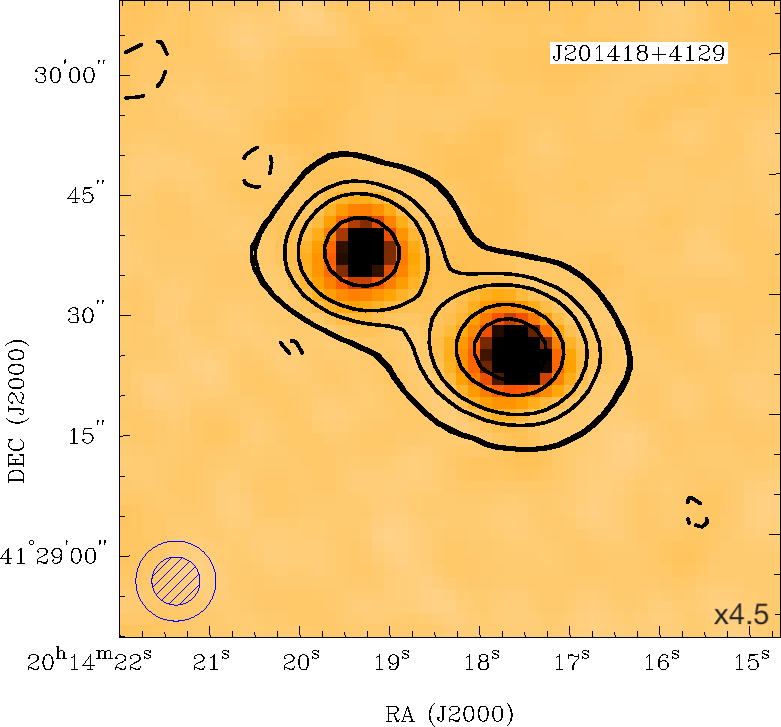}
    \includegraphics[width=0.25\textwidth]{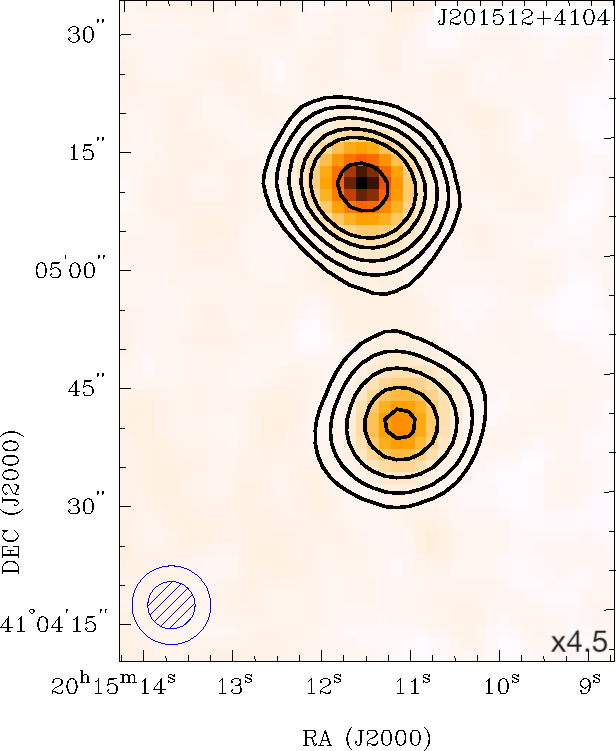}\\
    \includegraphics[width=0.25\textwidth]{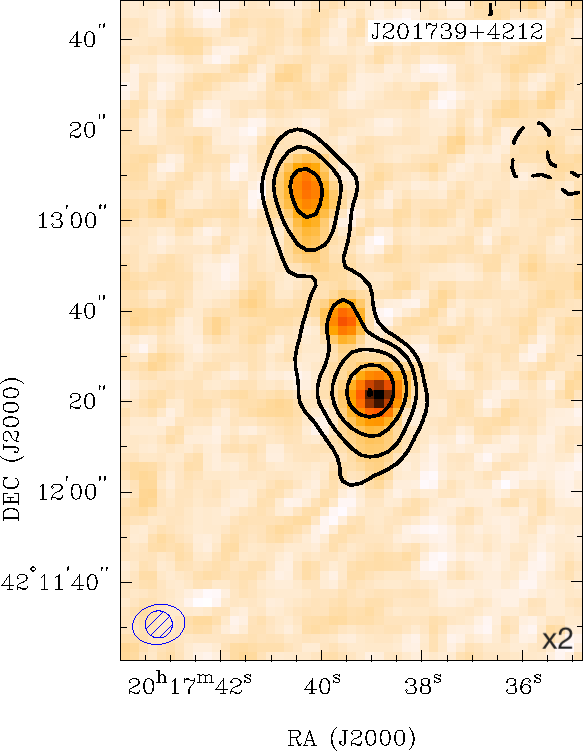} 
    \includegraphics[width=0.35\textwidth]{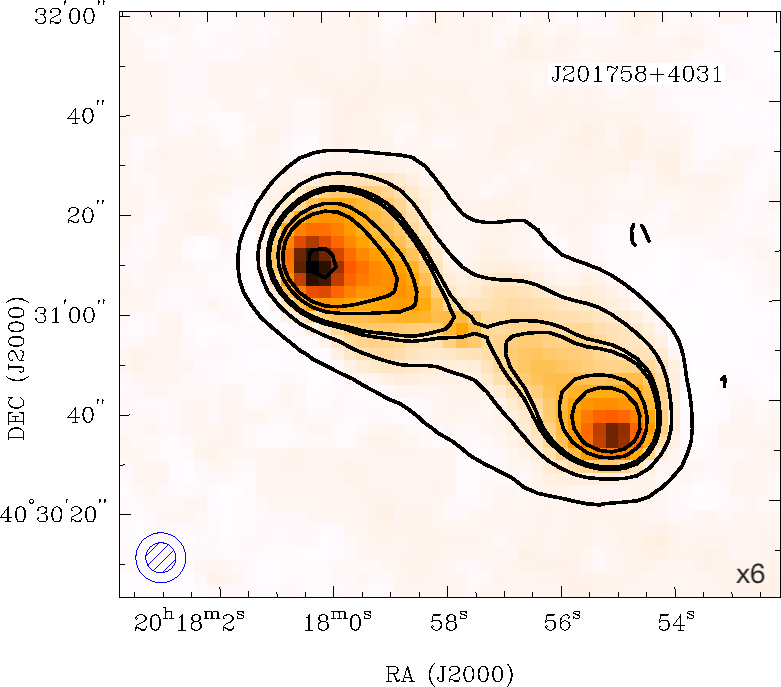} 
    \includegraphics[width=0.35\textwidth]{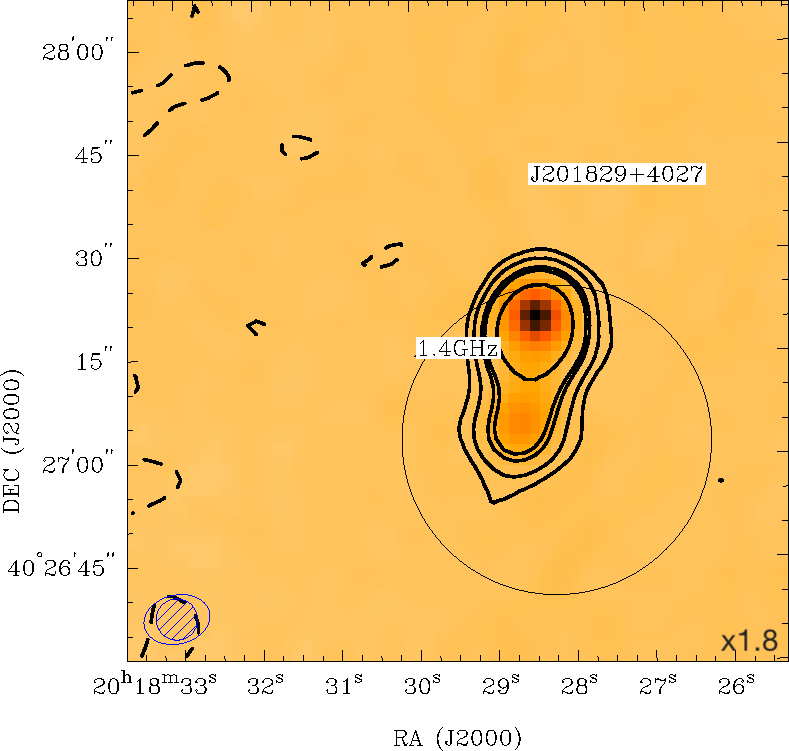}\\ 
    \includegraphics[width=0.3\textwidth]{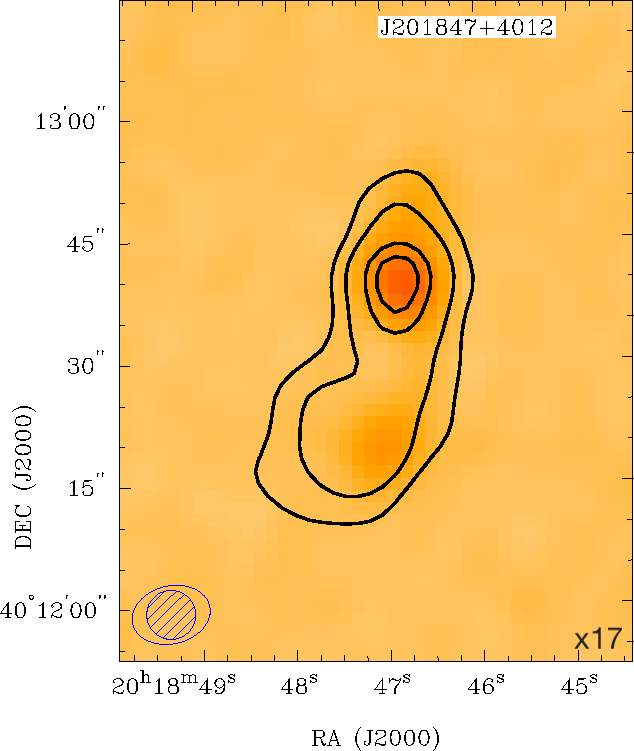} 
    \includegraphics[width=0.2\textwidth]{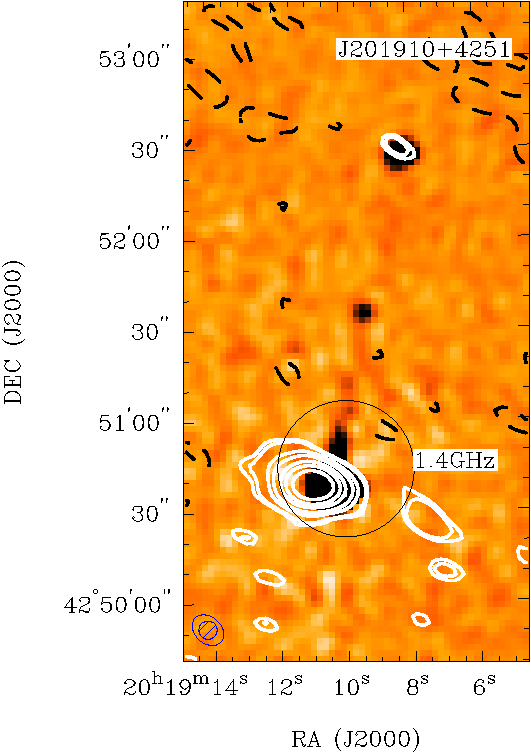}
    \includegraphics[width=0.25\textwidth]{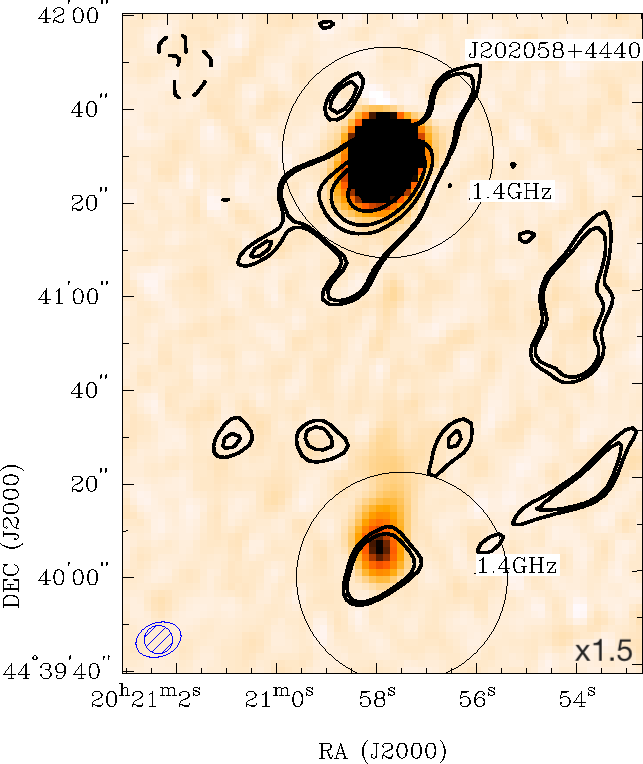} 
\end{center}
\caption{GMRT images of the double-lobed sources at 610~MHz (colour scale) and 325~MHz (contours). The contour levels and colour scale interval for  each source are as follows: 
\#1-J201255+404423: $-2.5$, 2.5, 3, 15, 27, 60 and 90, in units of $\sigma$ (=0.18~mJy\,beam$^{-1}$); 1.6, 20]~mJy\,beam$^{-1}$. 
\#2-J201418+412931: $-2.5$, 2.5, 3, 15, 27, 60, and 90, in units of $\sigma$ (=0.45~mJy\,beam$^{-1}$); [3, 13]~mJy\,beam$^{-1}$.
\#3-J201512+410457: $-2.5$, 2.5, 5, 9, 15, 21, and 41, in units of $\sigma$ (=0.45~mJy\,beam$^{-1}$); [4, 11]~mJy\,beam$^{-1}$.
\#4-J201739+421243: $-2.5$, 2.5, 5, 9, 21, and 27, in units of $\sigma$ (=0.2~mJy\,beam$^{-1}$); [0.4, 4]~mJy\,beam$^{-1}$.
\#5-J201758+403100: $-2.5$, 2.5, 15, 27, 30, 45, 55, and 100, in units of $\sigma$ (=0.6~mJy\,beam$^{-1}$); [0.6, 17]~mJy\,beam$^{-1}$.
\#6-J201828+402714: $-2.5$, 2.5, 3, 15, 27, 60, and 90, in units of $\sigma$ (=0.18~mJy\,beam$^{-1}$); [6, 21]~mJy\,beam$^{-1}$. 
\#7-J201847+401230: $-2.5$, 2.5, 5, 9, 11, and 21, in units of $\sigma$ (=1.7~mJy\,beam$^{-1}$); [6, 21]~mJy\,beam$^{-1}$. 
\#8-J201910+425136: $-2.5$, 2.5, 3, 6, 9, 15, 21, and 30, in units of $\sigma$ (=0.7~mJy\,beam$^{-1}$); [0.7, 9]~mJy\,beam$^{-1}$. 
\#9-J202058+444050: $-2.5$, 2.5, 3, 15, 27, 60, and 90, in units of $\sigma$ (=0.15~mJy\,beam$^{-1}$); [0.7, 9]~mJy\,beam$^{-1}$. 
The synthesised beams are shown at the bottom-left corner of each image.   NVSS detections are marked with a 45$''$ diameter circle and labelled as `1.4 GHz'. For the x factor of the bottom-right corner, see text.}
\label{fig:galaxitas1}  
\end{figure*}

\begin{figure*}
    \includegraphics[width=0.3\textwidth]{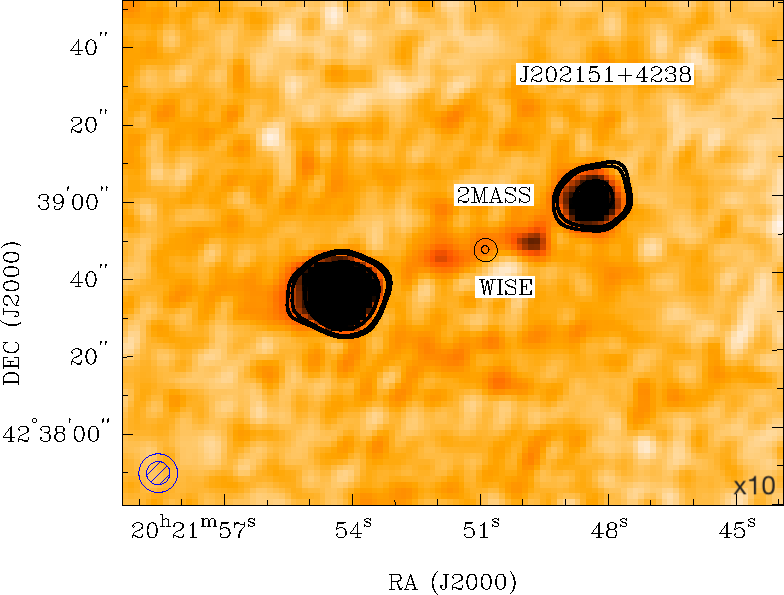}
    \includegraphics[width=0.3\textwidth]{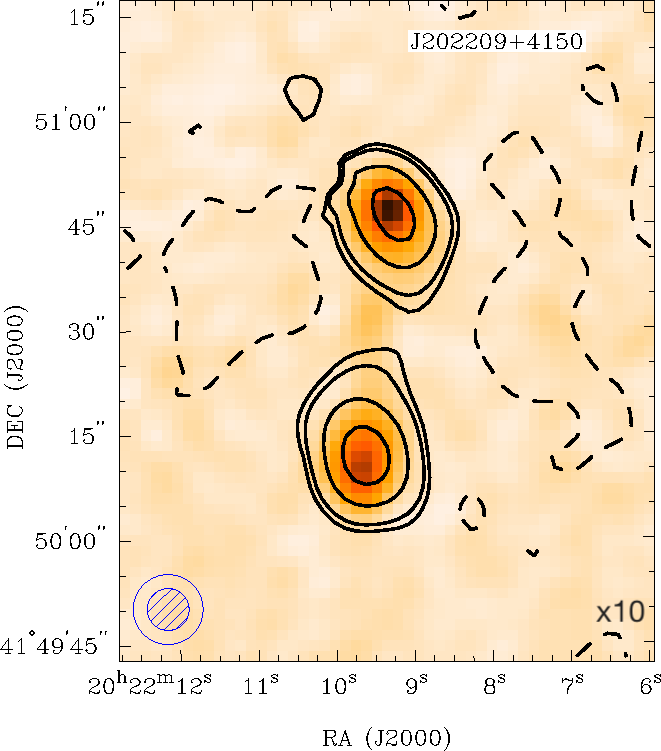}
    \includegraphics[width=0.3\textwidth]{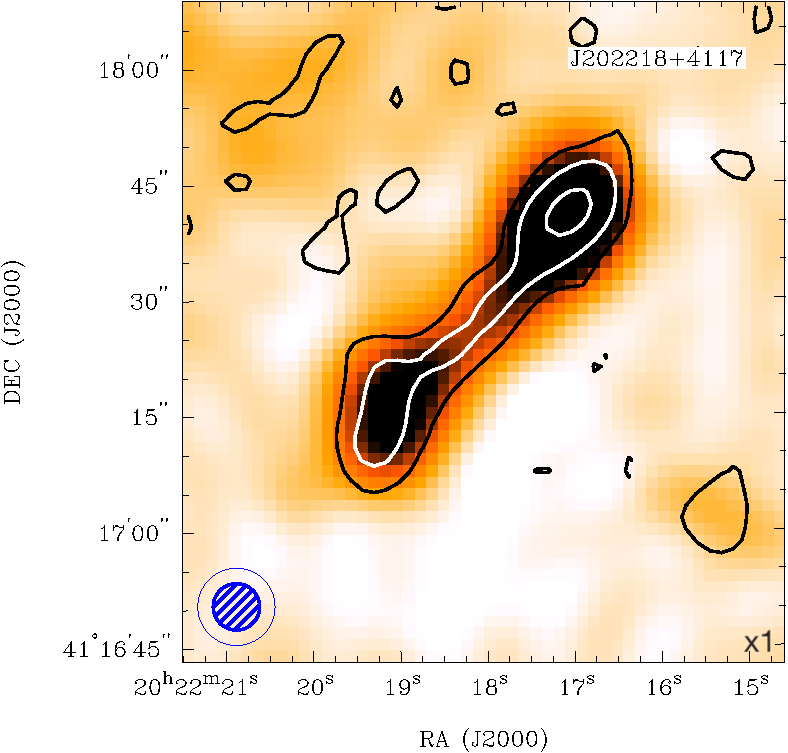}\\ 
    \includegraphics[width=0.3\textwidth]{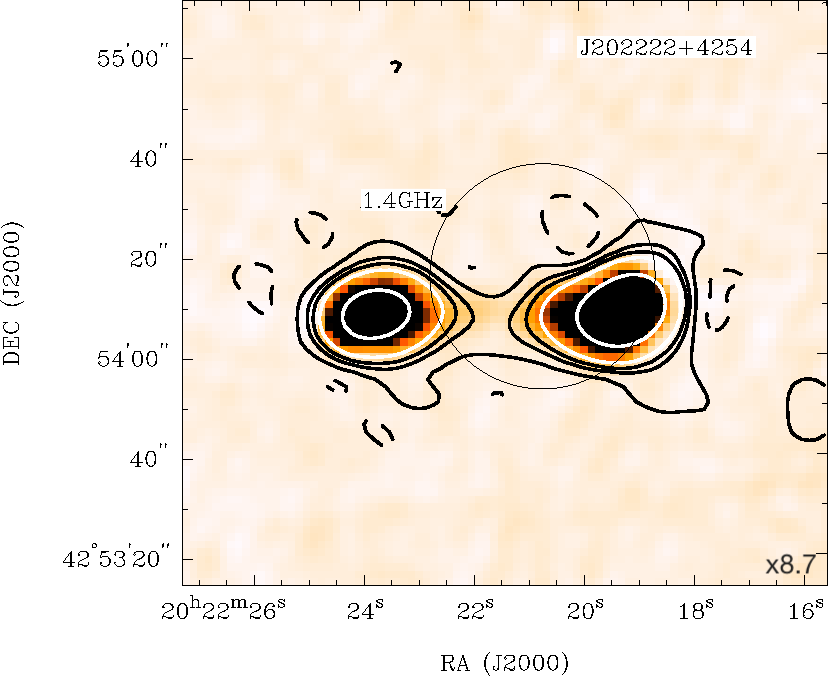} 
    \includegraphics[width=0.3\textwidth]{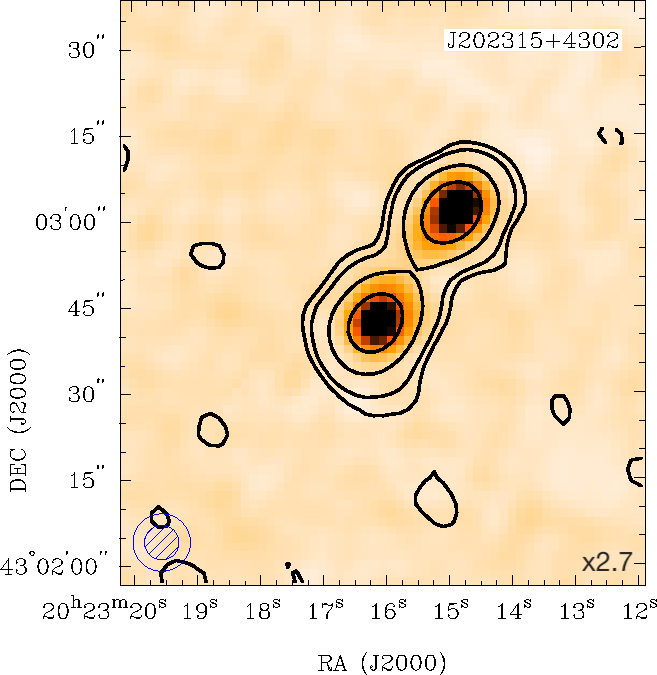} 
    \includegraphics[width=0.3\textwidth]{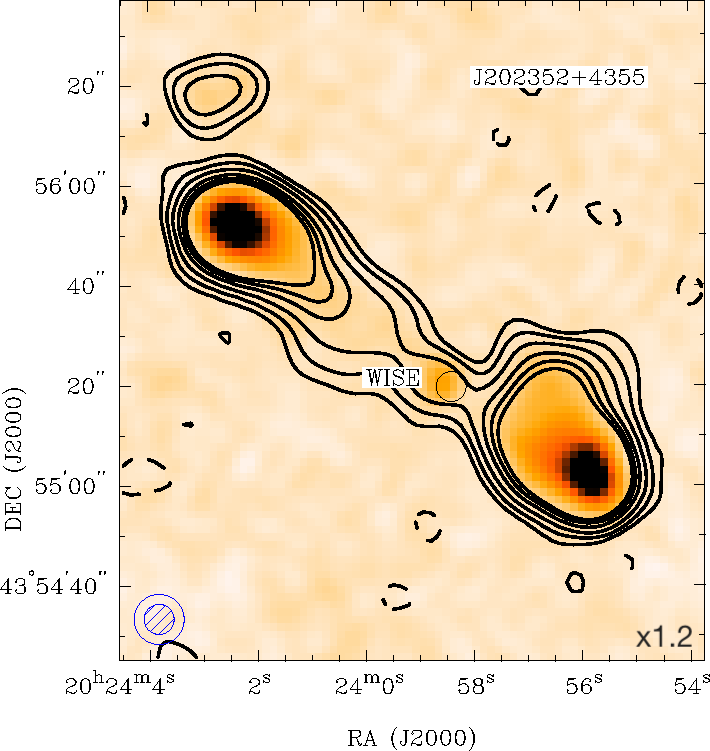}\\ 
    \includegraphics[width=0.3\textwidth]{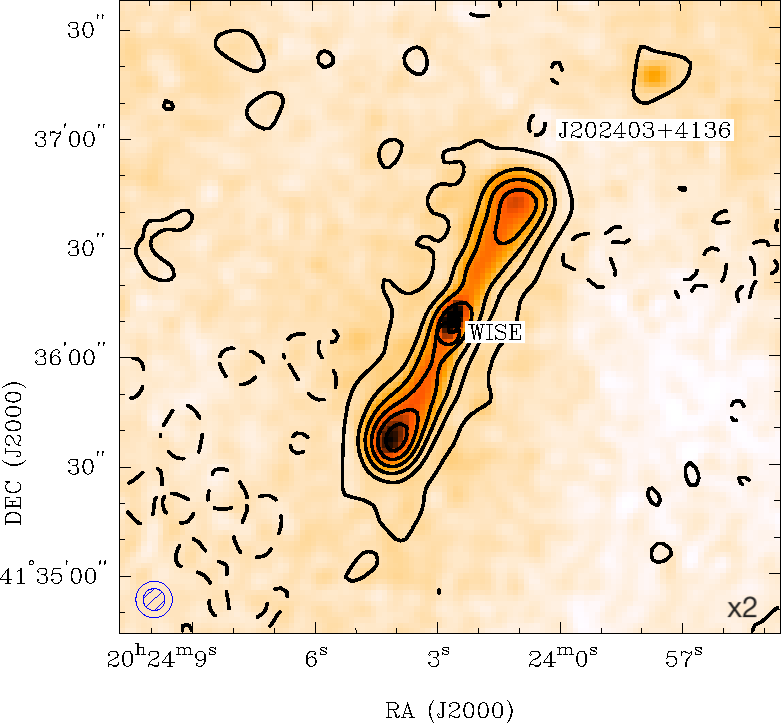} 
    \includegraphics[width=0.3\textwidth]{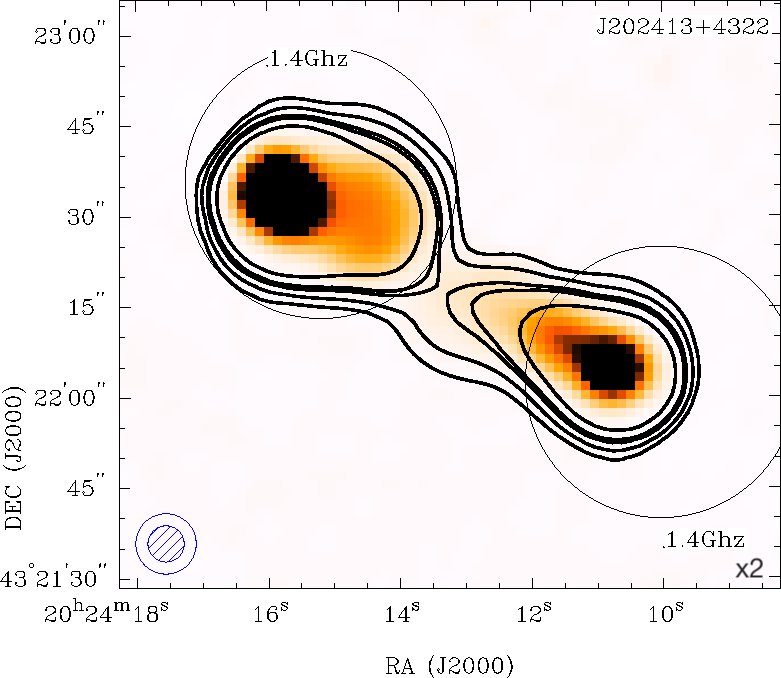} 
    \includegraphics[width=0.3\textwidth]{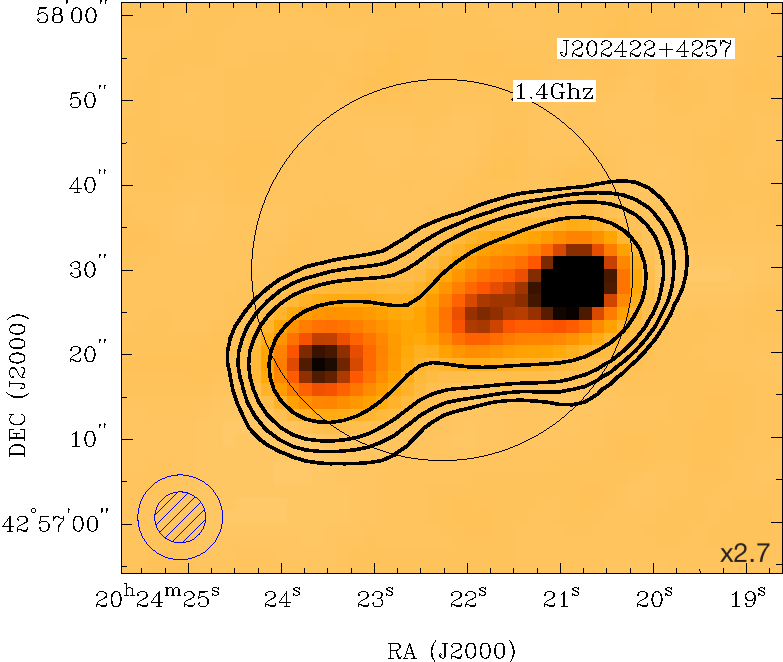} 
    \caption{GMRT images of the double-lobed sources at 610~MHz (colour scale) and 325~MHz (contours). The contour levels and colour scale interval for  each source are as follows: 
    \#10-J202151+423847: $-2.5$, 2.5, 3, and 5, in units of $\sigma$ (=1~mJy\,beam$^{-1}$); [0.3, 1]~mJy\,beam$^{-1}$.   
    \#11-J202209+415033: $-2.5$, 2.5, 3, 5, and 15, in units of $\sigma$ (=1~mJy\,beam$^{-1}$); [0.4, 3.9]~mJy\,beam$^{-1}$. 
    \#12-J202218+411728: $-2.5$, 2.5, 10, and 30, in units of $\sigma$ (=0.1~mJy\,beam$^{-1}$); [0.5, 3.8]~mJy\,beam$^{-1}$. 
    \#13-J202222+425409: $-2.5$, 2.5, 9, 15, 25, and 85, in units of $\sigma$(=0.87~mJy\,beam$^{-1})$; [0.3, 6.1]~mJy\,beam$^{-1}$. 
    \#14-J202315+430252:  $-2.5$, 2.5, 9, 15, and 25, in units of $\sigma$(=0.27~mJy\,beam$^{-1}$); [0.3, 7.9]~mJy\,beam$^{-1}$. 
    \#15-J202359+435525: $-2.5$, 2.5, 5, 9, 15, 21, and 27, in units of $\sigma$(=0.12~mJy\,beam$^{-1}$); [0.4, 4.2]~mJy\,beam$^{-1}$. 
    \#16-J202403+413611: $-2.5$, 2.5, 15, 25, 35, 45, and 55, in units of $\sigma$(=0.2~mJy\,beam$^{-1}$); [0.4, 4.2]~mJy\,beam$^{-1}$. 
    \#17-J202413+432220: $-2.5$, 2.5, 5, 15, 21, and 30, in units of $\sigma$(=0.2~mJy\,beam$^{-1}$; [1, 4.4]~mJy\,beam$^{-1}$. 
    $\#$18-J202422+425722: $-2.5$, 2.5, 5, 15, and 21, in units of $\sigma$(=0.27~mJy\,beam$^{-1}$); [6, 21]~mJy\,beam$^{-1}$. The synthesised beams are shown at the bottom-left corner of each image. 
     NVSS detections are marked with a 45$''$ diameter circle and labelled as `1.4 GHz'. WISE and 2MASS sources are marked and labelled as well.For the x factor of the bottom-right corner, see text.
    } 
    \label{fig:galaxitas2} 
    
\end{figure*}

 \begin{figure*}
    \includegraphics[width=0.33\textwidth]{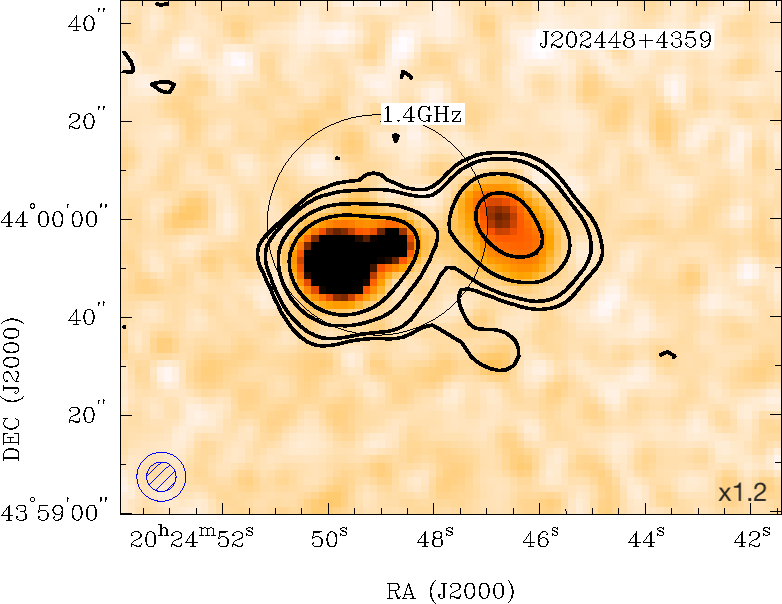}
    \includegraphics[width=0.3\textwidth]{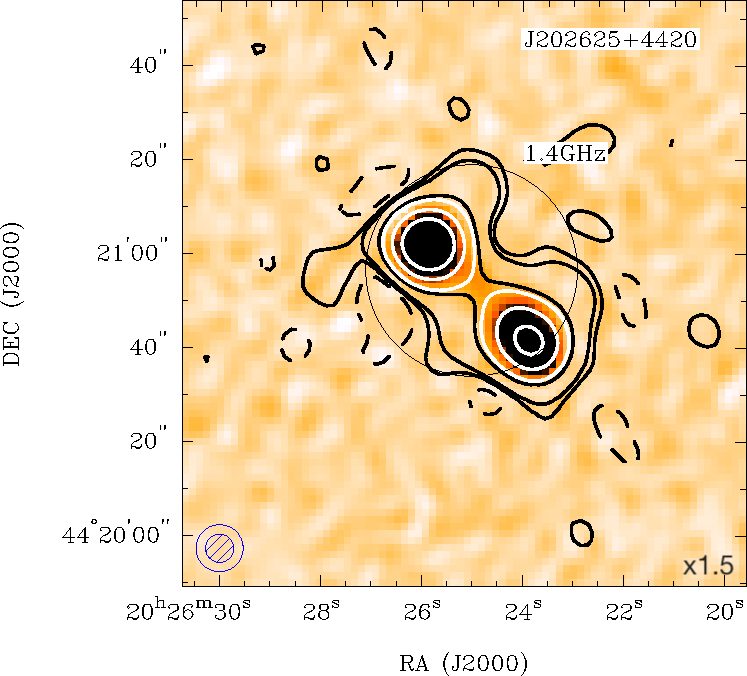} 
    \includegraphics[width=0.3\textwidth]{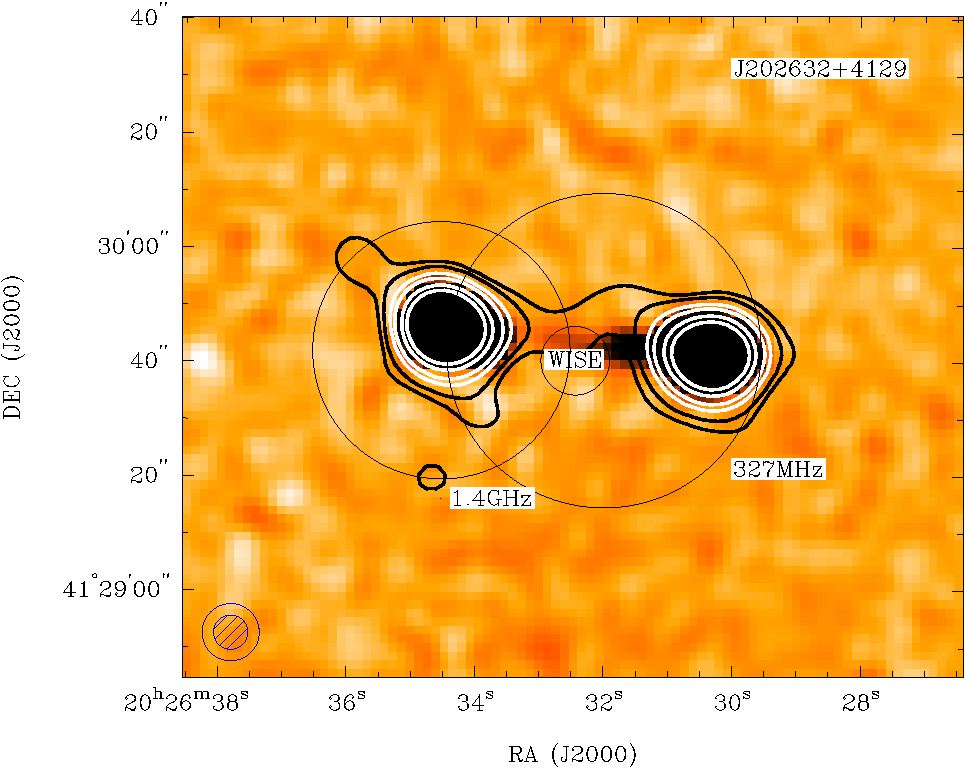} \\
    \includegraphics[width=0.27\textwidth]{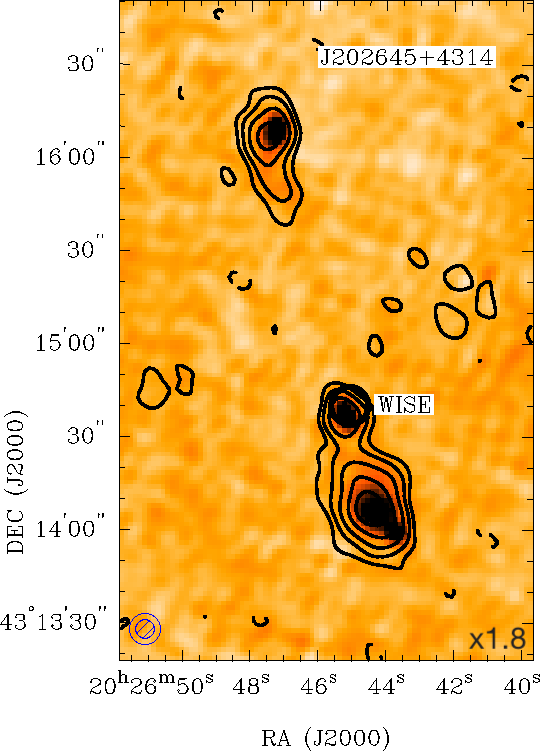} 
    \includegraphics[width=0.35\textwidth]{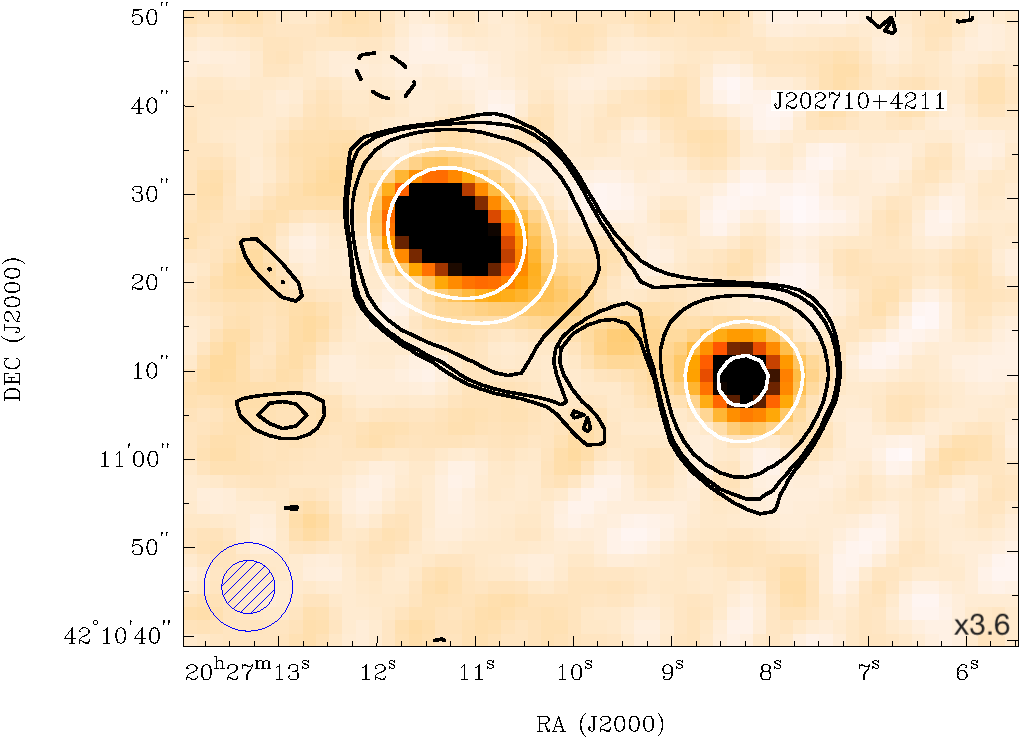} 
    \includegraphics[width=0.33\textwidth]{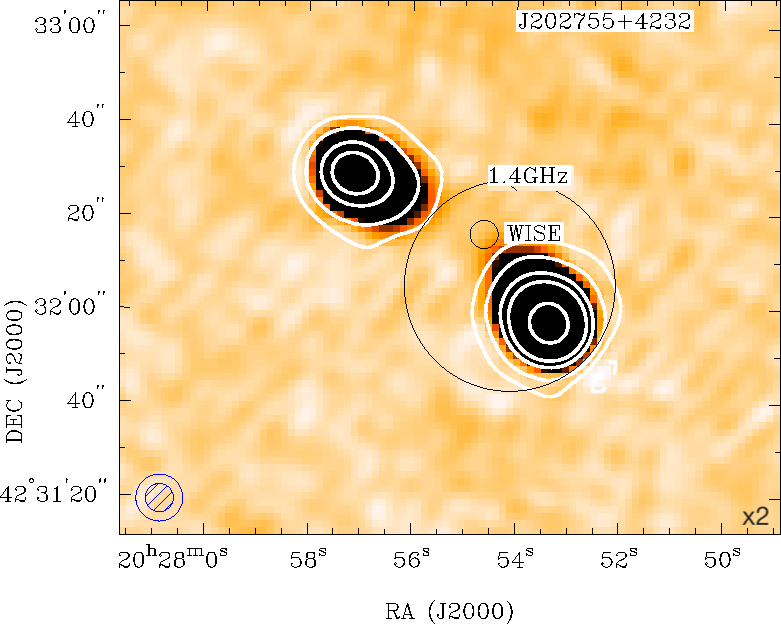} \\
    \includegraphics[width=0.3\textwidth]{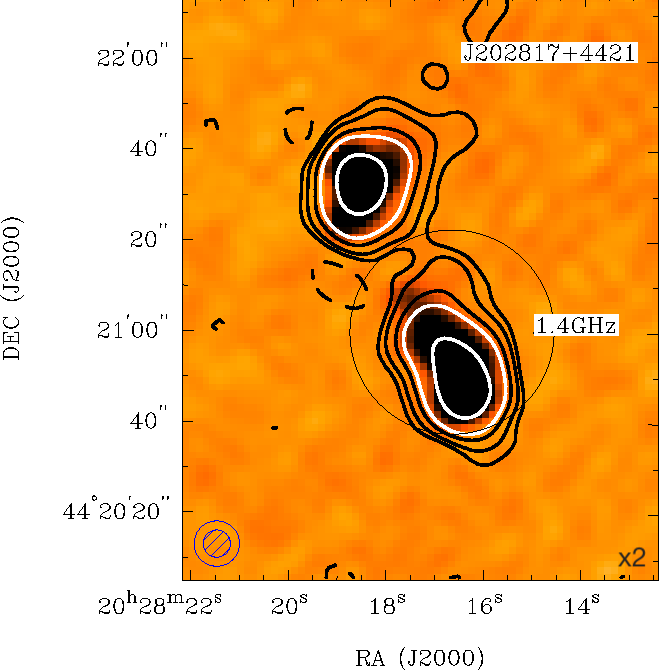} 
    \includegraphics[width=0.3\textwidth]{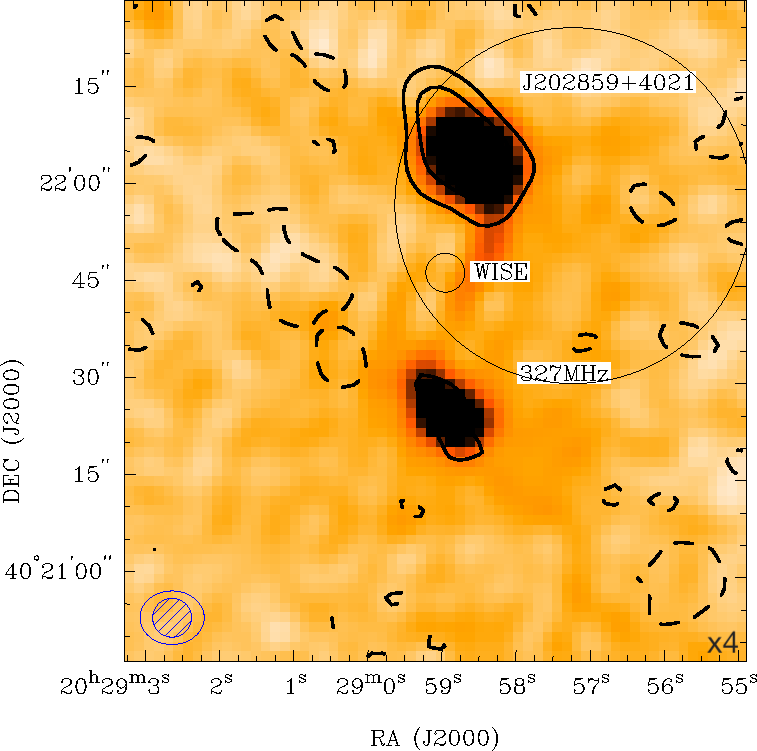} 
    \includegraphics[width=0.3\textwidth]{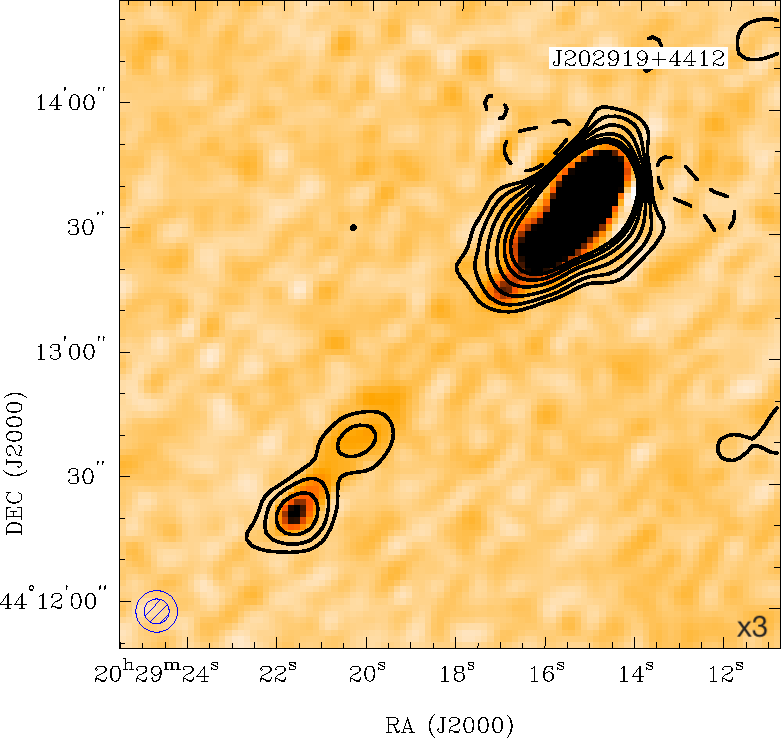} 
    \caption{GMRT images of the double-lobed sources at 610~MHz (colour scale) and 325~MHz (contours). The contour levels  and colour scale interval for  each source are as follows: 
    \#19-J202448+435956: $-2.5$, 2.5, 5, 10, and 27, in units of $\sigma$(=0.12~mJy\,beam$^{-1}$); [0.3, 2.6]~mJy\,beam$^{-1}$. 
    \#20-J202625+442048: $-2.5$,  2.5,  5, 25, 35, 75, and 135, in units of $\sigma$(=0.15~mJy\,beam$^{-1}$);[0.3, 2]~mJy\,beam$^{-1}$. 
    \#21-J202632+412942: $-2.5$, 2.5, 5, 10, 15, 25, and 35, in units of $\sigma$(=1~mJy\,beam$^{-1}$); [0.3, 0.6]~mJy\,beam$^{-1}$. 
    \#22-J202645+431435: $-2.5$, 2.5, 5, 9, 15, 21, and 27, in units of $\sigma$(=0.18~mJy\,beam$^{-1}$); [0.2, 0.7]~mJy\,beam$^{-1}$. 
    \#23-J202710+421113: $-2.5$, 2.5, 3, 5, 15, and 35, in units of $\sigma$(=0.36~mJy\,beam$^{-1}$); [0.3, 3.9]~mJy\,beam$^{-1}$. 
    \#24-J202755+423215: $-2.5$, 2.5, 30, 50, 130, and 310, in units of $\sigma$(=0.2~mJy\,beam$^{-1}$); [0.3, 3.9]~mJy\,beam$^{-1}$. 
    \#25-J202817+442114: $-2.5$, 2.5, 5, 10, 15, 35, and 85, in units of $\sigma$(=0.16~mJy\,beam$^{-1}$); [0.8, 1]~mJy\,beam$^{-1}$. 
    \#26-J202859+402141: $-2.5$, 2.5, 5, 9, 15, 21, and 27, in units of $\sigma$(=0.4~mJy\,beam$^{-1}$); [0.6, 1.8]~mJy\,beam$^{-1}$. 
    \#27-J202919+441341 $-2.5$, 2.5, 5, 9, 15, 21, and 27, in units of $\sigma$(=0.3~mJy\,beam$^{-1}$); [0.4, 2]~mJy\,beam$^{-1}$. The synthesised beams are shown at the bottom-left corner of each image. NVSS and WSRT detections are marked and labelled as `1.4 GHz'  and `327 MHz'. WISE and 2MASS sources are marked and labelled as well. For the x factor of the bottom-right corner, see text.} 
    \label{fig:galaxitas3}   
\end{figure*}

 \begin{figure*}
    \centering   
     \includegraphics[width=0.3\textwidth]{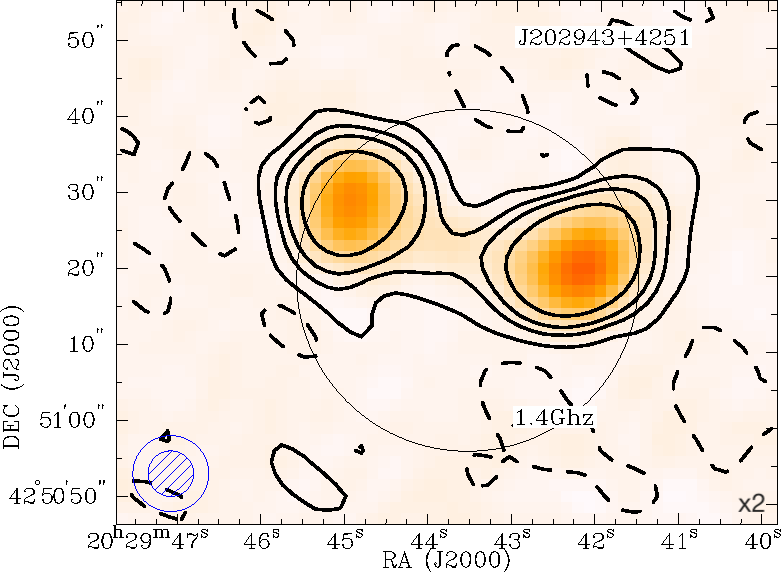} 
     \includegraphics[width=0.27\textwidth]{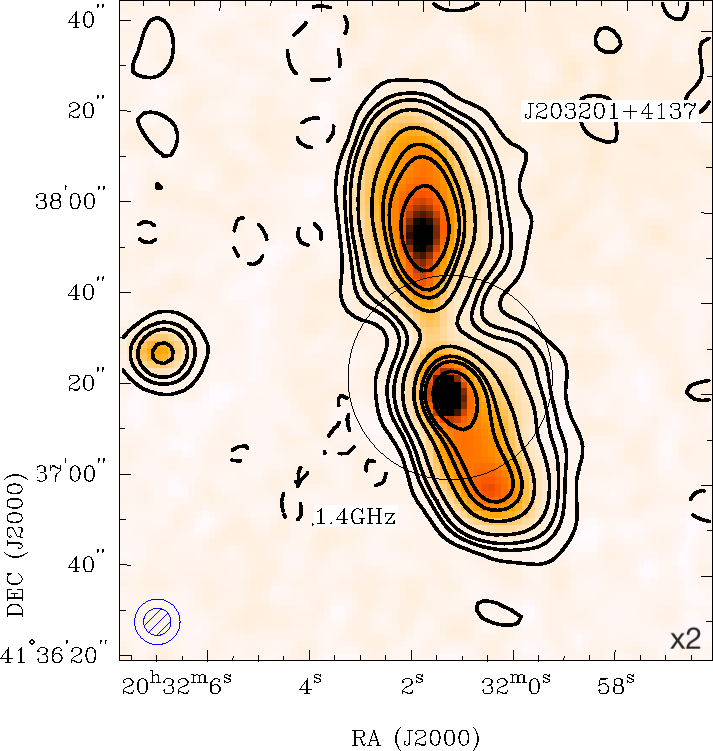} 
     \includegraphics[width=0.3\textwidth]{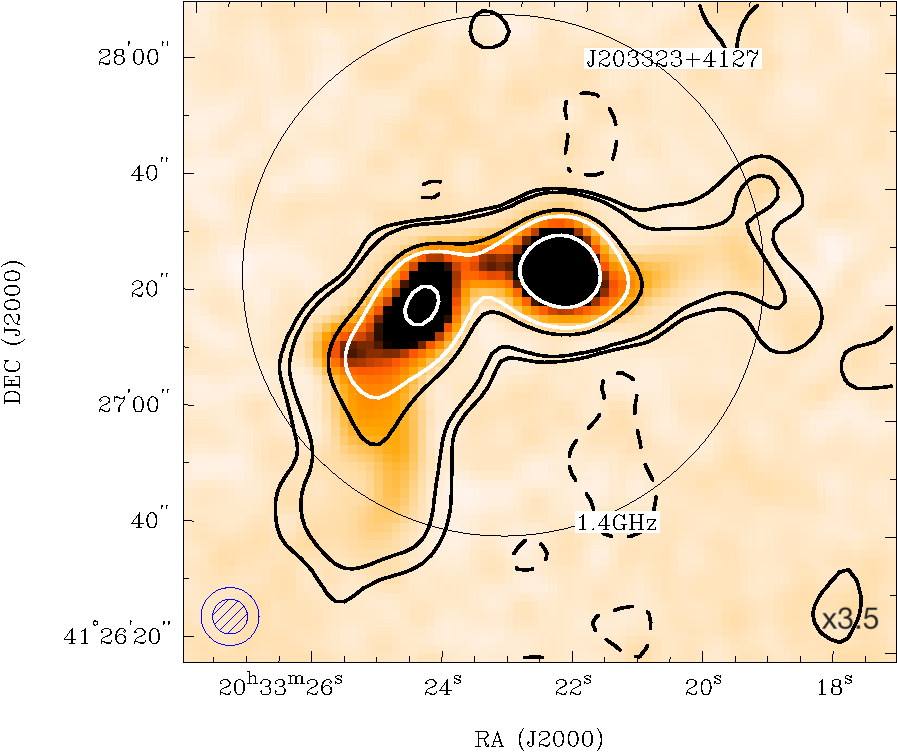} \\
      \includegraphics[width=0.3\textwidth]{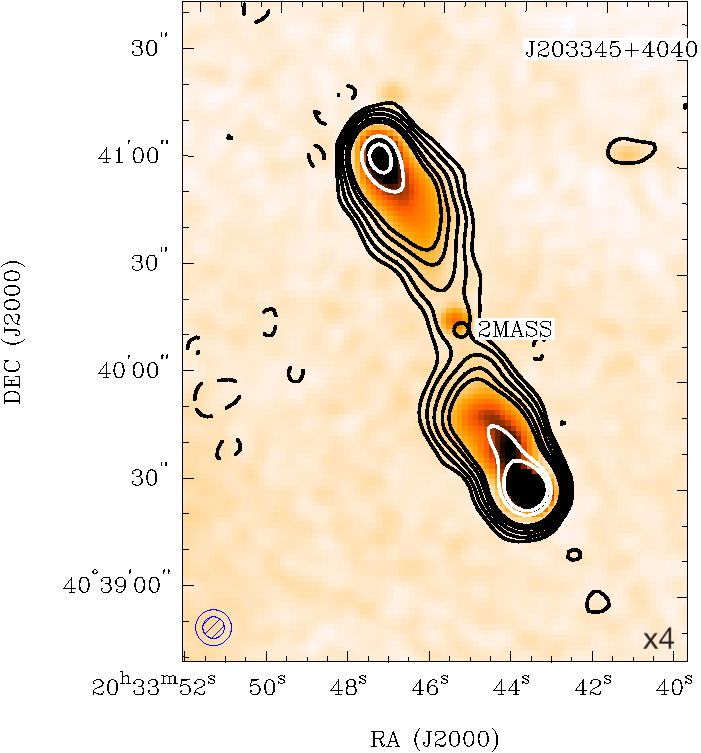} 
      \includegraphics[width=0.3\textwidth]{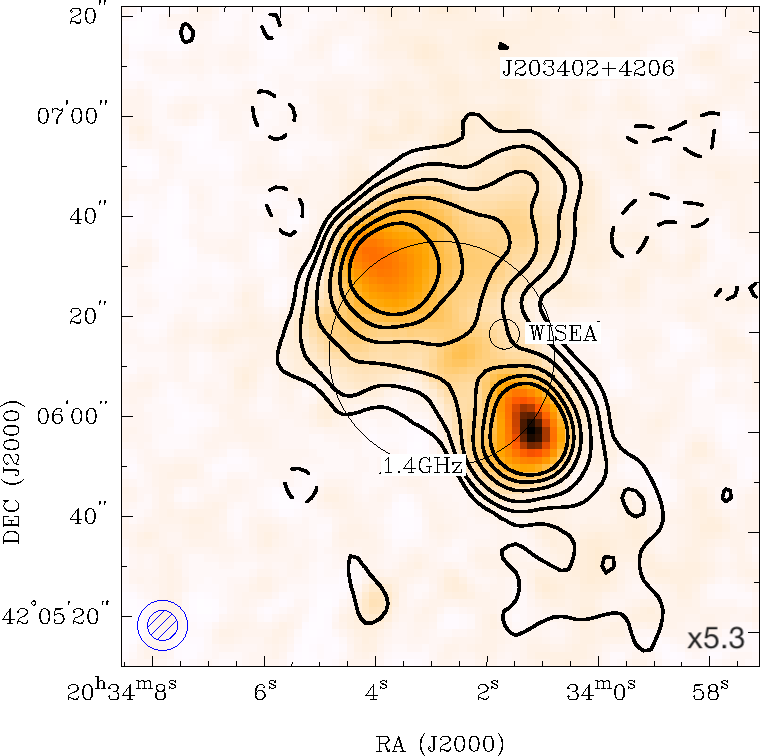} 
      \includegraphics[width=0.3\textwidth]{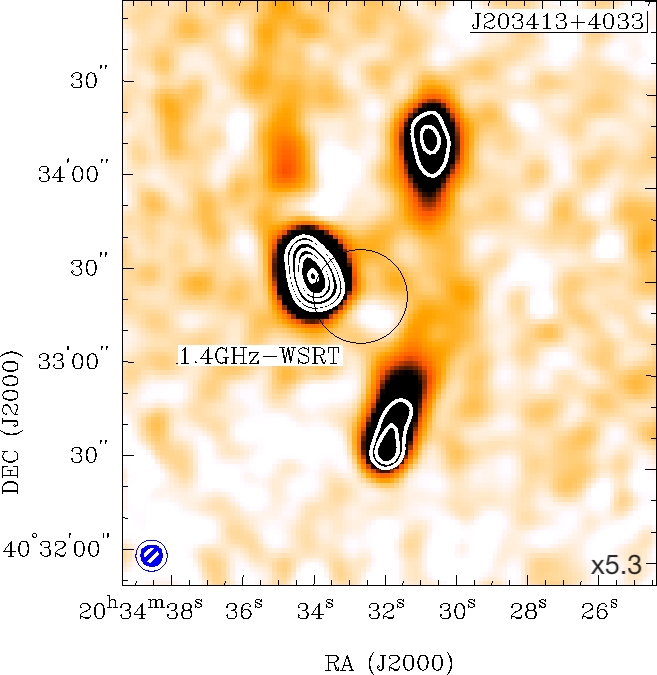}\\ 
      \includegraphics[width=0.3\textwidth]{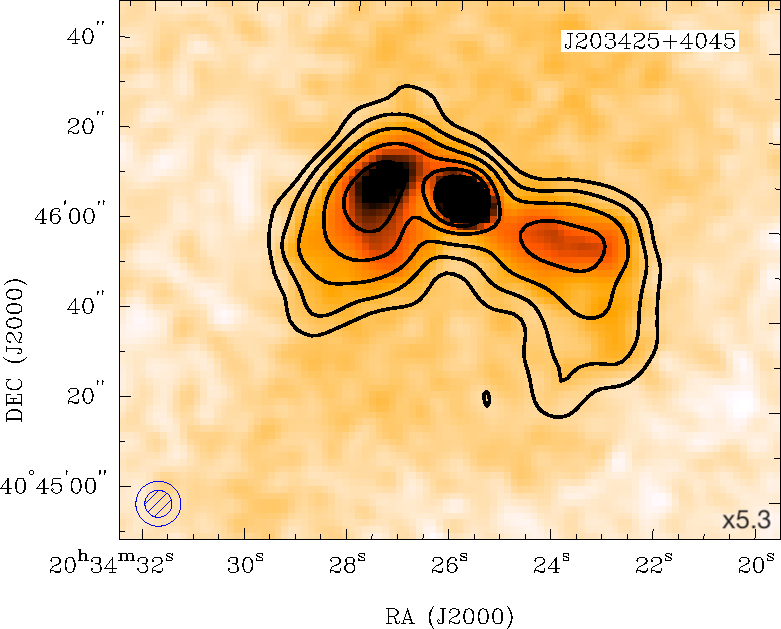} 
      \includegraphics[width=0.3\textwidth]{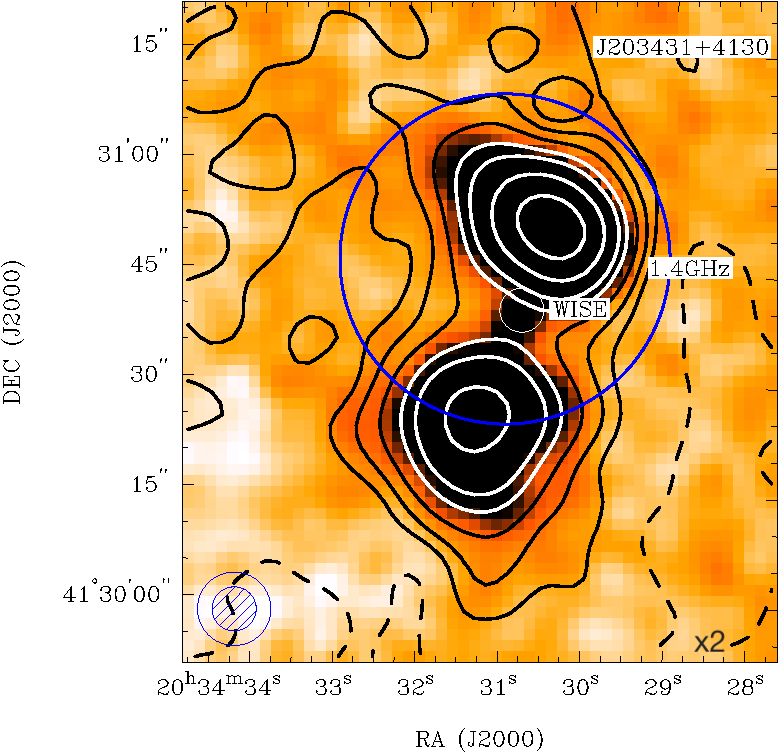}
      \includegraphics[width=0.3\textwidth]{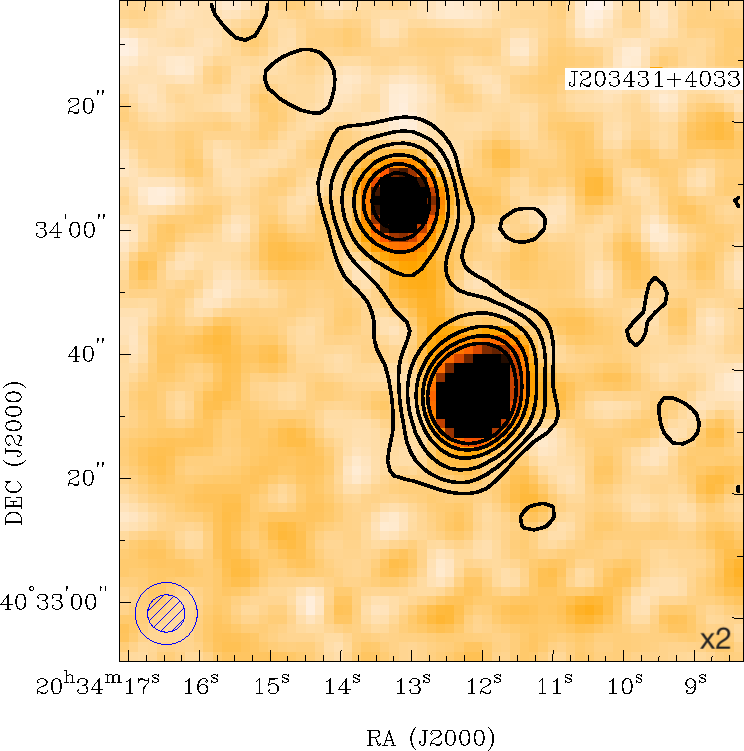} 
     
      \caption{GMRT images of the double-lobed sources at 610~MHz (colour scale) and 325~MHz (contours). The contour levels and colour scale interval for  each source are as follows: 
      \#28-J202943+425122: [$-2.5$, 2.5, 9, 15, 27] in units of $\sigma$(=0.2~mJy\,beam$^{-1}$); [0.4, 1]~mJy\,beam$^{-1}$. 
      \#29-J203201+413730: $-2.5$, 2.5, 9, 15, 27, 57, 87, 107, and 150, in units of $\sigma$(=0.2~mJy\,beam$^{-1}$); [0.4, 1]~mJy\,beam$^{-1}$. 
      \#30-J203323+412720: $-2.5$, 2.5, 5, 27, 40, and 100, in units of $\sigma$(=0.35~mJy\,beam$^{-1}$); [0.6, 5]~mJy\,beam$^{-1}$. 
      \#31-J203345+404015: $-2.5$, 2.5, 5, 9, 15, 21, 55, and 75, in units of $\sigma$(=0.4~mJy\,beam$^{-1}$); [0.5, 5]~mJy\,beam$^{-1}$. 
      \#32-J203402+420615: $-2.5$, 2.5, 5, 9, 15, 21, and 27, in units of $\sigma$(=0.53~mJy\,beam$^{-1}$); [0.4, 10]~mJy\,beam$^{-1}$. 
      \#33-J203413+403349: $-2.5$, 2.5, 5, 9, 15, 21, and 27, in units of $\sigma$(=0.53~mJy\,beam$^{-1}$); [0.4, 10]~mJy\,beam$^{-1}$. 
      \#34-J203425+404605: $-2.5$, 2.5, 5, 9, 15, 21, and 27, in units of $\sigma$(=0.53~mJy\,beam$^{-1}$); [0.3, 2.8]~mJy\,beam$^{-1}$. 
      \#35-J203431+413037: $-2.5$, 2.5, 5, 9, 15,  21,  35, and 55, in units of $\sigma$(=0.2~mJy\,beam$^{-1}$); [0.4, 14]~mJy\,beam$^{-1}$. 
      \#36-J203431+403332: $-2.5$, 2.5, 5, 9, 15, 21, and 27, in units of $\sigma$(=0.53~mJy\,beam$^{-1}$); [0.2, 0.7]~mJy\,beam$^{-1}$. 
      The synthesised beams are shown at the bottom-left corner of each image. NVSS and WSRT beams detection are marked with a circles and labelled as `1.4 GHz' and `1.4GHz-WSRT'. WISE and 2MASS position errors are marked with circles and labelled as well. For the x factor of the bottom-right corner, see text.}
      \label{fig:galaxitas4}   
 \end{figure*}

\begin{figure*}
    \centering
    \includegraphics[width=0.35\textwidth]{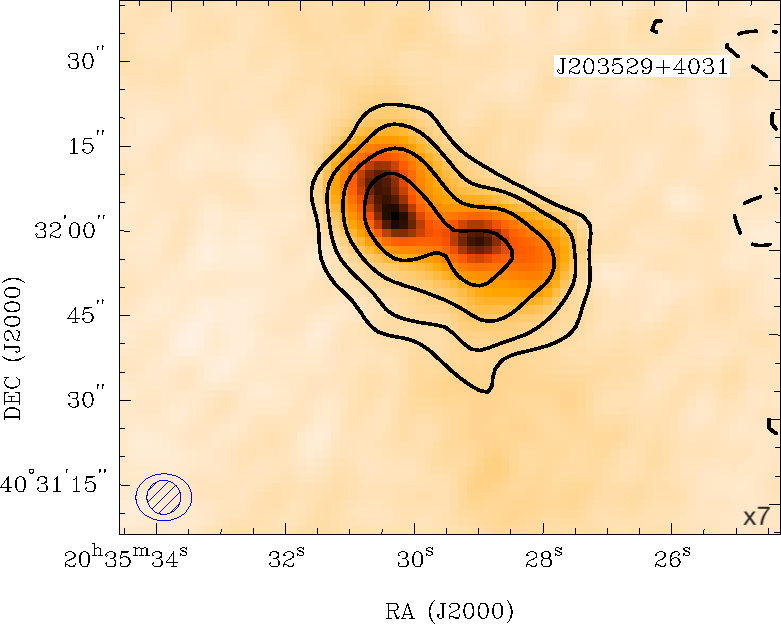}  
    \includegraphics[width=0.3\textwidth]{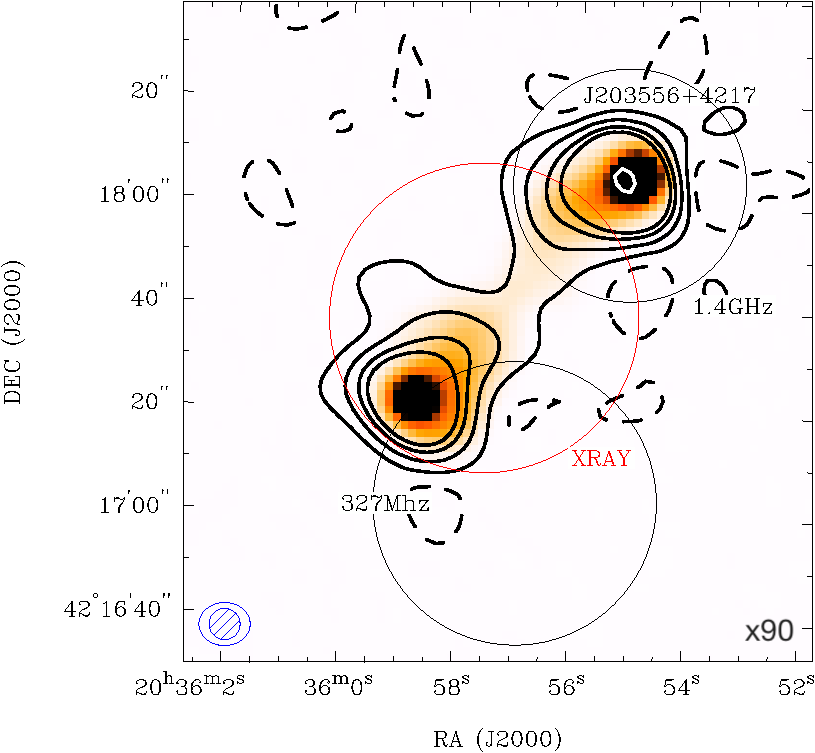} 
    \includegraphics[width=0.27\textwidth]{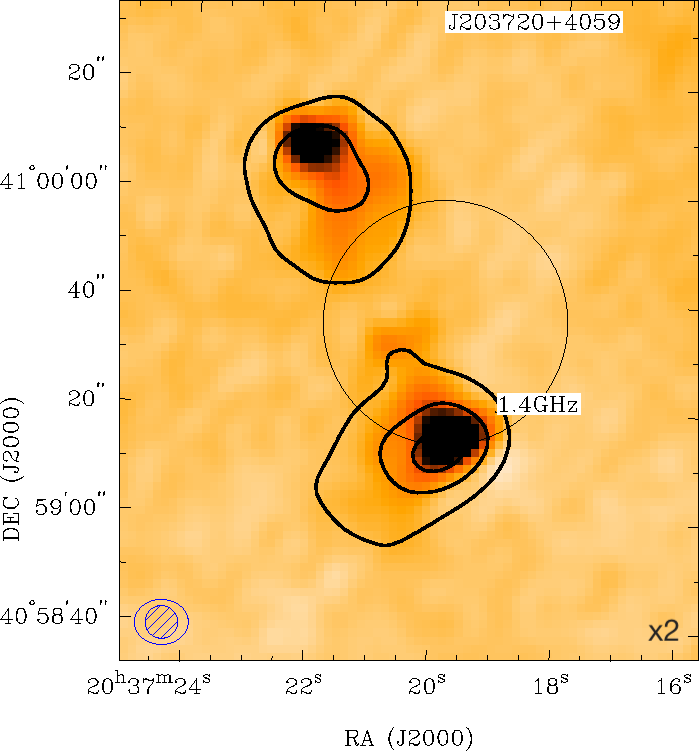} 
     \caption{GMRT images of the double-lobed sources at 610~MHz (colour scale) and 325~MHz (contours). The contour levels  and colour scale interval for  each source are as follows: 
    \#37-J203529+403201: $-2.5$, 2.5, 5, 9, 15, 21, and 27, in units of $\sigma$(=0.7~mJy\,beam$^{-1}$); [1, 8]~mJy\,beam$^{-1}$. 
     \#38-J203556+421744: $-2.5$, 2.5, 10, 20, 30, 120, 200, 300, and 500, in units of $\sigma$(=9~mJy\,beam$^{-1}$); [3, 237]~mJy\,beam$^{-1}$. 
     \#39-J203720+405941: $-2.5$, 2.5, and 5, in units of $\sigma$(=0.2~mJy\,beam$^{-1}$); [2, 7]~mJy\,beam$^{-1}$. 
     The synthesised beams are shown at the bottom-left corner of each image. NVSS beam detection is marked with a circle and labelled as `1.4 GHz' while the X-ray detection is shown with a red circle.
   For the x factor of the bottom-right corner, see text.
      }
    \label{fig:galaxitas5}
\end{figure*}

\begin{figure*}
    \centering
    \includegraphics[width=0.3\textwidth]{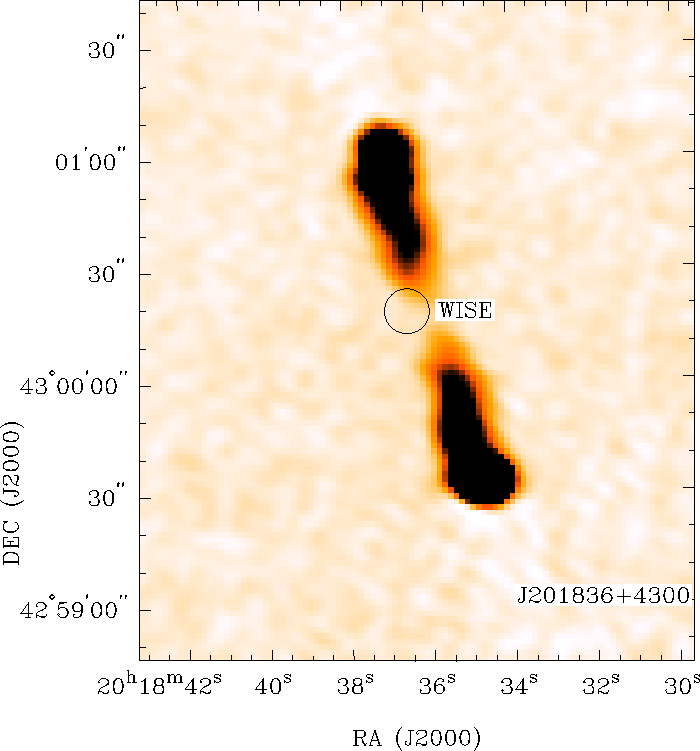} 
    \includegraphics[width=0.3\textwidth]{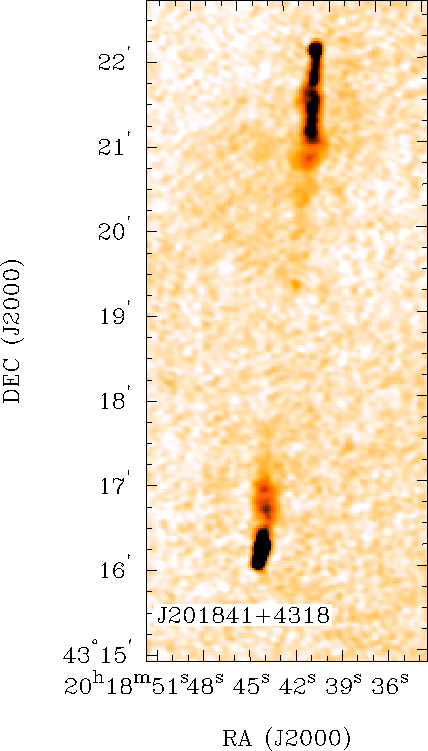}\\ 
    \includegraphics[width=0.3\textwidth]{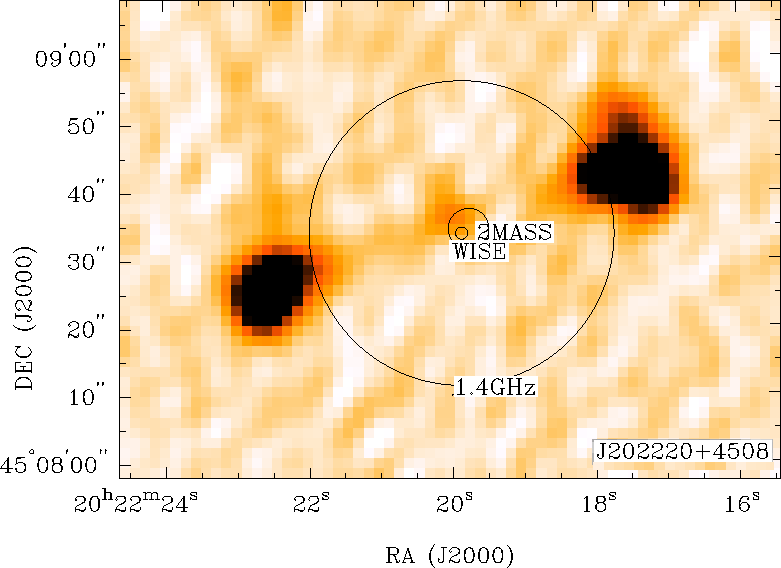} 
    \includegraphics[width=0.3\textwidth]{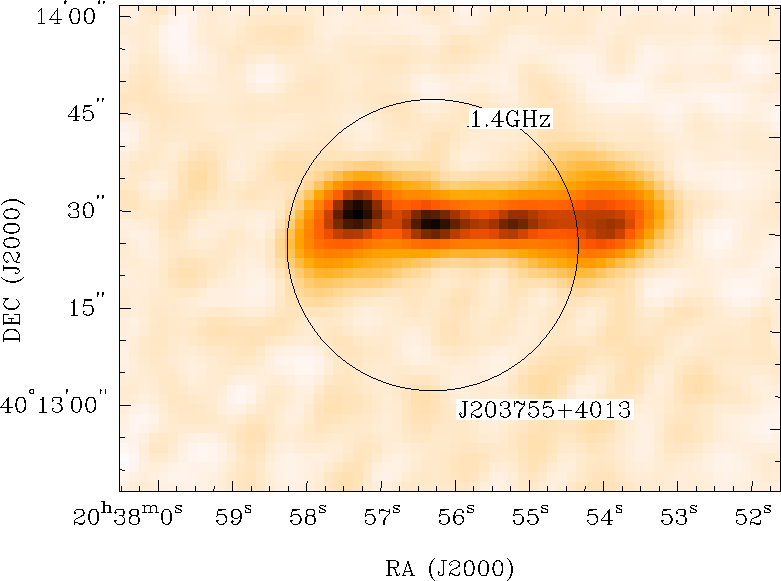} 

     \caption{GMRT images of the double-lobed sources at 610~MHz (colour scale). NVSS beam detection is marked with a circle and labelled as `1.4 GHz'. WISE and 2MASS position errors are marked with circles and labelled as well. The sources  and colour scale interval represented are: $\#$40-J201836+430018: [0.4, 5]~mJy\,beam$^{-1}$; \#41-J201841+431844: [0.2, 1.6]~mJy\,beam$^{-1}$; \#42-J202220+450835:  [0.4, 5.8]~mJy\,beam$^{-1}$; \#43-J203755+401328: [0.4, 5.8]~mJy\,beam$^{-1}$.
     }
    \label{fig:galaxitas6}
\end{figure*}

\setcounter{figure}{0}
\renewcommand{\thefigure}{B\arabic{figure}}

\begin{figure*}
\begin{center}
    \includegraphics[width=0.3\textwidth]{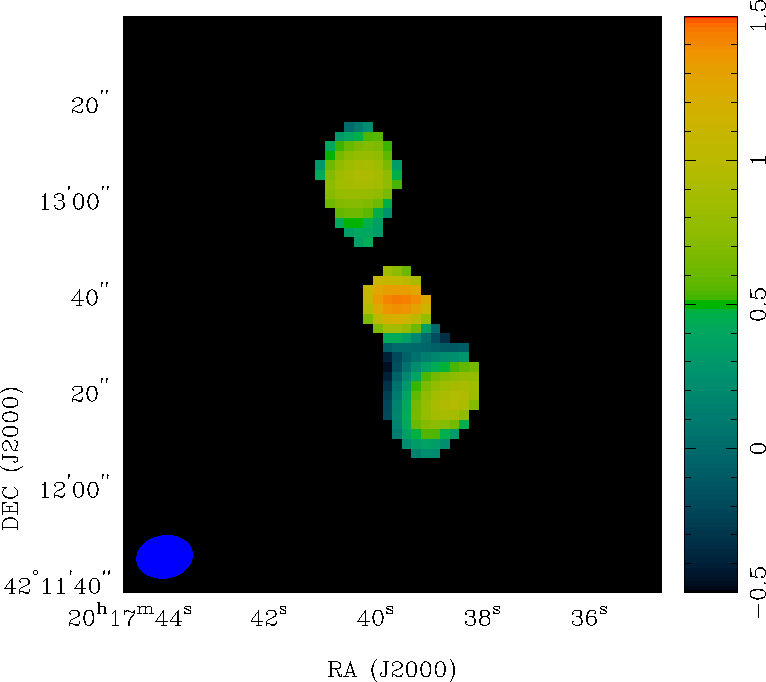} 
     \includegraphics[width=0.3\textwidth]{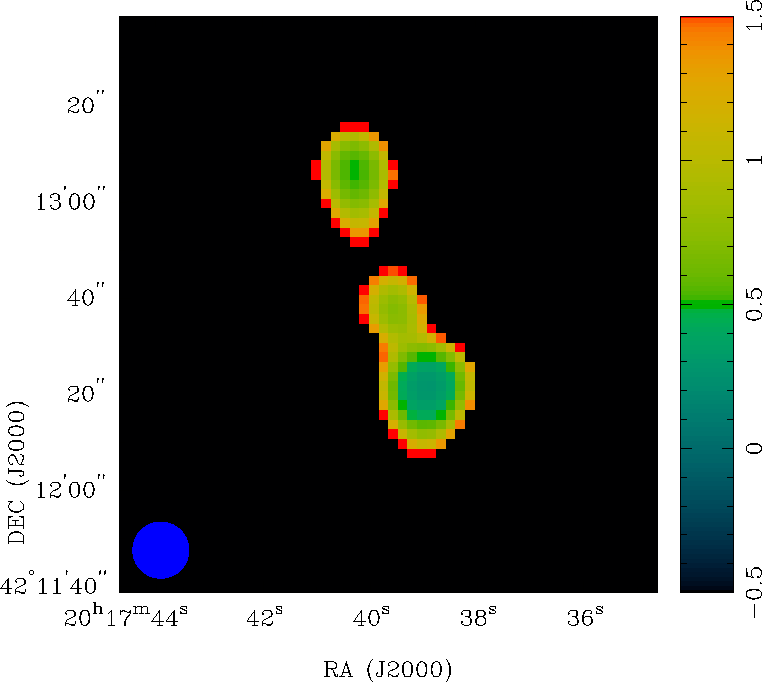}\\ 
    \includegraphics[width=0.3\textwidth]{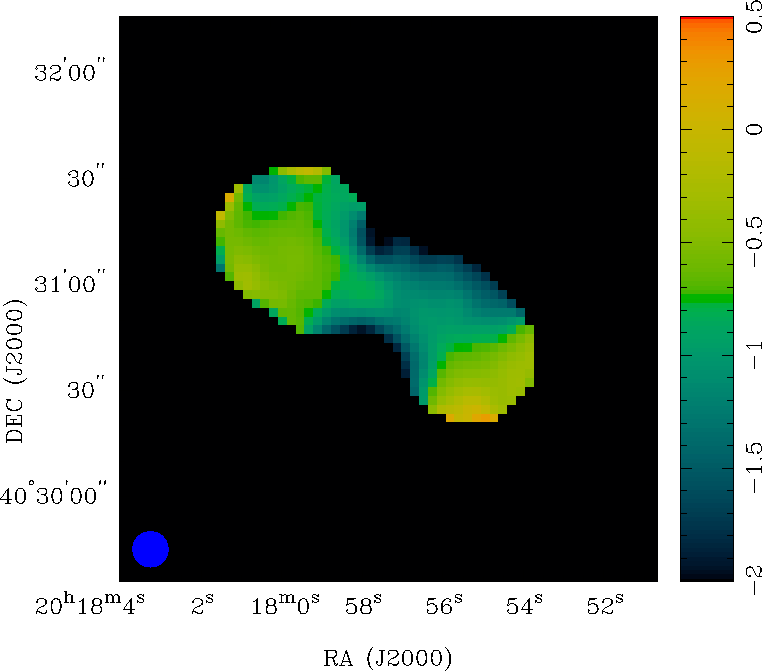} 
     \includegraphics[width=0.3\textwidth]{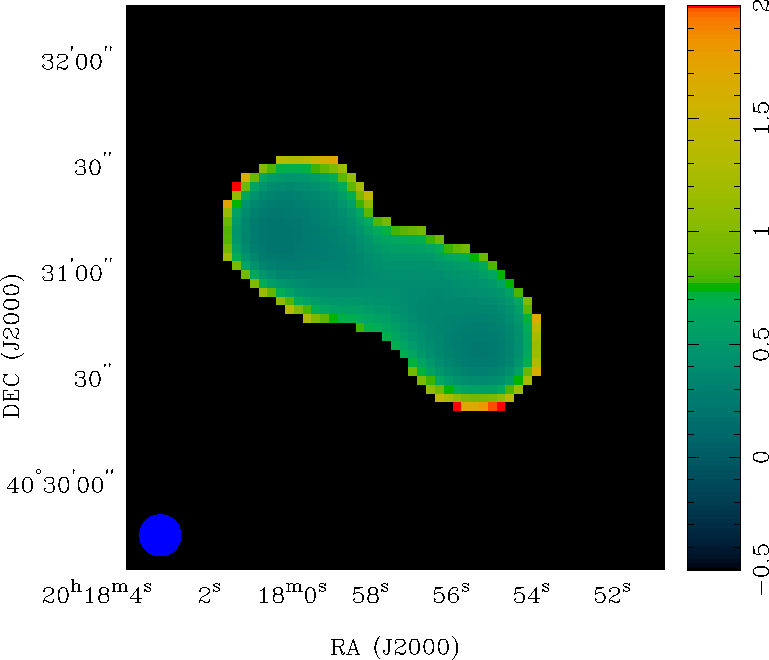}\\
     
    \includegraphics[width=0.3\textwidth]{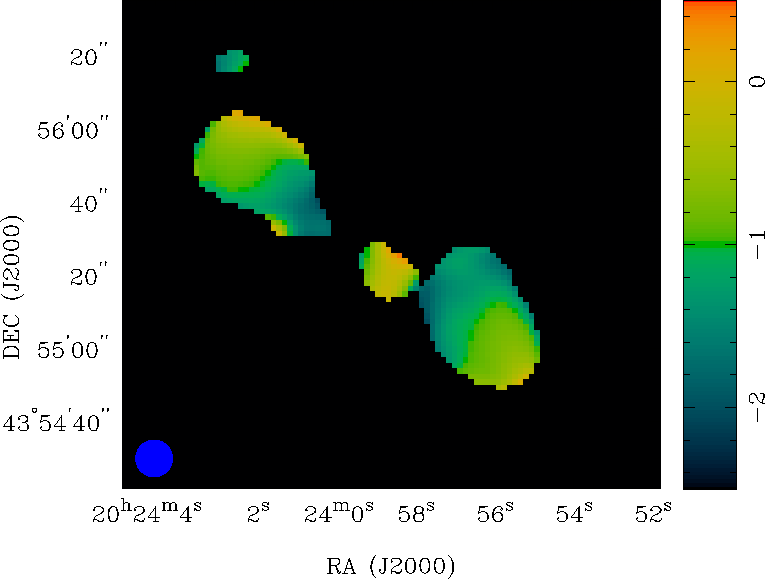} 
    \includegraphics[width=0.3\textwidth]{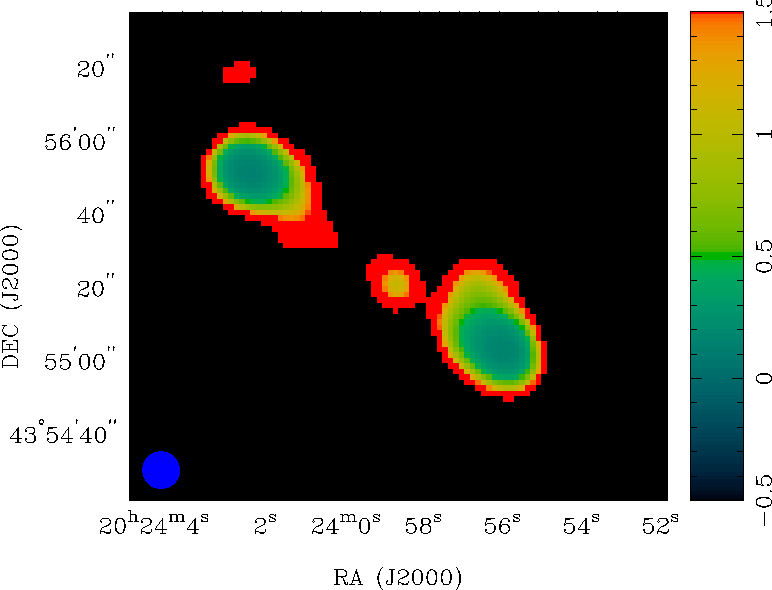}\\ 
    
    \includegraphics[width=0.3\textwidth]{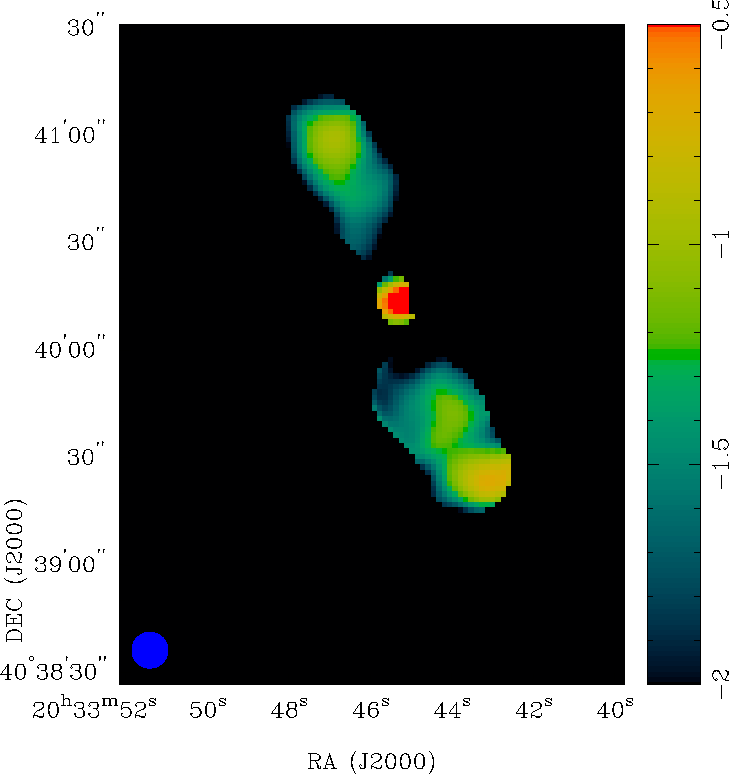} 
     \includegraphics[width=0.3\textwidth]{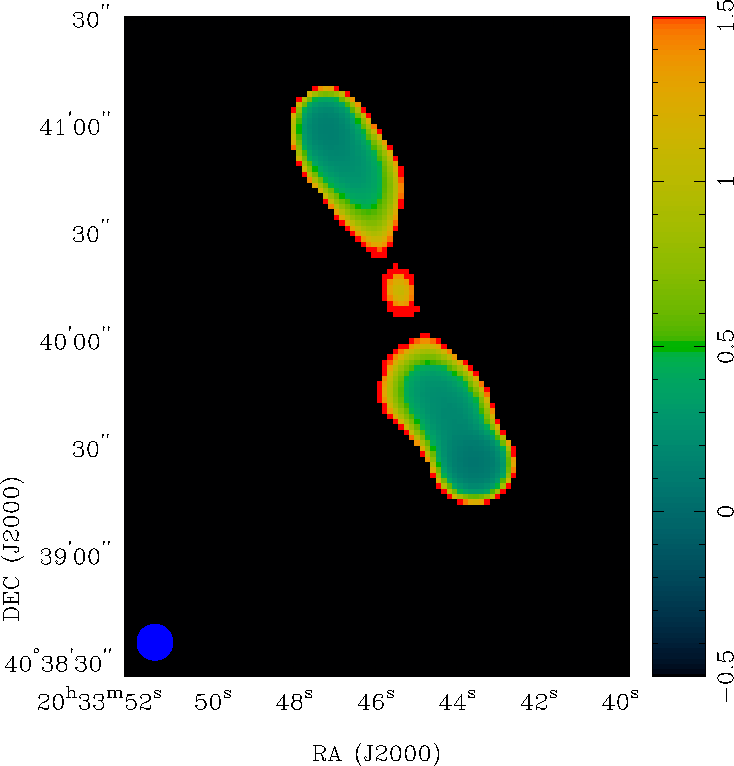} 
      \caption{Spectral index (left) and error maps (right) of J201739+421252 (\#4, first row), J201758+403100 (\#5, (second row), J202359+435525 (\#15, third  row), and J203345+404015 (\#31, fourth row) sources. The synthesised beam is shown at the bottom-left corner of each image.}
    \label{fig:sp-galaxies}
 \end{center}
\end{figure*}

\setcounter{figure}{0}
\renewcommand{\thefigure}{C\arabic{figure}}

\begin{figure*}
\begin{center}
    \includegraphics[width=0.3\textwidth]{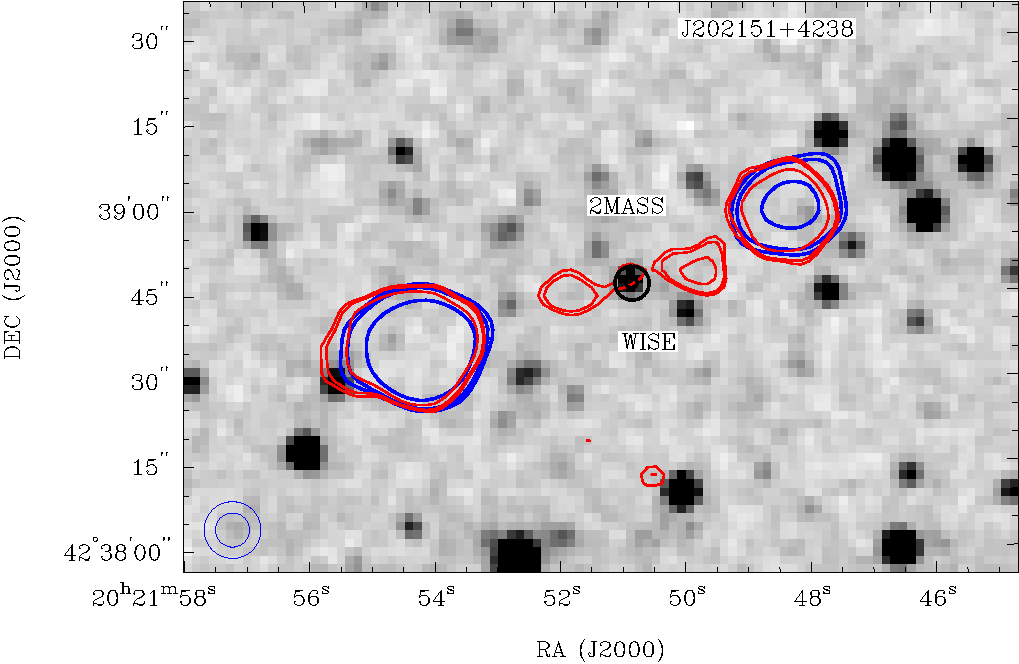}
    \includegraphics[width=0.3\textwidth]{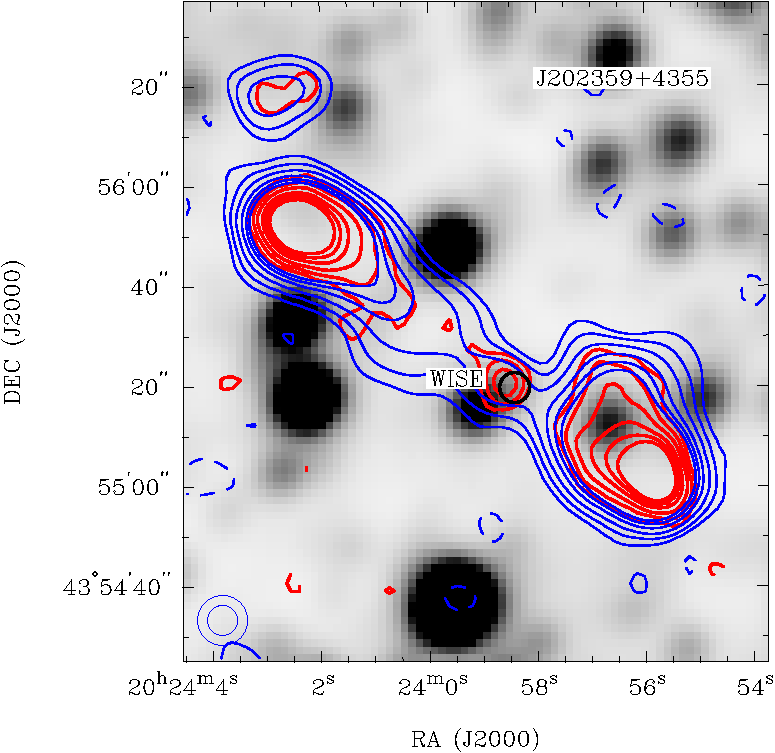}
    \includegraphics[width=0.3\textwidth]{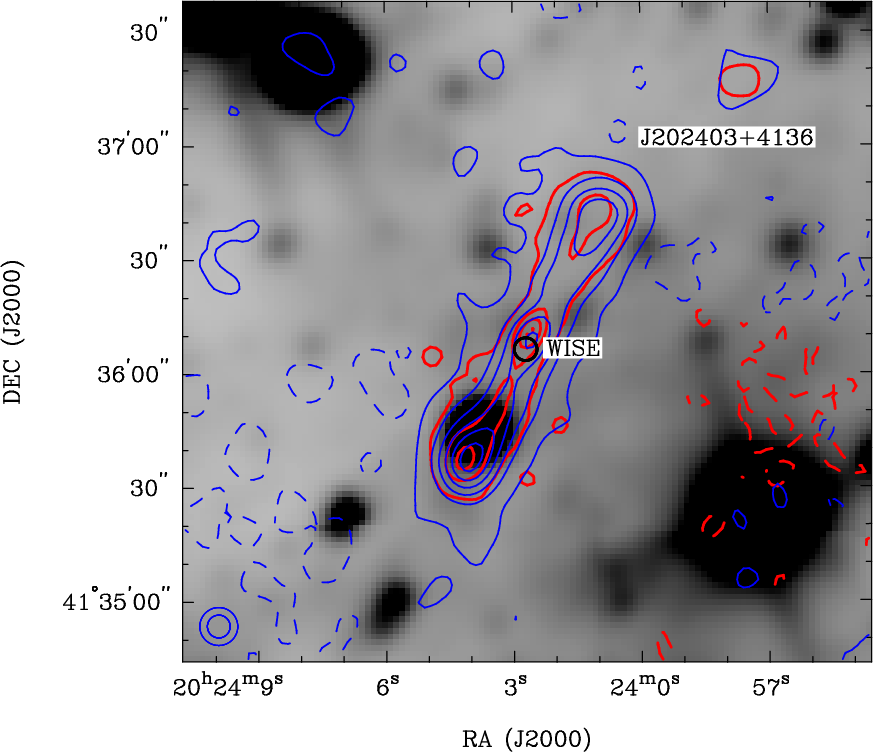}\\
    \includegraphics[width=0.3\textwidth]{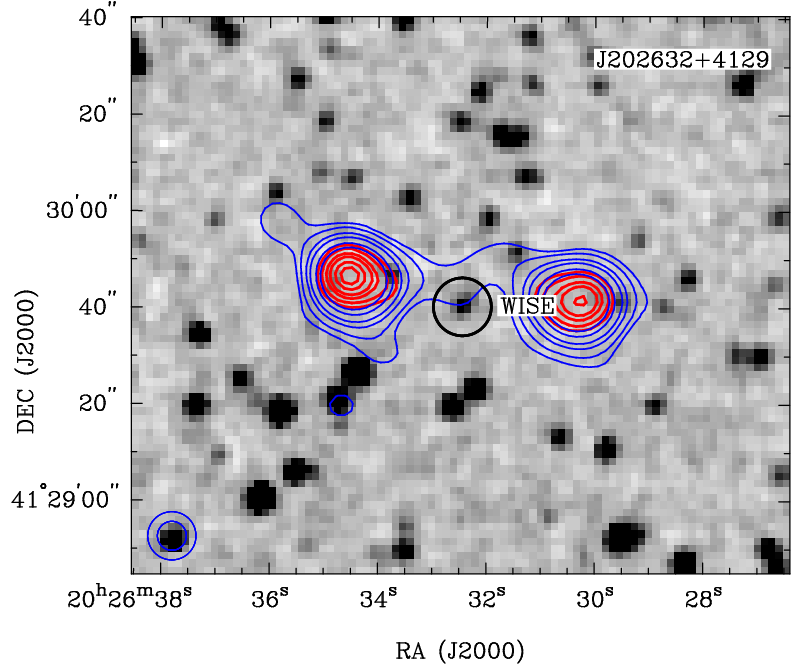}
    \includegraphics[width=0.3\textwidth]{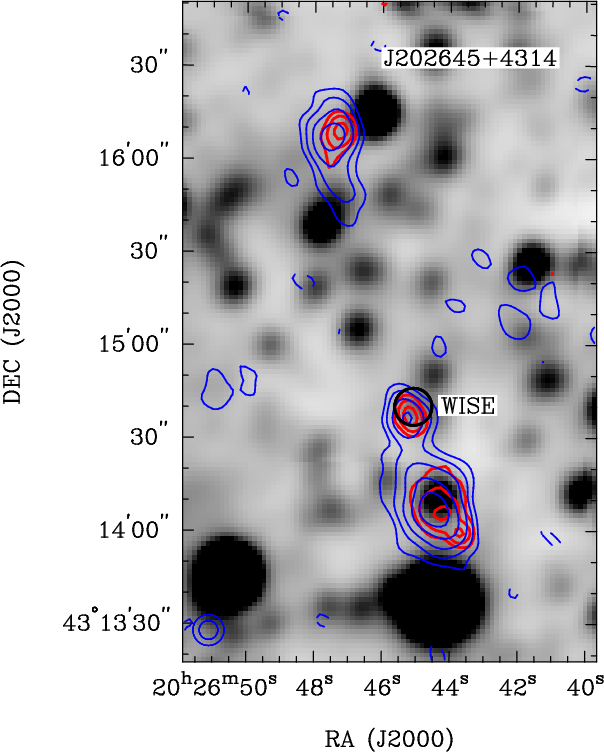}
    \includegraphics[width=0.3\textwidth]{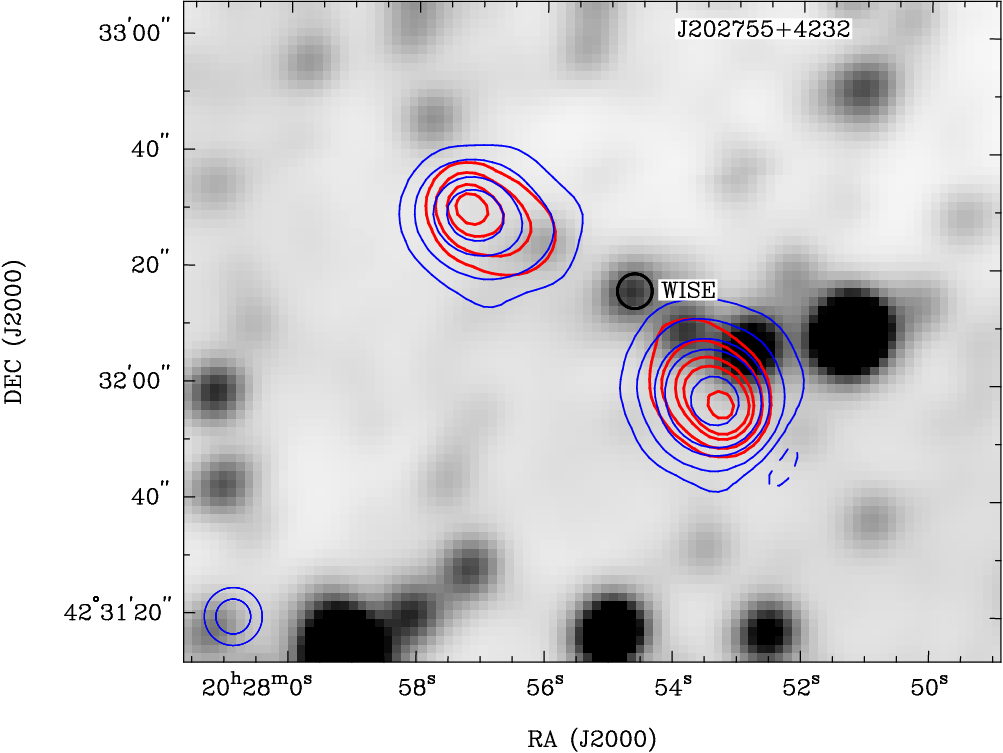}\\
    \includegraphics[width=0.3\textwidth]{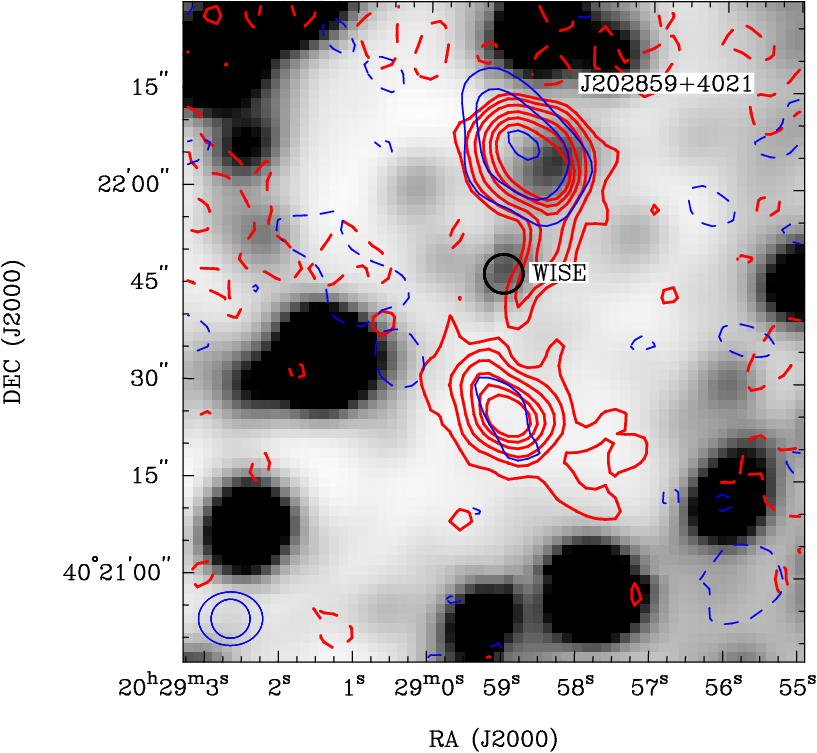}
    \includegraphics[width=0.3\textwidth]{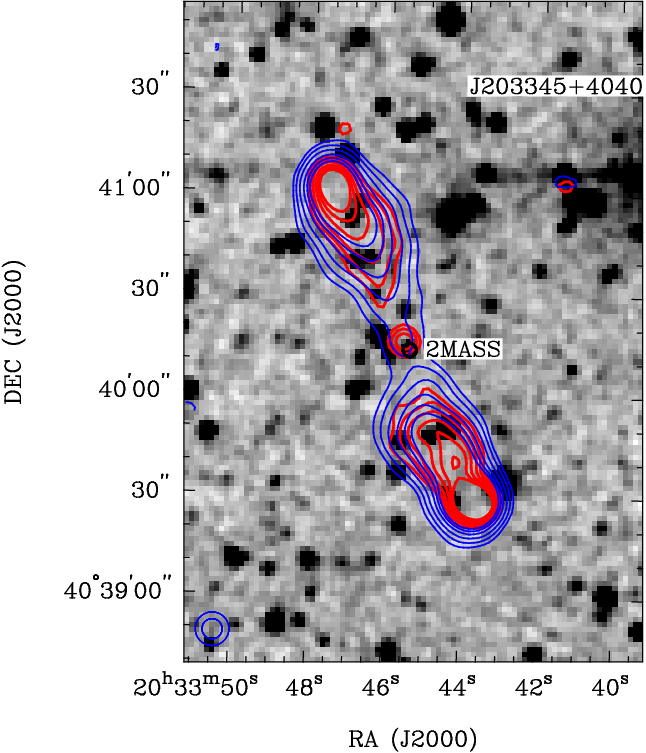}
    \includegraphics[width=0.3\textwidth]{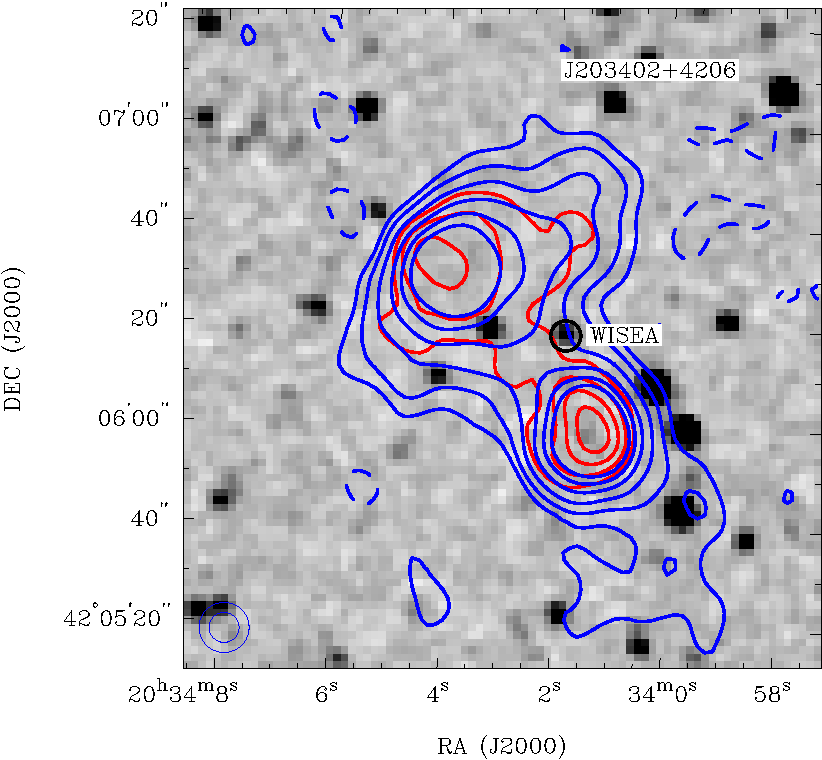}
       \caption{GMRT contours at 610~MHz (red) and 325~MHz (blue) overlaid on WISE (W1-band)/2MASS (3.4$\mu$m) infrared images. The contour levels at 325~MHz are the same shown in Figs.~\ref{fig:galaxitas1}-\ref{fig:galaxitas6}. The contour levels at 610~MHz are:
     \#10-J202151+423847: $-2.5$, 2.5, 3, and 5, in units of $\sigma$(=0.13~mJy\,beam$^{-1}$).
     \#15-J202352+435525: $-2.5$, 2.5, 5, 9, 15, 21, and 27, in units of $\sigma$(=0.1~mJy\,beam$^{-1}$).
     \#16-J202403+413611: $-2.5$, 2.5, 5, and  15, in units of $\sigma$(=0.15~mJy\,beam$^{-1}$).
     \#21-J202632+412942: $-2.5$, 2.5, 5, 10, 15, 25, and 35, in units of $\sigma$(=0.1~mJy\,beam$^{-1}$).
     \#22-J202645+431435: $-2.5$, 2.5, 5, 9, and 15, in units of $\sigma$(=0.1~mJy\,beam$^{-1}$).
     \#24-J202755+423215: $-2.5$, 2.5, 30, 50, and 130, in units of $\sigma$(=0.1~mJy\,beam$^{-1}$).
     \#26-J202859+402141: $-2.5$, 2.5, 9, 15, 21, and 27, in units of $\sigma$(=0.1~mJy\,beam$^{-1}$).
     \#31-J203345+404015: $-2.5$, 2.5, 9, 15, 21, and 27, in units of $\sigma$(=0.2~mJy\,beam$^{-1}$).
     \#32-J203402+420615: $-2.5$, 2.5, 9, 15, 21, and 27, in units of $\sigma$(=0.1~mJy\,beam$^{-1}$).
     The synthesised beams are shown at the bottom-left corner of each image.}
     
  \label{fig:IR+radio}
 \end{center}
\end{figure*}

\begin{figure*}
\begin{center} 
   \includegraphics[width=0.3\textwidth]{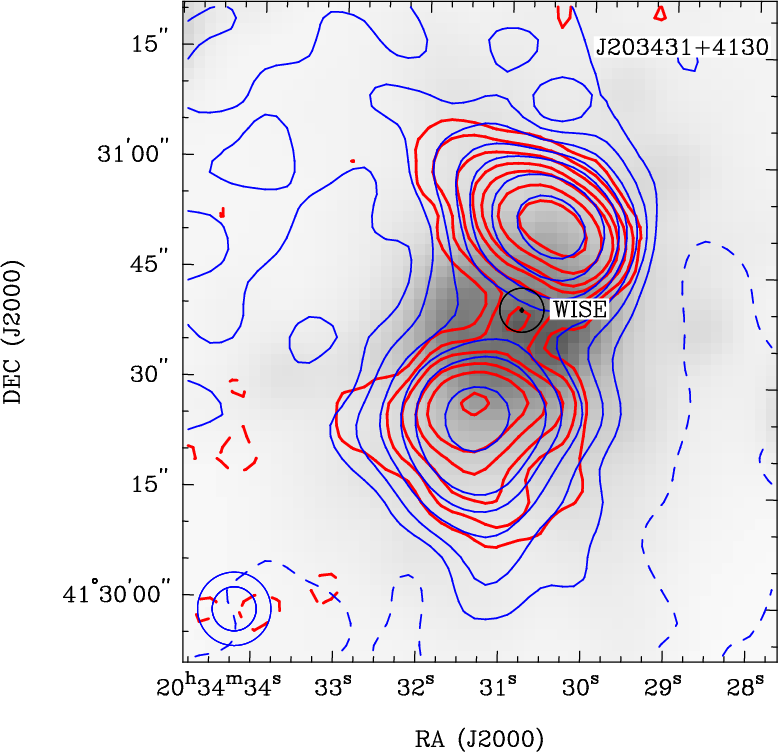}
    \includegraphics[width=0.3\textwidth]{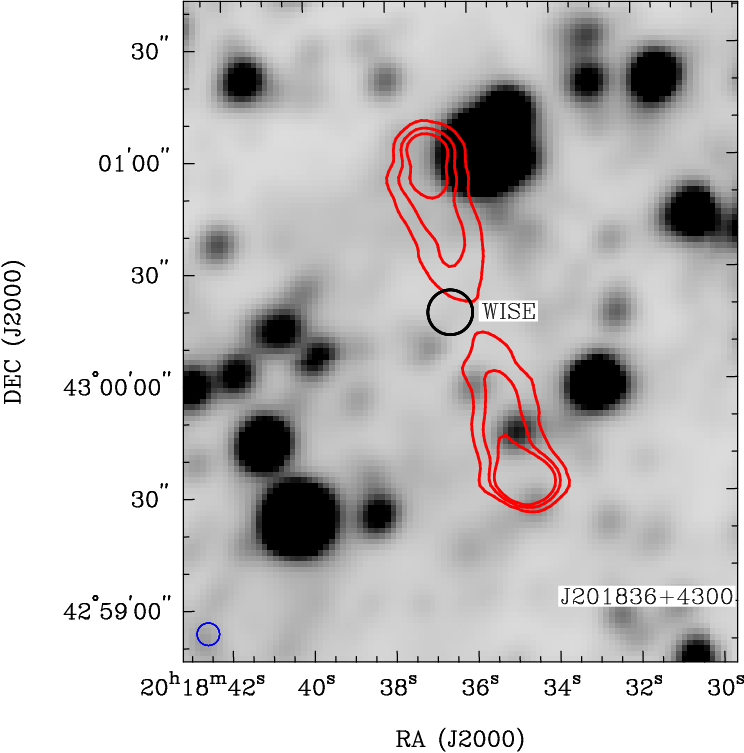}
     \includegraphics[width=0.3\textwidth]{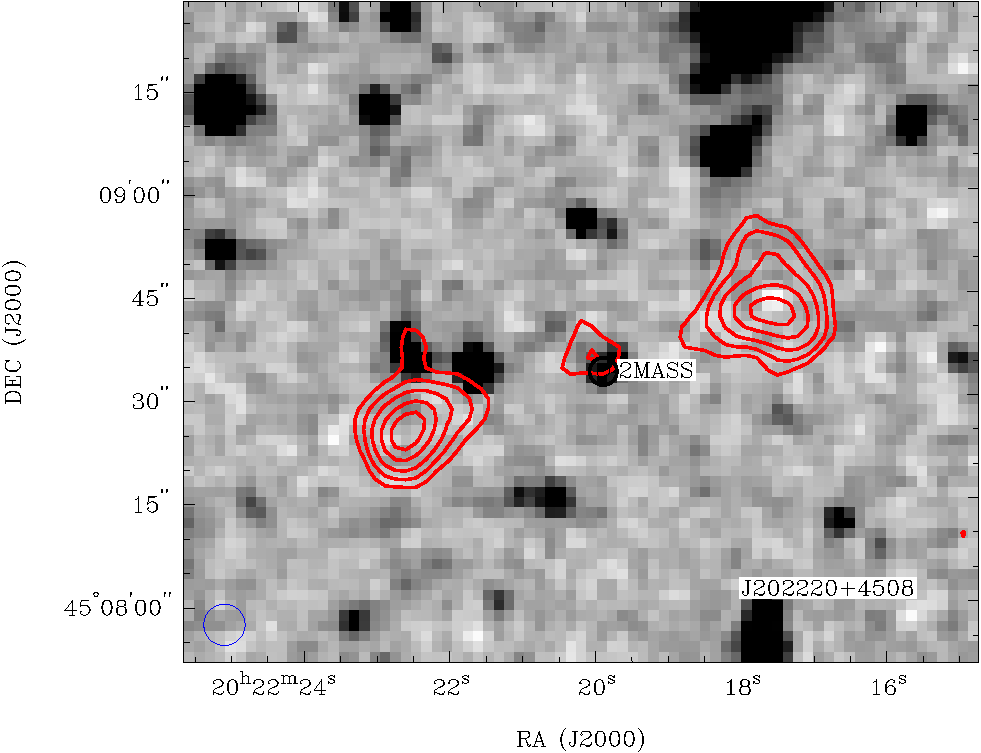}
     
      \caption{GMRT contours at 610~MHz (red) and 325~MHz (blue) overlaid on WISE/2MASS infrared images. The contour levels at 325~MHz are the same shown in Figs.~\ref{fig:galaxitas1}-\ref{fig:galaxitas6}. The contour levels at 610~MHz are:
    \#35-J203431+413037: $-2.5$, 2.5, 9, 15, 21, 35, and  55, in units of $\sigma$(=0.12~mJy\,beam$^{-1}$).
      \#40-J201836+430018: $-2.5$, 2.5, 10, and 20, in units of $\sigma$(=0.5~mJy\,beam$^{-1}$).
       \#42-J203431+413035: $-2.5$, 2.5, 5, 9, 15, 21, 35, 55, and 85, in units of $\sigma$(=0.15~mJy\,beam$^{-1}$).
     The synthesised beams are shown at the bottom-left corner of each image.}
 \label{fig:IR+radio1}
 \end{center}
\end{figure*}



\bibliographystyle{pasa-mnras}
\bibliography{dl-bib.bib}

\begin{thebibliography}{}
\makeatletter
\relax
\def\mn@urlcharsother{\let\do\@makeother \do\$\do\&\do\#\do\^\do\_\do\%\do\~}
\definecolor{darkblue}{rgb}{0,0,0.597656}
\def\mndoi{\begingroup\mn@urlcharsother \@ifnextchar [ {\mndoi@} {\mndoi@[]}}
\def\mndoi@[#1]#2{\def\@tempa{#1}\ifx\@tempa\@empty \href
  {http://dx.doi.org/#2} {\textcolor{darkblue}{doi:#2}}\else \href
  {http://dx.doi.org/#2} {\textcolor{darkblue}{#1}}\fi \endgroup}
\def\mn@eprint#1#2{\mn@eprint@#1:#2::\@nil}
\def\mn@eprint@arXiv#1{\href {http://arxiv.org/abs/#1} {{\tt arXiv:#1}}}
\def\mn@eprint@dblp#1{\href {http://dblp.uni-trier.de/rec/bibtex/#1.xml}
  {dblp:#1}}
\def\mn@eprint@#1:#2:#3:#4\@nil{\def\@tempa {#1}\def\@tempb {#2}\def\@tempc
  {#3}\ifx \@tempc \@empty \let \@tempc \@tempb \let \@tempb \@tempa \fi \ifx
  \@tempb \@empty \def\@tempb {arXiv}\fi \@ifundefined
  {mn@eprint@\@tempb}{\@tempb:\@tempc}{\expandafter \expandafter \csname
  mn@eprint@\@tempb\endcsname \expandafter{\@tempc}}}

\bibitem[\protect\citeauthoryear{{Albacete Colombo}, {Caramazza}, {Flaccomio},
  {Micela}  \& {Sciortino}}{{Albacete Colombo} et~al.}{2007}]{Albacete-2007}
{Albacete Colombo} J.~F.,  {Caramazza} M.,  {Flaccomio} E.,  {Micela} G.,
  {Sciortino} S.,  2007, VizieR Online Data Catalog, \href
  {https://ui.adsabs.harvard.edu/abs/2007yCat..34740495A} {pp J/A+A/474/495}

\bibitem[\protect\citeauthoryear{{Anderson}, {Armentrout}, {Johnstone},
  {Bania}, {Balser}, {Wenger}  \& {Cunningham}}{{Anderson}
  et~al.}{2015}]{Anderson-2015}
{Anderson} L.~D.,  {Armentrout} W.~P.,  {Johnstone} B.~M.,  {Bania} T.~M.,
  {Balser} D.~S.,  {Wenger} T.~V.,   {Cunningham} V.,  2015, \mndoi [\apjs]
  {10.1088/0067-0049/221/2/26}, \href
  {https://ui.adsabs.harvard.edu/abs/2015ApJS..221...26A} {221, 26}

\bibitem[\protect\citeauthoryear{{Anglada}, {Rodr{\'\i}guez}  \&
  {Carrasco-Gonz{\'a}lez}}{{Anglada} et~al.}{2018}]{Anglada2018}
{Anglada} G.,  {Rodr{\'\i}guez} L.~F.,   {Carrasco-Gonz{\'a}lez} C.,  2018,
  \mndoi [\aapr] {10.1007/s00159-018-0107-z}, \href
  {https://ui.adsabs.harvard.edu/abs/2018A&ARv..26....3A} {26, 3}

\bibitem[\protect\citeauthoryear{{Benaglia}, {De Becker}, {Ishwara-Chandra},
  {Intema}  \& {Isequilla}}{{Benaglia} et~al.}{2020a}]{Benaglia-2020a}
{Benaglia} P.,  {De Becker} M.,  {Ishwara-Chandra} C.~H.,  {Intema} H.~T.,
  {Isequilla} N.~L.,  2020a, \mndoi [\pasa] {10.1017/pasa.2020.21}, \href
  {https://ui.adsabs.harvard.edu/abs/2020PASA...37...30B} {37, e030}

\bibitem[\protect\citeauthoryear{{Benaglia}, {Ishwara-Chandra}, {Intema},
  {Colazo}  \& {Gaikwad}}{{Benaglia} et~al.}{2020b}]{Benaglia-2020b}
{Benaglia} P.,  {Ishwara-Chandra} C.~H.,  {Intema} H.,  {Colazo} M.~E.,
  {Gaikwad} M.,  2020b, \mndoi [\aap] {10.1051/0004-6361/202037916}, \href
  {https://ui.adsabs.harvard.edu/abs/2020A&A...642A.136B} {642, A136}

\bibitem[\protect\citeauthoryear{{Benaglia}, {Ishwara-Chandra}, {Paredes},
  {Intema}, {Colazo}  \& {Isequilla}}{{Benaglia} et~al.}{2021}]{Benaglia-2021a}
{Benaglia} P.,  {Ishwara-Chandra} C.~H.,  {Paredes} J.~M.,  {Intema} H.~T.,
  {Colazo} M.~E.,   {Isequilla} N.~L.,  2021, \mndoi [\apjs]
  {10.3847/1538-4365/abc891}, \href
  {https://ui.adsabs.harvard.edu/abs/2021ApJS..252...17B} {252, 17}

\bibitem[\protect\citeauthoryear{{Bennett}, {Lawrence}, {Burke}, {Hewitt}  \&
  {Mahoney}}{{Bennett} et~al.}{1986}]{Bennet-1986}
{Bennett} C.~L.,  {Lawrence} C.~R.,  {Burke} B.~F.,  {Hewitt} J.~N.,
  {Mahoney} J.,  1986, \mndoi [\apjs] {10.1086/191108}, \href
  {https://ui.adsabs.harvard.edu/abs/1986ApJS...61....1B} {61, 1}

\bibitem[\protect\citeauthoryear{{Brocksopp}, {Kaiser}, {Schoenmakers}  \& {de
  Bruyn}}{{Brocksopp} et~al.}{2011}]{Brocksopp-2011}
{Brocksopp} C.,  {Kaiser} C.~R.,  {Schoenmakers} A.~P.,   {de Bruyn} A.~G.,
  2011, \mndoi [\mnras] {10.1111/j.1365-2966.2010.17456.x}, \href
  {https://ui.adsabs.harvard.edu/abs/2011MNRAS.410..484B} {410, 484}

\bibitem[\protect\citeauthoryear{{Brown}, {Duncan}, {Landt}, {Kirk}, {Ricci},
  {Kamraj}, {Salvato}  \& {Ananna}}{{Brown} et~al.}{2019}]{Brown-2019}
{Brown} M.~J.~I.,  {Duncan} K.~J.,  {Landt} H.,  {Kirk} M.,  {Ricci} C.,
  {Kamraj} N.,  {Salvato} M.,   {Ananna} T.,  2019, \mndoi [\mnras]
  {10.1093/mnras/stz2324}, \href
  {https://ui.adsabs.harvard.edu/abs/2019MNRAS.489.3351B} {489, 3351}

\bibitem[\protect\citeauthoryear{{Butt}, {Combi}, {Drake}, {Finley},
  {Konopelko}, {Lister}, {Rodriguez}  \& {Shepherd}}{{Butt}
  et~al.}{2007}]{Butt-2007}
{Butt} Y.~M.,  {Combi} J.~A.,  {Drake} J.,  {Finley} J.~P.,  {Konopelko} A.,
  {Lister} M.,  {Rodriguez} J.,   {Shepherd} D.,  2007, in {Ritz} S.,
  {Michelson} P.,   {Meegan} C.~A.,  eds,  American Institute of Physics
  Conference Series Vol. 921, The First GLAST Symposium. pp 429--430
  (\mn@eprint {arXiv} {astro-ph/0703017}), \mndoi{10.1063/1.2757386}

\bibitem[\protect\citeauthoryear{{Carrasco-Gonz{\'a}lez}, {Rodr{\'\i}guez},
  {Anglada}, {Mart{\'\i}}, {Torrelles}  \& {Osorio}}{{Carrasco-Gonz{\'a}lez}
  et~al.}{2010}]{carrasco2010}
{Carrasco-Gonz{\'a}lez} C.,  {Rodr{\'\i}guez} L.~F.,  {Anglada} G.,
  {Mart{\'\i}} J.,  {Torrelles} J.~M.,   {Osorio} M.,  2010, \mndoi [Science]
  {10.1126/science.1195589}, \href
  {https://ui.adsabs.harvard.edu/abs/2010Sci...330.1209C} {330, 1209}

\bibitem[\protect\citeauthoryear{{Condon}, {Cotton}, {Greisen}, {Yin},
  {Perley}, {Taylor}  \& {Broderick}}{{Condon} et~al.}{1998}]{NVSS}
{Condon} J.~J.,  {Cotton} W.~D.,  {Greisen} E.~W.,  {Yin} Q.~F.,  {Perley}
  R.~A.,  {Taylor} G.~B.,   {Broderick} J.~J.,  1998, \mndoi [\aj]
  {10.1086/300337}, \href
  {https://ui.adsabs.harvard.edu/abs/1998AJ....115.1693C} {115, 1693}

\bibitem[\protect\citeauthoryear{{Douglas}, {Bash}, {Bozyan}, {Torrence}  \&
  {Wolfe}}{{Douglas} et~al.}{1996a}]{Doublges-1996}
{Douglas} J.~N.,  {Bash} F.~N.,  {Bozyan} F.~A.,  {Torrence} G.~W.,   {Wolfe}
  C.,  1996a, VizieR Online Data Catalog, \href
  {https://ui.adsabs.harvard.edu/abs/1996yCat.8042....0D} {p. VIII/42}

\bibitem[\protect\citeauthoryear{{Douglas}, {Bash}, {Bozyan}, {Torrence}  \&
  {Wolfe}}{{Douglas} et~al.}{1996b}]{Douglas-1996}
{Douglas} J.~N.,  {Bash} F.~N.,  {Bozyan} F.~A.,  {Torrence} G.~W.,   {Wolfe}
  C.,  1996b, \mndoi [\aj] {10.1086/117932}, \href
  {https://ui.adsabs.harvard.edu/abs/1996AJ....111.1945D} {111, 1945}

\bibitem[\protect\citeauthoryear{{Elvis}, {Maccacaro}, {Wilson}, {Ward},
  {Penston}, {Fosbury}  \& {Perola}}{{Elvis}
  et~al.}{1978}]{1978MNRAS.183..129E}
{Elvis} M.,  {Maccacaro} T.,  {Wilson} A.~S.,  {Ward} M.~J.,  {Penston} M.~V.,
  {Fosbury} R.~A.~E.,   {Perola} G.~C.,  1978, \mndoi [\mnras]
  {10.1093/mnras/183.2.129}, \href
  {https://ui.adsabs.harvard.edu/abs/1978MNRAS.183..129E} {183, 129}

\bibitem[\protect\citeauthoryear{{Evans} et~al.,}{{Evans}
  et~al.}{2010}]{Evans2010}
{Evans} I.~N.,  et~al., 2010, \mndoi [\apjs] {10.1088/0067-0049/189/1/37},
  \href {https://ui.adsabs.harvard.edu/abs/2010ApJS..189...37E} {189, 37}

\bibitem[\protect\citeauthoryear{{Fanaroff} \& {Riley}}{{Fanaroff} \&
  {Riley}}{1974}]{Fanaroff-1974}
{Fanaroff} B.~L.,  {Riley} J.~M.,  1974, \mndoi [\mnras]
  {10.1093/mnras/167.1.31P}, \href
  {https://ui.adsabs.harvard.edu/abs/1974MNRAS.167P..31F} {167, 31P}

\bibitem[\protect\citeauthoryear{{Ginzburg} \& {Syrovatskii}}{{Ginzburg} \&
  {Syrovatskii}}{1967}]{ginzburg1967}
{Ginzburg} V.~L.,  {Syrovatskii} S.~I.,  1967, in International Cosmic Ray
  Conference. p.~48

\bibitem[\protect\citeauthoryear{{Greisen}}{{Greisen}}{2003}]{greisen2003}
{Greisen} E.~W.,  2003, {AIPS, the VLA, and the VLBA}.
., p.~109, \mndoi{10.1007/0-306-48080-8_7}

\bibitem[\protect\citeauthoryear{{G{\"u}rkan}, {Hardcastle}  \&
  {Jarvis}}{{G{\"u}rkan} et~al.}{2014}]{gurkan2014}
{G{\"u}rkan} G.,  {Hardcastle} M.~J.,   {Jarvis} M.~J.,  2014, \mndoi [\mnras]
  {10.1093/mnras/stt2264}, \href
  {https://ui.adsabs.harvard.edu/abs/2014MNRAS.438.1149G} {438, 1149}

\bibitem[\protect\citeauthoryear{{Ibar}, {Ivison}, {Best}, {Coppin}, {Pope},
  {Smail}  \& {Dunlop}}{{Ibar} et~al.}{2010}]{ibar2010}
{Ibar} E.,  {Ivison} R.~J.,  {Best} P.~N.,  {Coppin} K.,  {Pope} A.,  {Smail}
  I.,   {Dunlop} J.~S.,  2010, \mndoi [\mnras]
  {10.1111/j.1745-3933.2009.00786.x}, \href
  {https://ui.adsabs.harvard.edu/abs/2010MNRAS.401L..53I} {401, L53}

\bibitem[\protect\citeauthoryear{{Intema}}{{Intema}}{2014}]{Intema-2014}
{Intema} H.~T.,  2014, in Astronomical Society of India Conference Series.
  p.~469 (\mn@eprint {arXiv} {1402.4889})

\bibitem[\protect\citeauthoryear{{Isequilla}, {Fern{\'a}ndez-L{\'o}pez},
  {Benaglia}, {Ishwara-Chandra}  \& {del Palacio}}{{Isequilla}
  et~al.}{2019}]{Isequilla-2019}
{Isequilla} N.~L.,  {Fern{\'a}ndez-L{\'o}pez} M.,  {Benaglia} P.,
  {Ishwara-Chandra} C.~H.,   {del Palacio} S.,  2019, \mndoi [\aap]
  {10.1051/0004-6361/201935179}, \href
  {https://ui.adsabs.harvard.edu/abs/2019A&A...627A..58I} {627, A58}

\bibitem[\protect\citeauthoryear{{Isequilla}, {Benaglia}, {Ishwara-Chandra}  \&
  {Intema}}{{Isequilla} et~al.}{2020}]{Isequilla-2020}
{Isequilla} N.~L.,  {Benaglia} P.,  {Ishwara-Chandra} C.~H.,   {Intema} H.,
  2020, Boletin de la Asociacion Argentina de Astronomia La Plata Argentina,
  \href {https://ui.adsabs.harvard.edu/abs/2020BAAA...61B.124I} {61B, 124}

\bibitem[\protect\citeauthoryear{{Jones}, {Garwood}  \& {Dickey}}{{Jones}
  et~al.}{1988}]{Garwood-1988}
{Jones} T.~J.,  {Garwood} R.,   {Dickey} J.~M.,  1988, \mndoi [\apj]
  {10.1086/166313}, \href
  {https://ui.adsabs.harvard.edu/abs/1988ApJ...328..559J} {328, 559}

\bibitem[\protect\citeauthoryear{{Kn{\"o}dlseder}, {Cervi{\~n}o}, {Le Duigou},
  {Meynet}, {Schaerer}  \& {von Ballmoos}}{{Kn{\"o}dlseder}
  et~al.}{2002}]{K-2002}
{Kn{\"o}dlseder} J.,  {Cervi{\~n}o} M.,  {Le Duigou} J.~M.,  {Meynet} G.,
  {Schaerer} D.,   {von Ballmoos} P.,  2002, \mndoi [\aap]
  {10.1051/0004-6361:20020799}, \href
  {https://ui.adsabs.harvard.edu/abs/2002A&A...390..945K} {390, 945}

\bibitem[\protect\citeauthoryear{{Laing}, {Riley}  \& {Longair}}{{Laing}
  et~al.}{1983}]{Laing1983}
{Laing} R.~A.,  {Riley} J.~M.,   {Longair} M.~S.,  1983, \mndoi [\mnras]
  {10.1093/mnras/204.1.151}, \href
  {https://ui.adsabs.harvard.edu/abs/1983MNRAS.204..151L} {204, 151}

\bibitem[\protect\citeauthoryear{{Lanzuisi} et~al.,}{{Lanzuisi}
  et~al.}{2013}]{2013MNRAS.431..978L}
{Lanzuisi} G.,  et~al., 2013, \mndoi [\mnras] {10.1093/mnras/stt222}, \href
  {https://ui.adsabs.harvard.edu/abs/2013MNRAS.431..978L} {431, 978}

\bibitem[\protect\citeauthoryear{Leahy \& Williams}{Leahy \&
  Williams}{1984}]{Leahy-1984}
Leahy J.~P.,  Williams A.~G.,  1984, \mndoi [Monthly Notices of the Royal
  Astronomical Society] {10.1093/mnras/210.4.929}, 210, 929

\bibitem[\protect\citeauthoryear{{Mart{\'\i}}, {Paredes}, {Ishwara Chandra}  \&
  {Bosch-Ramon}}{{Mart{\'\i}} et~al.}{2007}]{Marti-2007}
{Mart{\'\i}} J.,  {Paredes} J.~M.,  {Ishwara Chandra} C.~H.,   {Bosch-Ramon}
  V.,  2007, \mndoi [\aap] {10.1051/0004-6361:20077712}, \href
  {https://ui.adsabs.harvard.edu/abs/2007A&A...472..557M} {472, 557}

\bibitem[\protect\citeauthoryear{{Masetti} et~al.,}{{Masetti}
  et~al.}{2013}]{2013A&A...556A.120M}
{Masetti} N.,  et~al., 2013, \mndoi [\aap] {10.1051/0004-6361/201322026}, \href
  {https://ui.adsabs.harvard.edu/abs/2013A&A...556A.120M} {556, A120}

\bibitem[\protect\citeauthoryear{{Morford} et~al.,}{{Morford}
  et~al.}{2020}]{morford2020}
{Morford} J.~C.,  et~al., 2020, \mndoi [\aap] {10.1051/0004-6361/201731379},
  \href {https://ui.adsabs.harvard.edu/abs/2020A&A...637A..64M} {637, A64}

\bibitem[\protect\citeauthoryear{{M{\"u}nch} \& {Morgan}}{{M{\"u}nch} \&
  {Morgan}}{1953}]{Munch-1953A}
{M{\"u}nch} L.,  {Morgan} W.~W.,  1953, \mndoi [\apj] {10.1086/145737}, \href
  {https://ui.adsabs.harvard.edu/abs/1953ApJ...118..161M} {118, 161}

\bibitem[\protect\citeauthoryear{{Neugebauer} et~al.,}{{Neugebauer}
  et~al.}{1984}]{IRAS-1984}
{Neugebauer} G.,  et~al., 1984, \mndoi [\apjl] {10.1086/184209}, \href
  {https://ui.adsabs.harvard.edu/abs/1984ApJ...278L...1N} {278, L1}

\bibitem[\protect\citeauthoryear{{Ocran}, {Taylor}, {Vaccari}, {Ishwara-Chand
  ra}  \& {Prandoni}}{{Ocran} et~al.}{2020}]{ocran2020a}
{Ocran} E.~F.,  {Taylor} A.~R.,  {Vaccari} M.,  {Ishwara-Chand ra} C.~H.,
  {Prandoni} I.,  2020, \mndoi [\mnras] {10.1093/mnras/stz2954}, \href
  {https://ui.adsabs.harvard.edu/abs/2020MNRAS.491.1127O} {491, 1127}

\bibitem[\protect\citeauthoryear{{Osterbrock}}{{Osterbrock}}{1989}]{osterbrock1989}
{Osterbrock} D.~E.,  1989, {Astrophysics of gaseous nebulae and active galactic
  nuclei}.
.

\bibitem[\protect\citeauthoryear{{Pal} \& {Kumari}}{{Pal} \&
  {Kumari}}{2021}]{pal-2021}
{Pal} S.,  {Kumari} S.,  2021, arXiv e-prints, \href
  {https://ui.adsabs.harvard.edu/abs/2021arXiv210315199P} {p. arXiv:2103.15199}

\bibitem[\protect\citeauthoryear{{Pratap} \& {McIntosh}}{{Pratap} \&
  {McIntosh}}{2005}]{Pratap2005}
{Pratap} P.,  {McIntosh} G.,  2005, \mndoi [American Journal of Physics]
  {10.1119/1.1858485}, \href
  {https://ui.adsabs.harvard.edu/abs/2005AmJPh..73..399P} {73, 399}

\bibitem[\protect\citeauthoryear{{Reddish}, {Lawrence}  \& {Pratt}}{{Reddish}
  et~al.}{1966}]{Reddish-1966}
{Reddish} V.~C.,  {Lawrence} L.~C.,   {Pratt} N.~M.,  1966, Publications of the
  Royal Observatory of Edinburgh, \href
  {https://ui.adsabs.harvard.edu/abs/1966PROE....5..111R} {5, 111}

\bibitem[\protect\citeauthoryear{{Reipurth} \& {Schneider}}{{Reipurth} \&
  {Schneider}}{2008}]{Reipurth-2008}
{Reipurth} B.,  {Schneider} N.,  2008, {Star Formation and Young Clusters in
  Cygnus}.
., p.~36

\bibitem[\protect\citeauthoryear{{Saikia}, {Thomasson}, {Roy}, {Pedlar}  \&
  {Muxlow}}{{Saikia} et~al.}{2004}]{Saikia2004}
{Saikia} D.~J.,  {Thomasson} P.,  {Roy} S.,  {Pedlar} A.,   {Muxlow} T.~W.~B.,
  2004, \mndoi [\mnras] {10.1111/j.1365-2966.2004.08244.x}, \href
  {https://ui.adsabs.harvard.edu/abs/2004MNRAS.354..827S} {354, 827}

\bibitem[\protect\citeauthoryear{{Saxena} et~al.,}{{Saxena}
  et~al.}{2018}]{saxena-2018}
{Saxena} A.,  et~al., 2018, \mndoi [\mnras] {10.1093/mnras/sty1996}, \href
  {https://ui.adsabs.harvard.edu/abs/2018MNRAS.480.2733S} {480, 2733}

\bibitem[\protect\citeauthoryear{Schoenmakers, de Bruyn, Röttgering, van~der
  Laan  \& Kaiser}{Schoenmakers et~al.}{2000}]{Arno-2000}
Schoenmakers A.~P.,  de Bruyn A.~G.,  Röttgering H. J.~A.,  van~der Laan H.,
  Kaiser C.~R.,  2000, \mndoi [Monthly Notices of the Royal Astronomical
  Society] {10.1046/j.1365-8711.2000.03430.x}, 315, 371

\bibitem[\protect\citeauthoryear{{Schulte}}{{Schulte}}{1956a}]{Schulte-1956-b}
{Schulte} D.~H.,  1956a, \mndoi [\apj] {10.1086/146156}, \href
  {https://ui.adsabs.harvard.edu/abs/1956ApJ...123..250S} {123, 250}

\bibitem[\protect\citeauthoryear{{Schulte}}{{Schulte}}{1956b}]{Schulte-1956a}
{Schulte} D.~H.,  1956b, \mndoi [\apj] {10.1086/146256}, \href
  {https://ui.adsabs.harvard.edu/abs/1956ApJ...124..530S} {124, 530}

\bibitem[\protect\citeauthoryear{{Schulte}}{{Schulte}}{1958}]{Schulte-1958}
{Schulte} D.~H.,  1958, \mndoi [\apj] {10.1086/146513}, \href
  {https://ui.adsabs.harvard.edu/abs/1958ApJ...128...41S} {128, 41}

\bibitem[\protect\citeauthoryear{{Setia Gunawan}, {de Bruyn}, {van der Hucht}
  \& {Williams}}{{Setia Gunawan} et~al.}{2003}]{Gunawan-2003}
{Setia Gunawan} D. Y.~A.,  {de Bruyn} A.~G.,  {van der Hucht} K.~A.,
  {Williams} P.~M.,  2003, \mndoi [\apjs] {10.1086/377598}, \href
  {https://ui.adsabs.harvard.edu/abs/2003ApJS..149..123S} {149, 123}

\bibitem[\protect\citeauthoryear{{Skrutskie} et~al.,}{{Skrutskie}
  et~al.}{2006}]{2mass-2006}
{Skrutskie} M.~F.,  et~al., 2006, \mndoi [\aj] {10.1086/498708}, \href
  {https://ui.adsabs.harvard.edu/abs/2006AJ....131.1163S} {131, 1163}

\bibitem[\protect\citeauthoryear{{Taylor}, {Goss}, {Coleman}, {van Leeuwen}  \&
  {Wallace}}{{Taylor} et~al.}{1996}]{Taylor-1996}
{Taylor} A.~R.,  {Goss} W.~M.,  {Coleman} P.~H.,  {van Leeuwen} J.,   {Wallace}
  B.~J.,  1996, \mndoi [\apjs] {10.1086/192363}, \href
  {https://ui.adsabs.harvard.edu/abs/1996ApJS..107..239T} {107, 239}

\bibitem[\protect\citeauthoryear{{Werner} et~al.,}{{Werner}
  et~al.}{2004}]{spitzer-2004}
{Werner} M.~W.,  et~al., 2004, \mndoi [\apjs] {10.1086/422992}, \href
  {https://ui.adsabs.harvard.edu/abs/2004ApJS..154....1W} {154, 1}

\bibitem[\protect\citeauthoryear{{Williams}, {Dougherty}, {Davis}, {van der
  Hucht}, {Bode}  \& {Setia Gunawan}}{{Williams} et~al.}{1997}]{Williams1997}
{Williams} P.~M.,  {Dougherty} S.~M.,  {Davis} R.~J.,  {van der Hucht} K.~A.,
  {Bode} M.~F.,   {Setia Gunawan} D.~Y.~A.,  1997, \mndoi [\mnras]
  {10.1093/mnras/289.1.10}, \href
  {https://ui.adsabs.harvard.edu/abs/1997MNRAS.289...10W} {289, 10}

\bibitem[\protect\citeauthoryear{{Wright} et~al.,}{{Wright}
  et~al.}{2010}]{WISE-2010A}
{Wright} E.~L.,  et~al., 2010, \mndoi [\aj] {10.1088/0004-6256/140/6/1868},
  \href {https://ui.adsabs.harvard.edu/abs/2010AJ....140.1868W} {140, 1868}

\bibitem[\protect\citeauthoryear{{Zoonematkermani}, {Helfand}, {Becker},
  {White}  \& {Perley}}{{Zoonematkermani} et~al.}{1990}]{zoon-1990}
{Zoonematkermani} S.,  {Helfand} D.~J.,  {Becker} R.~H.,  {White} R.~L.,
  {Perley} R.~A.,  1990, \mndoi [\apjs] {10.1086/191496}, \href
  {https://ui.adsabs.harvard.edu/abs/1990ApJS...74..181Z} {74, 181}

\makeatother
\end{thebibliography}

\end{document}